\documentclass[12pt]{article}
\usepackage{amsfonts}
\usepackage{float}
\usepackage{placeins}
\usepackage{amsmath}
\usepackage{amssymb}
\usepackage[normalem]{ulem}
\usepackage{amsthm}
\usepackage{bbm}
\usepackage{dsfont}
\usepackage{adjustbox}
\usepackage{xcolor}
\usepackage{microtype}
\usepackage{graphicx}
\usepackage{booktabs} 
\usepackage{algorithm}
\usepackage{algpseudocode}
\usepackage{mathtools}
\usepackage{wrapfig}
\usepackage{subcaption}
\usepackage{multirow}

\usepackage{tikz}
\usepackage[left=1.12in, right=1.12in, top=1in, bottom=1in]{geometry}
\usepackage{natbib}
\usepackage{setspace}
\usepackage{hyperref}
\usepackage{ifthen}
\hypersetup{
  colorlinks=true,        
  linkcolor=blue,         
  citecolor=blue,         
  urlcolor=magenta        
}

\makeatletter
\def\BState{\State\hskip-\ALG@thistlm}
\makeatother

\newcommand{\tocentriesoff}{\renewcommand{\addcontentsline}[3]{}}
\newcommand{\tocentrieson}{\let\addcontentsline\SavedAddContentsLine}

\numberwithin{equation}{section}

\theoremstyle{plain}
\newtheorem{proposition}{Proposition}
\newtheorem{definition}{Definition}
\newtheorem{lemma}{Lemma}
\newtheorem{theorem}{Theorem}

\newtheorem{assumption}{Assumption}
\newtheorem{remark}{Remark}
\newtheorem*{theorem*}{Theorem}
\newtheorem*{proposition*}{Proposition}
\newtheorem*{assumption*}{Assumption}
\newtheorem*{definition*}{Definition}
\providecommand{\theHassumption}{\arabic{assumption}}
\providecommand{\theHtheorem}{\arabic{theorem}}
\newcommand{\LTVnum}{\renewcommand{\theassumption}{V.\arabic{assumption}}\renewcommand{\thetheorem}{V.\arabic{theorem}}\renewcommand{\theHassumption}{V.\arabic{assumption}}\renewcommand{\theHtheorem}{V.\arabic{theorem}}}
\newcommand{\LTInum}{\renewcommand{\theassumption}{I.\arabic{assumption}}\renewcommand{\thetheorem}{I.\arabic{theorem}}\renewcommand{\theHassumption}{I.\arabic{assumption}}\renewcommand{\theHtheorem}{I.\arabic{theorem}}}
\newcommand{\COMnum}{\renewcommand{\theassumption}{\arabic{assumption}}\renewcommand{\thetheorem}{\arabic{theorem}}\renewcommand{\theHassumption}{\arabic{assumption}}\renewcommand{\theHtheorem}{\arabic{theorem}}}

\theoremstyle{definition}
\newtheorem{example}{Example}

\usetikzlibrary{shapes,decorations,arrows,calc,arrows.meta,fit,positioning}
\tikzset{
    -Latex,auto,node distance =1 cm and 1 cm,semithick,
    state/.style ={ellipse, draw, minimum width = 0.7 cm},
    point/.style = {circle, draw, inner sep=0.04cm,fill,node contents={}},
    bidirected/.style={Latex-Latex,dashed},
    el/.style = {inner sep=2pt, align=left, sloped}
}

\DeclareRobustCommand{\Rb}{\mathbb{R}}

\DeclareRobustCommand{\bB}{\boldsymbol{B}}

\newcommand{\bC}{\boldsymbol{C}}

\DeclareRobustCommand{\Cc}{\mathcal{C}}

\DeclareRobustCommand{\Ic}{\mathcal{I}}

\DeclareRobustCommand{\Rc}{\mathcal{R}}

\newcommand{\LFc}{\mathcal{LF}}

\DeclareRobustCommand{\Rc}{\mathcal{R}}

\DeclareRobustCommand{\Rb}{\mathbb R}
\DeclareRobustCommand{\Nb}{\mathbb N}

\DeclareRobustCommand{\EE}{\mathbb E}

\DeclareMathOperator*{\argmax}{\arg\!\max}

\newcommand{\distas}[1]{\mathbin{\overset{#1}{\kern\z@\sim}}}%
\newsavebox{\mybox}\newsavebox{\mysim}
\newcommand{\distras}[1]{%
  \savebox{\mybox}{\hbox{\kern3pt$\scriptstyle#1$\kern3pt}}%
  \savebox{\mysim}{\hbox{$\sim$}}%
  \mathbin{\overset{#1}{\kern\z@\resizebox{\wd\mybox}{\ht\mysim}{$\sim$}}}%
}

\newcommand{\hEx}{\widehat{\Ex}}

\newcommand{\overbar}[1]{\mkern 1.3mu\overline{\mkern-1.3mu#1\mkern-1.3mu}\mkern 1.3mu}

\newcommand{\Ex}{\mathbb{E}}

\newcommand{\ldot}[2]{\left\langle #1, #2 \right\rangle}

\newcommand{\tZero}{\tilde{0}}

\makeatletter
\newcommand*{\independent}{%
  \mathbin{%
    \mathpalette{\@indep}{}%
  }%
}
\newcommand*{\notindep}{%
  \mathbin{
    \mathpalette{\@indep}{/}%
  }%
}
\newcommand*{\@indep}[2]{%
  \sbox0{$#1\perp\m@th$}
  \sbox2{$#1=$}
  \sbox4{$#1\vcenter{}$}
  \rlap{\copy0}
  \dimen@=\dimexpr\ht2-\ht4-.2pt\relax
  \kern\dimen@
  \ifx\\#2\\%
  \else
    \hbox to \wd2{\hss$#1#2\m@th$\hss}%
    \kern-\wd2 %
  \fi
  \kern\dimen@
  \copy0 
}
\makeatother

\newcommand{\PP}{\mathbb{P}}

\newcommand{\cX}{\mathcal{X}}
\newcommand{\cA}{\mathcal{A}}
\newcommand{\PiC}{\Pi}

\newcommand{\Nat}{\mathrm{Nat}}

\usepackage{color-edits}
\addauthor{aa}{purple}
\addauthor{vs}{blue}
\addauthor{sh}{teal}
\addauthor{ds}{brown}
\addauthor{hy}{red}

\title{Synthetic Blips: \\ Generalizing Synthetic Controls for Dynamic Treatment Effects\thanks{Vasilis Syrgkanis was supported by NSF Award IIS-2337916. Haeyeon Yoon
was supported by National Research Foundation of Korea (NRF) Grant
2024S1A5A8022044; she is also affiliated with the Nam Duck Woo Economic
Research Institute, Seoul, South Korea. We thank seminar and conference
participants for helpful feedback and Sukyung Joo for excellent research
assistance. Claude Code and ChatGPT were used to assist with coding and
editing. We are grateful to the Export-Import Bank of Korea, the Korea Trade
Insurance Corporation, Statistics Korea, and the Korea Statistics Promotion
Institute for providing the data. All results have been reviewed to ensure
that no confidential information is disclosed. All errors are ours.}}

\author{
Anish Agarwal\\
Columbia University\\
\url{aa5194@columbia.edu}
\and
Sukjin Han\\
University of Bristol\\
\url{vincent.han@bristol.ac.uk}
\and
Dwaipayan Saha\\
Columbia University\\
\url{ds4386@columbia.edu}
\and
Vasilis Syrgkanis\\
Stanford University\\
\url{vsyrgk@stanford.edu}
\and
Haeyeon Yoon\\
University of Bristol\\
\url{haeyeon.yoon@bristol.ac.uk}
}

\date{\today} 
\date{September 10, 2026}

\begin{document}

\let\SavedAddContentsLine\addcontentsline
\tocentriesoff

\maketitle

\begin{abstract}
We propose a generalization of the synthetic control methods to the setting with dynamic treatment effects, in which each unit receives multiple treatments sequentially, according to an adaptive policy that depends on a latent, endogenously time-varying confounding state. Under a low-rank latent factor model assumption, which admits linear time-varying and time-invariant dynamic triangular systems as special cases, we develop an identification strategy for any unit-specific mean outcome under any sequence of interventions. Our method, which we term \emph{synthetic blips}, is a backward induction process in which the blip effect of a treatment at each period for a target unit is recursively expressed as a linear combination of the blip effects of other units that received the designated treatment, avoiding the combinatorial donor requirements of naive synthetic control extensions. We provide easy-to-implement estimation algorithms that yield consistent estimators. Using unique Korean firm-level panel data, we estimate individualized dynamic treatment effects and optimal allocation rules in the context of financial support for exporting firms.

\vspace{0.1in}



\end{abstract}


\section{Introduction}\label{sec:intro}

In many observational studies, units undergo multiple treatments sequentially over time---for example, patients receive multiple therapies, customers are exposed to multiple advertising campaigns, and governments implement multiple policies. The treatment sequence often follows a general pattern beyond staggered adoption, and interventions typically occur in a data-adaptive manner (e.g., via a recommendation algorithm), whereby treatment assignment depends on the treated unit's current (potentially unobserved) confounding state, as well as its past outcomes and treatments. Furthermore, temporal spillovers across treatments and intermediate outcomes make treatment effects inherently dynamic.
A common policy question is what the expected outcome would have been under an alternative policy or course of action.

%
One approach for identification with time-varying treatments is to impose sequential unconfoundedness, under which the treatment decision at each period is exogenous conditional on the history of outcomes and treatments \citep{hernan2020causal}. However, most observational datasets are plagued by unobserved confounding, and endogeneity can take complex forms especially in dynamic settings.
Many techniques exist for addressing unobserved confounding in one-shot treatment settings, such as instrumental variables, difference-in-differences (DiD), regression discontinuity designs, and synthetic controls (SC), some of which have been extended to dynamic contexts. For example, event studies and DiD have been generalized to accommodate sequences of treatments (see below), but most studies assume staggered designs (i.e., irreversible treatment sequences), with a few important exceptions \citetext{\citealp{shahn2022structural}; \citealp{de2024difference}, \citeyear{de2025treatment}}.\footnote{Instrumental variables and regression discontinuity have also been extended to dynamic settings \citep{han2021identification, han2024optimal, hsu2024dynamic, sojitra2024dynamic, han2023semiparametric}, which requires the existence of sequences of instruments or running variables over time.} Beyond these contributions, methods for handling unobserved confounding in settings with general time-varying treatments remain largely underexplored.

%
In this work, we develop an estimation framework for dynamic treatment effects through recursive generation of synthetic counterfactuals.
The SC methods \citep{abadie1, abadie2} are widely used empirical approaches for handling unobserved confounding from observational panel data. 
However, the existing literature typically assumes that units are treated only once (with non-reversal) or in a non-adaptive manner. 
Our framework can be viewed as a generalization of the SC framework that enables identification of mean counterfactual outcomes under arbitrary treatment sequences, even when the observational data arise from an adaptive dynamic treatment policy.
Following SC, we assume that the panel data stem from a low-rank data-generating model, with latent factors capturing unobserved confounding signals. 
However, a naive extension of SC would require unrealistic donor complexity and cannot accommodate adaptive treatment sequences. The key idea of our identification strategy is to express the mean outcome for a unit under a sequence of interventions as an additive function of \emph{blip effects} corresponding to that sequence, where each blip effect is the incremental treatment effect of an intervention at a given period while holding interventions for all other periods fixed.
Subsequently, under our low-rank assumption and by applying a recursive argument, we can identify the blip effect of each treatment for each unit and time period. Our procedure can be viewed as a dynamic programming approach, in which a synthetic-control-type procedure is used to compute \emph{synthetic blip effects} at each step of the dynamic program from a carefully chosen subgroup of other units. These step-specific causal quantities are then combined to build the overall counterfactual outcome of any unit under any sequence of interventions. Motivated by the low-rank factor model and our identification analysis, we
estimate the weights that define the synthetic outcomes and blips via principal
component regression, which inherently de-noises the observed outcomes through
its low-rank approximation.

%
%
The proposed framework allows us to conduct individualized policy evaluation and policy learning. We illustrate the usefulness of the proposed framework by estimating individual dynamic treatment effects and optimal treatment allocation rules in the context of providing financial support to exporting firms. Using novel Korean firm-level data, we first estimate the effects of insurance and loans as two distinct treatments on firm performance, measured by export values. In particular, we recover individualized counterfactual outcomes for policy-relevant hypothetical intervention sequences. Aggregating across firms yields average effects, which reveal that the sequencing of treatments matters for improving export values over time. For example, we find that whether concentrating interventions in early or late periods is more effective on average than smoothing them across periods depends on the type of support. Firm-specific counterfactual outcome estimates enable us to learn targeting rules that maximize expected firm-level performance and compare the rules with the observed rules implemented by the agencies to select firms---a step we call \emph{retrospective policy learning}. We show that these rules can substantially improve outcomes while requiring \emph{less} public spending than the agencies' observed policies and that there is value in coordination between the agencies. Finally, we train interpretable decision trees to guide policymakers in selecting new and not-yet-treated firms for financial support and in scheduling interventions---that is, \emph{prospective policy learning}.

The paper is organized as follows. We close this section by discussing related work and introducing the setting and notation. Section~\ref{sec:model} presents the latent factor model for time-varying treatments, and Section~\ref{sec:identification_SI} shows the limitations of a naive SC extension in our setting. Section~\ref{sec:time_varying_systems} introduces our main models---the time-varying and time-invariant latent factor models---which involve modeling trade-offs, and establishes identification. Section~\ref{subsection:SBE-Varying-Estimator} develops estimation algorithms and shows consistency of the resulting estimators. Section~\ref{sec:application} contains our empirical application, and Section~\ref{sec:conclusion} concludes. The Appendix collects the estimation and consistency results for the time-invariant model, and the Supplemental Appendix (SA) includes all proofs and additional remarks on the models and assumptions.

\subsection{Related Work}\label{sec:related_work}

{\bf Panel data methods in econometrics.}
Prominent approaches for panel data settings include DiD \citep{ashenfelter1984using, bertrand2004much, harmless_econometrics} and SC \citep{abadie1, abadie2, Hsiao12, imbens16, LiBell17, xu_2017, rsc, mrsc, Li18, ark, bai2020matrix, asc, Kwok20, chernozhukov2020practical, fernandezval2020lowrank, agarwal2020robustness, agarwal2020principal, athey1}.
These frameworks estimate what would have happened to a treated unit had it instead remained under control, even in the presence of unobserved confounding.
Recently, the DiD literature has advanced by taking heterogeneity seriously under staggered designs \citep[among others]{de2020two,callaway2021difference,sun2021estimating,borusyak2024revisiting}. Staggered interventions have also been studied in the SC literature  \citep{shaikh2021randomization,ben2022synthetic,powell2022synthetic,cattaneo2025uncertainty} and synthetic DiD \citep{arkhangelsky2024sequential}. The staggered design allows these approaches to estimate the counterfactual trajectory of treated units had they remained not-yet-treated. DiD approaches with time-invariant and time-varying treatments explicitly or implicitly assume strict exogeneity, which rules out adaptivity of treatments \citep{chabe2015analysis, de2022not, ghanem2022selection, marx2024parallel} or equivalently \emph{feedback} \citep{bonhomme2025back}. SC approaches typically allow more flexible forms of unobserved confounders as shown by \cite{arkhangelsky2023large,imbens2023identification}, who also develop asymptotic theory in the case of a binary treatment.\footnote{Also see the recent survey on panel data methods by \cite{arkhangelsky2024causal}.}

Both one-shot and staggered designs can be viewed as special cases of the problem we study in this paper: estimating counterfactual outcomes for a unit under {\em any} hypothetical sequence of interventions over $T$ time steps.
%
%
A critical aspect underlying the above methods is the structure imposed across units and over time under ``control.'' 
One elegant way of encoding this structure is through latent factor models (also known as interactive fixed effects models) \citep{chamberlain, arellano, bai03, bai09, pesaran, moon_15, moon_weidner_2017}.
Since the goal in these works is to estimate outcomes under ``control,'' no structure is imposed on the potential outcomes under intervention.\footnote{One extension of such factor models is the class of dynamic factor models, originally proposed in \cite{geweke1976dynamic}. We refer the reader to \cite{stock_watson_DFM, feedback_DFM_chamberlain} for extensive surveys, and to \cite{imbens2021controlling} for a recent analysis of time-varying factor models in the context of SC. Like ours, these models allow outcomes to depend autoregressively on their lags. However, their objective differs: identifying the latent factors or forecasting, rather than estimating causal effects.}

\cite{SI, agarwal2021causal} extend the latent factor model to factorize across interventions as well---hence the name synthetic interventions (SI)---allowing identification and estimation of mean counterfactual outcomes under any intervention, not only under control.
That framework, however, is designed for static regimes and faces two limitations in the dynamic setting: 
(i) it does not allow adaptive treatment assignment over time, and
(ii) with $A$ possible interventions at each of $T$ time steps, the sample complexity of the SI estimator scales as $A^T$ to estimate all intervention sequences.
%
In contrast, we show that by imposing that each intervention has an additive effect on future outcomes (i.e., an additive latent factor model), we achieve significant gains in identification and estimation.
We study time-varying and time-invariant variants, which respectively nest linear time-varying and time-invariant dynamic triangular models as special cases.
We establish identification results and propose associated estimators to infer all $A^T$ counterfactual trajectories per unit.
Importantly, our identification result allows the interventions to be selected in an adaptive manner (in a specific sense that will become clear later), and the sample complexity of the estimator no longer grows exponentially in $T$; see Table~\ref{tab:sample-complexity}.
\renewcommand{\arraystretch}{1.4}
\begin{table}[t]
    \centering
    \footnotesize
    \begin{tabular}{@{}p{6.4cm}p{2.5cm}p{5.8cm}@{}}
        \toprule
        Latent Factor Models\newline (LFM) & Donor Sample\newline Complexity & Adaptivity of\newline Intervention Policy \\
        \midrule
        Naive LFM (Synthetic Interventions) & $O(A^{T})$ & Non-adaptive \\
        Additive Time-Varying LFM (This Work) & $O(A \times T)$ & Adaptive after some periods (i.e., staggered adoption of adaptive policy) \\
        Additive Time-Invariant LFM (This Work) & $O(A)$ & Adaptive after period 1 \\
        \bottomrule
        \\
    \end{tabular}
    \caption{Comparison of Donor Sample Complexity and Adaptivity Across Models}
    \label{tab:sample-complexity}
\end{table}
\renewcommand{\arraystretch}{1.0}

Our work can also be viewed as extending the g-estimation framework for structural nested mean models  \citep{Robins2004,Vansteelandt2014,lewis2020double} in the biostatistics literature to accommodate unobserved confounding under a low-rank structure. In doing so, our work helps connect the econometric literature on SC with this literature on structural nested mean models.

\noindent{\bf Public financial support for exports.}
Financial frictions play a central role in shaping firms' export performance, particularly in times of crisis \citep{Amiti2011, Chor2012, Paravisini2015}. To mitigate financing barriers and sustain exports, governments provide public financial support via export credit agencies (ECAs), mainly in the form of insurance and loans. Financial support can generate different effects depending not only on its scale but also on how it is allocated and structured \citep{criscuolo2019some, rotemberg2019equilibrium}. Empirical studies of ECAs are typically limited to a single treatment (mostly insurance) due to data constraints \citep[among others]{felbermayr2013export}, leaving the broader impact of combined support largely unexplored. This paper considers the entire set of support programs and analyzes how the timing and sequencing of interventions influence firm performance.

\subsection{Setting and Notation}\label{sec:formal_setting}
{\bf Notation.}
$[R]$ denotes $\{1, \dots, R\}$ for $R \in \Nb$.
$[R_1, R_2]$ denotes $\{R_1, \dots, R_2\}$ for $R_1, R_2 \in \Nb \cup \{0\}$, with $R_1 < R_2$.
%
%
For a vector $a$, we denote its transpose by $a^{\top}$.
For vectors $a, b \in \Rb^d$, we define the inner product of $a$ and $b$ as $\ldot{a}{b}=a^{\top}b=\sum_{\ell=1}^{d}a_\ell b_\ell$.
For a matrix $M\in \Rb^{m \times n}$, we denote its Frobenius norm as $\|M\|_F$. 
Let $\|\cdot\|_{\psi_2}$ denote the $\psi_2$-Orlicz norm.\footnote{For a r.v. $X$, the $\psi_2$-Orlicz norm $\|X\|_{\psi_2}$ is defined as $\|X\|_{\psi_2} \coloneqq \inf\bigl\{ t > 0 : \mathbb{E}\bigl[\exp\bigl(X^2/t^2\bigr)\bigr] \le 2 \bigr\}$.}
Let $O_p$ and $o_p$ denote the stochastic big-$O$ and little-$o$ notation.

\noindent{\bf Setup.}
Consider $N$ heterogeneous units, each observed over $T$ time steps.

{\em Observed outcomes.}
For each unit $n$ and time period $t$, we observe $Y_{n, t} \in \Rb$, which is the outcome of interest.

{\em Treatments.}
For each $n \in [N]$ and $t \in [T]$, we observe a treatment action $D_{n, t} \in [A]$, where $A \in \Nb$.
We allow $D_{n, t}$ to be categorical, i.e., it can simply serve as a unique identifier for the action chosen.
%
%
Denote sequences of actions $\bar{d}^{t}=(d_1, \dots, d_t)\in [A]^t$ and $\underline{d}^t=(d_t, \dots, d_T)\in [A]^{T - t+1}$.
Define $\bar{D}_n^{t}, \underline{D}_n^t$ analogously to $\bar{d}^{t}, \underline{d}^t$, respectively, but now with respect to the observed sequence of actions $D_{n, t}$.
%

{\em Control and intervention periods.}
For each $n$, we assume there exists $t^*_n \in [T]$ before which it is in ``control''.
We denote the control action at $t$ as $0_t \in [A]$.\footnote{The notation $0_t$ is introduced to allow a general control action that is not necessarily ``no treatment.''}
Note $0_{\ell}$ and $0_{t}$ for $\ell \neq t$, do not necessarily equal each other (i.e., the control state can vary in time).
For $t \in [T]$, denote $\bar{0}^t = (0_1, \dots, 0_t)$ and $\underline{0}^t = (0_t, \dots, 0_T)$.
For $t < t^*_n$, we assume $D_{n, t} = 0_t$, i.e., $\bar{D}_n^{t^*_n - 1} = \bar{0}^{t^*_n - 1}$: during the control period all units are under a common sequence of actions. For $t \geq t^*_n$, each unit $n$ can undergo a sequence that may differ across units, denoted by $\underline{D}^{t^*_n}_n$.
If $t^*_n = 1$, then unit $n$ is never in the control period.

{\em Counterfactual outcomes.}
We let $Y_{n, t}^{(\bar{d}^t)}$ denote the potential outcome if unit $n$ had undergone $\bar{d}^t$.
More generally, we denote the potential outcome $Y_{n, t}^{(\bar{D}^\ell_n, \underline{d}^{\ell + 1})}$ if unit $n$ receives the observed sequence of actions $\bar{D}^\ell_n$ until time step $\ell$, and then instead undergoes $\underline{d}^{\ell + 1}$ for the remaining $t - \ell$ time steps.

We make the standard stable unit treatment value assumption (SUTVA) and the consistency assumption as follows.
\begin{assumption}[Sequential Action SUTVA]\label{assumption:SUTVA}
For all $ n\in [N], t\in [T], \ell \in [t], \bar{d}^T \in [A]^{T}$:
\begin{align*}
    Y^{(\bar{D}^\ell_n, \underline{d}^{\ell + 1})}_{n, t} = \sum_{\bar{\delta}^\ell \in [A]^{\ell}} Y^{(\bar{\delta}^\ell, \underline{d}^{\ell + 1})}_{n, t} \cdot \boldsymbol{1}(\bar{D}^\ell_n  = \bar{\delta}^\ell).
\end{align*}
Further, for all $\bar{D}^t_n \in [A]^t$, $Y^{(\bar{D}^t_n)}_{n, t} = Y_{n, t}$.
%
%
As an immediate implication, $Y^{(\bar{d}^\ell, \underline{d}^{\ell + 1})}_{n, t} \mid \bar{D}^\ell_n = \bar{d}^\ell$ equals 
$Y^{(\bar{D}^\ell_n, \underline{d}^{\ell + 1})}_{n, t} \mid \bar{D}^\ell_n = \bar{d}^\ell$, and $Y^{(\bar{d}^t)}_{n, t} \mid \bar{D}^t_n = \bar{d}^t$ equals $Y_{n, t} \mid \bar{D}^t_n = \bar{d}^t$.
%
\end{assumption}

\noindent{\bf Goal.}
Our goal is to estimate the potential outcome if a given unit $n$ had undergone $\bar{d}^T$ (instead of the observed sequence $\bar{D}_n^T$), for any given sequence of actions $\bar{d}^T$ over $T$ time steps.
We more formally define the target causal parameter in Section~\ref{sec:model}.

\section{Latent Factor Model for Dynamic Treatments}\label{sec:model}
We now present a novel latent factor model for causal inference with dynamic treatments.
To that end, we first define the collection of latent factors that are of interest.

\begin{definition}[Latent factors]\label{def:latent_factors}
For a given unit $n$ and time step $t$, denote its latent factor by $v_{n, t}$.
For a given sequence of actions over $t$ time steps, $\bar{d}^t$, denote its associated latent factor by $w_{\bar{d}^t}$.
Denote the collection of latent factors by 
\begin{align*}
\LFc 
\coloneqq 
\left\{ v_{n, t} \right\}_{n \in [N], t \in [T]} 
\cup 
\left\{ w_{\bar{d}^t} \right\}_{\bar{d}^t \in [A]^{t}, \ t \in [T]}.
\end{align*}
Here $v_{n, t}, w_{\bar{d}^t} \in \Rb^{m(t)}$, where $m(t)$ is allowed to depend on $t$.
\end{definition}

\begin{assumption}[General factor model]\label{assumption:general_latent_factor_model}
Assume $\forall \ n \in [N]$, $t \in [T], \bar{d}^t \in [A]^t$,
\begin{align}
Y^{(\bar{d}^t)}_{n, t} 
&= \ldot{v_{n, t}}{w_{\bar{d}^t}} + \varepsilon^{(\bar{d}^t)}_{n, t}, \label{eq:general_factor_model}
\end{align}
and $\Ex[\varepsilon^{(\bar{d}^t)}_{n, t} \mid \LFc] = 0$.
\end{assumption}
In \eqref{eq:general_factor_model}, the key assumption made is that $v_{n, t}$ does not depend on the action sequence $\bar{d}^t$, while $w_{\bar{d}^t}$ does not depend on unit $n$.
That is, $v_{n, t}$ captures the unit-$n$-specific latent heterogeneity in determining the expected conditional potential outcome $\Ex[Y^{(\bar{d}^t)}_{n, t} \mid  \LFc]$; $w_{\bar{d}^t}$ follows a similar intuition but with respect to the  action sequence $\bar{d}^t$. Importantly, the factors can be correlated with the observed sequence $\bar{D}^t_n$, making them unobserved confounders.
This latent factorization will be key in our identification and estimation algorithms, and the associated theoretical results.
Here $\varepsilon^{(\bar{d}^t)}_{n, t}$ represents the component of the potential outcome $Y^{(\bar{d}^t)}_{n, t}$ that is not factorizable into the latent factors represented by $\LFc$. It also helps model the inherent randomness in the potential outcomes $Y^{(\bar{d}^t)}_{n, t}$.
%

\paragraph{Target causal parameter.}
Our target causal parameter to estimate is, for all units $n \in [N]$ and any action sequence $\bar{d}^T \in [A]^T$, 
\begin{align}
    \Ex[Y^{(\bar{d}^T)}_{n, T} ~|~ \LFc], \tag{target parameter} \label{eq:target_causal_param}
\end{align}
i.e., the expected potential outcome conditional on the latent factors, $\LFc$.
In total this amounts to estimating $N \times A^T$ different (expected) potential outcomes.
%




\subsection{Limitations of Existing Approaches}\label{sec:identification_SI}\label{sec:SI_complexity}
Given the setting, it is natural to ask how far existing factor-model methods designed for the static regime---in particular, the SI framework \citep{SI}, a generalization of the SC framework---can be applied in our setting.
In SA~\ref{app:SI} we adapt the SI identification argument to dynamic treatments (Theorem~\ref{thm:newID1}): for each sequence $\bar{d}^T$, one selects donor units that received exactly that sequence, requires that their potential outcomes be conditionally mean independent of the assigned sequence (i.e., non-adaptivity), and expresses the target unit's mean counterfactual outcome as a linear combination of the donors' observed outcomes under a condition of well-supported factors.
This strategy faces two limitations in the dynamic regime.
First, \emph{donor sample complexity}: a sufficiently large donor set is needed for every sequence $\bar{d}^T \in [A]^T$, so the number of donor units must scale at the order of $A^T$, which grows exponentially in $T$.
Second, \emph{donor exogeneity}: the entire action sequence of each donor must be non-adaptive, i.e., chosen at $t = 0$ conditional on the latent factors, which rules out adaptive treatment assignments common in many settings (as discussed in the introduction and as in our empirical application).
In the next section we study how additional structure on the latent factor model gives rise to novel identification strategies, which allow us to reduce the donor sample complexity and avoid the stringent exogeneity requirements.


\section{Identification: Additive Latent Factor Models}\label{sec:time_varying_systems}
Motivated by the limitations of the identification strategy in Section~\ref{sec:identification_SI}, we now impose additional structure on the latent factor model, where an intervention at a given period has an additive effect on future outcomes. We study two variants: a linear \emph{time-varying} (LTV) factor model and a linear \emph{time-invariant} (LTI) factor model. The latter further restricts treatment effects to depend only on the elapsed time since treatment. The additional structure of the LTV and LTI models leads to a substantial reduction in donor sample complexity and much weaker exogeneity requirements on the donor units (i.e., adaptivity is allowed); see Table~\ref{tab:sample-complexity}. Throughout this section we develop the LTV case in full and describe the LTI case through its differences, deferring details to the SA.

\subsection{Two Latent Factor Models}\label{sec:models_additive}\label{sec:time_invariant_systems}
\LTVnum\begin{assumption}[Linear time-varying (LTV) factor model]\label{assumption:LTV_factor_model}
Assume $\forall \ n \in [N]$, $t \in [T], \bar{d}^t \in [A]^t$,
\begin{align}
Y^{(\bar{d}^t)}_{n, t} &= \sum^{t}_{\ell = 1} \ldot{\psi^{t, \ell}_{n}}{w_{d_\ell}} + \varepsilon^{(\bar{d}^t)}_{n, t},\label{eq:LTV_factor_model}
\end{align}
where $\psi^{t, \ell}_{n}, w_{d_\ell} \in \Rb^m$ for $\ell \in [t]$.
Further, let $\LFc = \{\psi^{t, \ell}_{n} \}_{n \in [N], t \in [T],  \ell \in [t]} \cup \{w_{d} \}_{d \in [A]}$.
Also assume $\Ex[\varepsilon^{(\bar{d}^t)}_{n, t} \mid \LFc] = 0$.
\end{assumption}

\begin{remark}\label{rem:LTV_nests_general}
Note Assumption~\ref{assumption:LTV_factor_model} implies Assumption~\ref{assumption:general_latent_factor_model} holds with
\begin{align*}
v_{n, t} = [\psi^{t, 1}_{n}, \dots, \psi^{t, t}_{n}], \qquad
w_{\bar{d}^t} = [w_{d_1}, \dots, w_{d_t}],
\end{align*}
and $m(t) = m \times t$ for $m(t)$ in Definition~\ref{def:latent_factors}.
\end{remark}
Assumption~\ref{assumption:LTV_factor_model} imposes additional structure in the latent factors: the effect of action $d_\ell$ on $Y^{(\bar{d}^t)}_{n, t}$ for $\ell \in [t]$ is additive, given by $\langle \psi^{t, \ell}_{n}, w_{d_\ell} \rangle$.
Intuitively, $\psi_n^{t, \ell}$ captures the latent unit-specific heterogeneity in the potential outcome for unit $n$ at time step $t$ for an action taken at time step $\ell \le t$, while $w_{d_\ell}$ captures the latent effect of the action itself.
This structure will be central to the identification strategy in Section~\ref{sec:LTV_identification_strategy}.

The time-invariant variant additionally imposes that the unit factor $\psi^{t, \ell}_{n}$ depends on $(t, \ell)$ only through the lag $t - \ell$.
\addtocounter{assumption}{-1}\LTInum\begin{assumption}[Linear time-invariant (LTI) factor model]\label{assumption:LTI_factor_model}
Assume $\forall \ n \in [N]$, $t \in [T], \bar{d}^t \in [A]^t$,
\begin{align}
     Y^{(\bar{d}^t)}_{n, t} &= \sum^{t}_{\ell = 1} \ldot{\psi^{t - \ell}_{n}}{w_{d_\ell}} + \varepsilon^{(\bar{d}^t)}_{n, t}, \label{eq:LTI_factor_model}
\end{align}
where $\psi^{t - \ell}_{n}, w_{d_\ell} \in \Rb^m$ for $\ell \in [t]$.
Further, let $\LFc = \{\psi^{t - \ell}_{n} \}_{n \in [N], t \in [T],  \ell \in [t]} \cup \{w_{d} \}_{d \in [A]}$.
Also assume $\Ex[\varepsilon^{(\bar{d}^t)}_{n, t} \mid \LFc] = 0$. Finally, assume the control sequence is time invariant:\label{assumption:LTI_control_sequence} there exists $\tZero \in [A]$ such that $0_t = \tZero$ for all $t \in [T]$.
\end{assumption}
Assumption~\ref{assumption:LTI_factor_model} is the special case of Assumption~\ref{assumption:LTV_factor_model} in which $\psi^{t, \ell}_{n} = \psi^{t - \ell}_{n}$; it implies Assumption~\ref{assumption:general_latent_factor_model} with $v_{n, t} = [\psi^{t - 1}_{n}, \dots, \psi^{0}_{n}]$ as in Remark~\ref{rem:LTV_nests_general}. The effect of an action taken at time $\ell$ on the outcome at time $t$ is only a function of the lag $t - \ell$; hence the name ``time-invariant.'' The additivity of Assumptions~\ref{assumption:LTV_factor_model} and \ref{assumption:LTI_factor_model} can be relaxed; see Remark~\ref{rem:hybrid} below for hybrid models. Even under additivity, temporal spillovers are still allowed.

\begin{example}[Dynamic triangular models with adaptive treatments]\label{ex:dyn_triangular_model}
Assumption~\ref{assumption:LTV_factor_model} is implied by a dynamic triangular system of equations with a latent state and adaptive
treatment assignment.\footnote{In engineering terminology, this relates to a linear time-varying dynamical system.} Let each
action $d \in [A]$ have a latent encoding $w_d \in \Rb^m$, with $z_{n, 0} = w_{D_{n, 0}} = 0$.
Suppose for all $t \in [T]$, $n \in [N]$, actions $D_{n, t} \in [A]$ and counterfactual
outcomes $Y^{(\bar{D}_n^t)}_{n, t}$ obey the dynamic triangular model
\begin{align}
    Y^{(\bar{D}_n^t)}_{n, t} &= \ldot{\theta_{n, t}}{z^{(\bar{D}_n^t)}_{n, t}} + \ldot{\tilde{\theta}_{n, t}}{w_{D_{n, t}}} + \tilde{\eta}_{n, t},\label{eq:dyn_triangular_2}
    \\
    D_{n, t} &= h_n(w_{D_{n, t-1}}, \ z^{(\bar{D}_n^{t - 1})}_{n, t - 1}),\label{eq:dyn_triangular_1}
\end{align}
where the latent state $z^{(\bar{D}_n^t)}_{n, t} \in \Rb^m$ of unit $n$ evolves as
$z^{(\bar{D}_n^t)}_{n, t} = \bB_{n, t} \, z^{(\bar{D}_n^{t - 1})}_{n, t - 1} + \bC_{n, t} \, w_{D_{n, t}} + \eta_{n, t}$;
$\eta_{n, t} \in \Rb^m, \tilde{\eta}_{n, t} \in \Rb$ are independent mean-zero
innovations; and the time-variation of $\bB_{n, t}, \bC_{n, t} \in \Rb^{m \times m}$ and
$\theta_{n, t}, \tilde{\theta}_{n, t} \in \Rb^{m}$ is what makes the system LTV.
For instance, in advertising, $z_{n, t}$ is customer $n$'s goodwill stock (i.e., brand capital), $w_d$
encodes campaign $d$'s spend profile, $\bB_{n, t}$ governs carryover, and $Y_{n, t}$
is purchases, as in \cite{nerlove1962optimal}.
The policy $h_n$ maps the previous action and state into the next action; through
this dependence on $z^{(\bar{D}_n^{t-1})}_{n, t-1}$, the action sequence is
\emph{adaptive}: $\eta_{n, \ell}$ can be correlated with $D_{n, t}$ for $\ell < t$,
i.e., treatments feed back on past shocks.
\begin{proposition}\label{lemma:LTV_representation}
Under \eqref{eq:dyn_triangular_2}--\eqref{eq:dyn_triangular_1}, for all $\bar{d}^t \in [A]^t$,
\begin{align}
    Y^{(\bar{d}^t)}_{n, t} = \sum^{t}_{\ell = 1} \Big(\ldot{\psi^{t, \ell}_{n}}{w_{d_\ell}} + \varepsilon_{n, t, \ell} \Big),
\label{eq:LTV_factor_model_representation}
\end{align}
where $\psi^{t, \ell}_{n} \in \Rb^m$ is an explicit function of
$(\{\bB_{n, k}\}_{k = \ell + 1}^{t}, \bC_{n, \ell}, \theta_{n, t}, \tilde{\theta}_{n, t})$ and
$\varepsilon_{n, t, \ell}$ is linear in the innovations
$(\eta_{n, \ell}, \tilde{\eta}_{n, \ell})$; expressions are given in
SA~\ref{app:LTV_example_details}.
\end{proposition}
Hence Assumption~\ref{assumption:LTV_factor_model} holds, with the additional
structure that $\varepsilon^{(\bar{d}^t)}_{n, t} = \sum^{t}_{\ell = 1} \varepsilon_{n, t, \ell}$
is additive and does not depend on $\bar{d}^t$---the first sufficient condition for
Assumption~\ref{assumption:LTV_seq_exog}.
The linear time-\emph{invariant} system---constant $\bB_{n}, \bC_{n}, \theta_{n},
\tilde{\theta}_{n}$---analogously satisfies the factor-model representation in Assumption~\ref{assumption:LTI_factor_model} (its time-invariant control-sequence clause is a separate condition on the environment);
see SA~\ref{sec:LTI_example}.
\end{example}

\subsection{Identification}\label{sec:LTV_identification_strategy}\label{sec:LTI_identification_strategy}
In this section we identify $\Ex[Y_{n,T}^{(\bar{d}^T)}\mid \LFc]$, that is, we represent this expected potential outcome for a target unit $n$ and action sequence $\bar{d}^T$ as some function of observed outcomes. We first introduce the essential ingredients: the blip effects, baseline outcomes, and donor sets.

\noindent{\bf Blip effects and baselines.}
For any unit $n \in [N]$, define
\begin{align}
         \gamma_{n, T, t}(d) := \ldot{\psi^{T, t}_{n}}{w_{d} - w_{0_t}},
         \qquad
         b_{n, T} := \sum^T_{t = 1} \ldot{\psi^{T, t}_{n}}{w_{0_t}}. \nonumber
\end{align}
The quantity $\gamma_{n, T, t}(d)$ can be interpreted as a \emph{blip effect}: under Assumption~\ref{assumption:LTV_factor_model}, $\gamma_{n, T, t}(d_t) = \Ex[Y^{(\bar{d}^t, \underline{0}^{t + 1})}_{n, T} - Y^{(\bar{d}^{t -1}, \underline{0}^{t})}_{n, T} \mid \LFc ]$ is the expected difference in potential outcomes between two sequences that differ only at time $t$.
Likewise, under Assumption~\ref{assumption:LTV_factor_model}, $b_{n, T} = \Ex[ Y^{(\bar{0}^T)}_{n, T} \mid \LFc ]$ is the expected potential outcome if unit $n$ remains under the control sequence $\bar{0}^T$. Moreover, Assumption~\ref{assumption:LTV_factor_model} implies that the blips and baselines themselves have the latent factor structure.
In the LTI model, under Assumption~\ref{assumption:LTI_control_sequence}, the corresponding blip depends only on the lag: for $t \in [T]$,
\begin{align}
     \gamma_{n, T - t}(d) := \ldot{\psi^{T - t}_{n}}{w_{d} - w_{\tZero}},
     \qquad
     b_{n, t} := \sum^{t}_{\ell = 1} \ldot{\psi^{t - \ell}_{n}}{w_{\tZero}}, \nonumber
\end{align}
with the analogous interpretations.

\noindent{\bf Donor sets.}
In the LTV setting, we define subsets of units based on: (i) the treatment sequence they receive up to $t$; (ii) the correlation between their potential outcomes and the chosen actions: for $t\in[T]$
\begin{align}
\Ic^d_t
\coloneqq
\{j \in [N]:
& \ (i) \ \bar{D}^t_{j} = (0_1, \dots, 0_{t -1}, d), \nonumber
\\ &(ii) \ \Ex[Y^{(\bar{0}^T)}_{j, T} \mid \LFc, \bar{D}^t_{j}= (0_1, \dots, 0_{t -1}, d)] = \Ex[Y^{(\bar{0}^T)}_{j, T} \mid \LFc]
\}. \label{eq:LTV_donor}
\end{align}
The donor set $\Ic^d_t$ contains units that remain under the control sequence $(0_1, \dots, 0_{t - 1})$ until time step $t - 1$, receive action $d$ at time step $t$ (i.e., $t^*_j \ge t$), and \emph{any} actions after $t$. In this way, the donor requirement is substantially relaxed: donors need only switch from control to action $d$ at time $t$ and may receive any actions after $t$, so far more units qualify than if the entire sequence were prescribed.
Further, we require that for these particular units, the action sequence $\bar{D}^t_{j}= (0_1, \dots, 0_{t -1}, d)$ was chosen such that the \emph{untreated} potential outcomes are conditionally mean independent of it. Therefore, the adaptivity is restricted for these donors in this specific sense for the control sequence and the first treatment action, but not for the subsequent actions. 
In the context of Example~\ref{ex:dyn_triangular_model}, the dynamic selection equation \eqref{eq:dyn_triangular_1} can be taken to hold from period $t+1$ onward, with the timing and identity of the first adopted action chosen at time $0$. This structure arises, for instance, under a predetermined pilot policy followed by adaptive follow-up decisions.
Note that, given Assumption~\ref{assumption:LTV_factor_model}, condition (ii) can be equivalently stated as $\Ex[\varepsilon^{(\bar{0}^T)}_{j, T} \mid \LFc, \bar{D}^t_{j}] = \Ex[\varepsilon^{(\bar{0}^T)}_{j, T} \mid \LFc] = 0$ for these donors.

In the LTI setting, the lag structure allows much coarser donor sets, indexed by the first action alone: for $t\in[T]$
\begin{align}
\Ic^d
\coloneqq
\{j \in [N]:
& \ (i) \ \bar{D}^{t^*_j}_{j} = (\tZero, \dots, \tZero, d), \nonumber
\\ (ii) \  & \forall \ell \in [t^*_j,T], \ \Ex[Y^{(\bar{\tZero}^\ell)}_{j, \ell} \mid \bar{D}^{t^*_j}_{j}= (\tZero, \dots, \tZero, d),\LFc] = \Ex[Y^{(\bar{\tZero}^\ell)}_{j, \ell} \mid \LFc]
\}, \label{eq:LTI_donor}
\\ \Ic^{0}_t
\coloneqq
\{j \in [N]:
& \ (i) \ \bar{D}^{t}_{j} = (\tZero, \dots, \tZero), \nonumber
\\ &(ii) \ \Ex[Y^{(\bar{\tZero}^t)}_{j, t} \mid \bar{D}^{t}_{j}=(\tZero, \dots, \tZero),\LFc] = \Ex[Y^{(\bar{\tZero}^t)}_{j, t} \mid \LFc]
\}. \label{eq:LTI_donor_control}
\end{align}
The donor set $\Ic^d$ contains units that remain under the control sequence $(\tZero, \dots, \tZero)$ until time step $t^*_j - 1$ and receive action $d$ at time step $t^*_j$, with the \emph{weaker} conditional mean independence requirement holding only up to their (unit-specific) first treatment time $t^*_j$; see Figure~\ref{fig:Dags} below for comparisons to the LTV setting. The donor set $\Ic^{0}_t$ collects units under control through time step $t$.

We now introduce identifying assumptions.
\LTVnum\begin{assumption}[Well-supported factors]\label{assumption:LTV_well_supported_factors}
For $n \in [N]$, let $v_{n, T} := [\psi^{T, 1}_{n}, \dots, \psi^{T, T}_{n}]$.
We assume that for all $n \in [N]$, $v_{n, T}$ satisfies a well-supported condition with respect to the various donor sets, i.e., for all $d \in [A]$ and $t \in [T]$, there exists $\beta^{n,\Ic_{t}^d} \in \Rb^{|\Ic_{t}^d|}$ such that
\begin{align}
    v_{n, T} = \sum_{k \in \Ic_{t}^d} \beta_k^{n,\Ic_{t}^d} v_{k, T}. \label{eq:LTV_well_supported}
\end{align}
\end{assumption}
Assumption~\ref{assumption:LTV_well_supported_factors} requires that, for units $n \in [N]$, their latent factors are expressible as a linear combination of those of the units in the donor sets $\Ic_{t}^d$.
Note that if the $\{v_{k, T}\}_{k \in [N]}$ are sampled as independent, mean zero, sub-Gaussian vectors and if $|\Ic_{t}^d| \gg m(T)$, then $\text{span}\{v_{k,T} : k \in \Ic_{t}^d\} = \Rb^{m(T)}$ with high probability as $|\Ic_{t}^d|$ grows (recall $m(T)$ is the dimension of $v_{n, T}$) \citep[][Theorem 4.6.1]{vershynin2018high}.

\addtocounter{assumption}{-1}\LTInum\begin{assumption}[Well-supported factors]\label{assumption:LTI_well_supported_factors}
For $n \in [N]$, let $v_{n, T} := [\psi^{T - 1}_{n}, \dots, \psi^{0}_{n}]$.
We assume that for all $n \in [N]$, $v_{n, T}$ satisfies a well-supported condition with respect to the various donor sets, i.e., for all $d \in [A]$ and $t\in[T]$ there exist $\beta^{n,\Ic^d} \in \Rb^{|\Ic^d|}$ and $\beta^{n, \Ic^{0}_t} \in \Rb^{|\Ic^{0}_t|}$ such that
\begin{align}
    v_{n, T} = \sum_{k \in \Ic^d} \beta_k^{n,\Ic^d} v_{k, T} = \sum_{k \in \Ic^{0}_t} \beta_k^{n, \Ic^{0}_t} v_{k, T}. \label{eq:LTI_well_supported}
\end{align}
\end{assumption}
The discussion under Assumption~\ref{assumption:LTV_well_supported_factors} justifying such an assumption applies here as well when $\Ic^d$ and $\Ic^{0}_t$ are sufficiently large.

\LTVnum\begin{assumption}[Sequential mean exogeneity of blips]\label{assumption:LTV_seq_exog}
For all $n \in [N], t \in [T], \bar{d}^t \in [A]^t$,
\begin{align*}
\Ex\left[Y_{n, T}^{(\bar{d}^t, \underline{0}^{t + 1})} - Y_{n, T}^{(\bar{d}^{t - 1}, \underline{0}^{t})} \mid \bar{D}_n^{t} = \bar{d}^t, \LFc \right] = \gamma_{n, T, t}(d_{t}) \mid \LFc.
\end{align*}
\end{assumption}
Note that given Assumption~\ref{assumption:LTV_factor_model}, this condition can be equivalently written as
\begin{align*}
\Ex\left[\varepsilon_{n, T}^{(\bar{d}^t, \underline{0}^{t + 1})} - \varepsilon_{n, T}^{(\bar{d}^{t - 1}, \underline{0}^{t})} \mid \bar{D}_n^{t} = \bar{d}^t, \LFc \right]
= 0.
\end{align*}
Two sufficient conditions for Assumption~\ref{assumption:LTV_seq_exog} are (i) a noise that is not action-dependent, as in Example \ref{ex:dyn_triangular_model} (dynamic triangular models with adaptivity), and (ii) an additive action-dependent noise whose per-period innovations are independent of the action history through the current period (no anticipation); see SA~\ref{app:seq-exog-sufficient} for details.
The LTI analogue is the following.
\addtocounter{assumption}{-1}\LTInum\begin{assumption}[Sequential mean exogeneity of blips]\label{assumption:LTI_seq_exog}
For all $n \in [N]$, $t \in [T]$, $\ell \in [t]$, and $\bar{\delta}^{\ell} \in [A]^{\ell}$,
\begin{align*}
\Ex\left[\varepsilon^{(\bar{\delta}^{\ell},\, \underline{\tZero}^{\ell+1})}_{n, t} - \varepsilon^{(\bar{\delta}^{\ell-1},\, \underline{\tZero}^{\ell})}_{n, t} \mid \bar{D}^{\ell}_n = \bar{\delta}^{\ell}, \ \LFc \right] = 0,
\end{align*}
where $\underline{\tZero}^{\ell+1}$ denotes the control sequence from period $\ell+1$ through the outcome date $t$ (the empty sequence when $\ell = t$).
\end{assumption}

\begin{remark}[Connection to Models for Dynamic Treatment Effects in Biostatistics]
In SA~\ref{app:snmm_assm} we show that our two key conditions, Assumptions~\ref{assumption:LTV_factor_model} and \ref{assumption:LTV_seq_exog}, are implied by a sequential conditional exogeneity condition together with factor-model representations of the blip effects and the baseline outcomes. The former condition appears in the structural nested mean model (SNMM) and marginal structural model (MSM) literature on dynamic treatment effects \citep{hernan2020causal}. 
Unlike a full SNMM specification, our blip condition in Assumption~\ref{assumption:LTV_seq_exog} only requires that the blip effect is not modified by past actions except through the latent confounding states, on which treatment assignment may depend arbitrarily; the full SNMM, by contrast, precludes any dependence of blips on these states.
An analogous result holds in the LTI setting by an identical argument.
\end{remark}

Given these assumptions, we now present the main identification results. We first illustrate the key intuition in a two-period LTV setting ($T=2$), suppressing the terminal subscript $T$ and the conditioning on $\LFc$. For a unit $n$ and an action sequence $(d_{1},d_{2})\in[A]^2$, the expected potential outcome decomposes into two blip effects and a baseline:
\begin{equation}\label{eq:illustration}
\mathbb{E} \left[Y_{n}^{(d_{1},d_{2})}\right]
= \underbrace{\mathbb{E} \left[Y_{n}^{(d_{1},d_{2})}\right]
              - \mathbb{E} \left[Y_{n}^{(0_{1},d_{2})}\right]}_{\mathrm{Blip}_{1}(d_{1})}
+
\underbrace{\mathbb{E} \left[Y_{n}^{(0_{1},d_{2})}\right]
            - \mathbb{E} \left[Y_{n}^{(0_{1},0_{2})}\right]}_{\mathrm{Blip}_{2}(d_{2})}
+
\underbrace{\mathbb{E} \left[Y_{n}^{(0_{1},0_{2})}\right]}_{\text{Baseline}},
\end{equation}
where $\mathrm{Blip}_{t}(d_{t}) = \gamma_{n,2,t}(d_t)$ and $\text{Baseline} = b_{n,2}$. Under Assumption~\ref{assumption:LTV_factor_model}, $\mathrm{Blip}_{2}(d_{2})= \langle \psi^{2,2}_n,\, w_{d_2}-w_{0_2} \rangle$, $\mathrm{Blip}_{1}(d_{1})= \langle \psi^{2,1}_n,\, w_{d_1}-w_{0_1} \rangle$, and $\text{Baseline}= \sum_{\ell=1}^{2}  \langle \psi^{2,\ell}_n,\, w_{0_\ell} \rangle$, so that each blip does not depend on the action taken in the other period. Under Assumption~\ref{assumption:LTV_well_supported_factors}, blips and baselines of different units are linear combinations of one another, which makes the identification problem akin to those in SC and SI.

We identify the three components in \eqref{eq:illustration} recursively from three types of donors, listed below with their observed treatment sequences.
\begin{center}
    \footnotesize
    \begin{tabular}{@{}ccc@{}}
        \toprule
        Donor Type & Donor Set & Treatment Sequence \\
        \midrule
        1 & $\Ic^0_2$ & $(0_1, 0_2)$ \\
        2 & $\Ic^{d_2}_2$ & $(0_1,d_2)$ \\
        3 & $\Ic^{d_1}_1$ & $(d_1,D_2)$ \\
        \bottomrule
    \end{tabular}
\end{center}
Let $\mathbb{E}[\,\cdot \mid \Ic\,]$ denote the expectation conditional on $n \in \Ic$.

\emph{Step 1.} For Type 1 donors the entire sequence $(0_1,0_2)$ is fixed by the donor set, so Assumption~\ref{assumption:SUTVA} and condition (ii) of \eqref{eq:LTV_donor} for $\Ic^0_2$ give $\mathbb{E}[Y_{n} \mid \Ic^0_2] = \mathbb{E}\big[Y_{n}^{(0_{1},0_{2})} \,\big|\, \Ic^0_2\big] = \mathbb{E}\big[Y_{n}^{(0_{1},0_{2})}\big]$. Hence the baseline is ``observed'' for Type 1 donors. A linear combination of these baselines (Assumption~\ref{assumption:LTV_well_supported_factors}) identifies the baseline of \emph{every} unit.

\emph{Step 2.} For Type 2 donors the entire sequence $(0_1,d_2)$ is likewise fixed by the donor set, so $\mathbb{E}[Y_n \mid \Ic^{d_2}_2]=\mathbb{E}[Y_n^{(0_1,d_2)}\mid \Ic^{d_2}_2]$. Assumption~\ref{assumption:LTV_seq_exog} at $t=2$, given $(D_1,D_2)=(0_1,d_2)$, together with condition (ii) of \eqref{eq:LTV_donor} for $\Ic^{d_2}_2$, then gives $\mathbb{E}[Y_n^{(0_1,d_2)}\mid \Ic^{d_2}_2]=\mathrm{Blip}_2(d_2)+\mathbb{E}[Y_n^{(0_1,0_2)}]$, so $\mathrm{Blip}_2(d_2)=\mathbb{E}[Y_n\mid\Ic^{d_2}_2]-\mathbb{E}[Y_n^{(0_1,0_2)}]$ is ``observed'' for Type 2 donors, because their baseline is identified in Step 1. A linear combination of these blips identifies $\mathrm{Blip}_{2}(d_{2})$ for \emph{every} unit.

\emph{Step 3.} For Type 3 donors only the first action is fixed by the donor set; the second action $D_2$ is observed but may be adaptive. Adding and subtracting $Y_{n}^{(d_{1},0_{2})}$ and $Y_{n}^{(0_{1},0_{2})}$, and using Assumption~\ref{assumption:SUTVA},
\begin{align*}
\mathbb{E}[Y_{n} \mid \Ic^{d_1}_1]
&= \mathbb{E}\big[Y_{n}^{(d_{1},D_{2})} - Y_{n}^{(d_{1},0_{2})} \,\big|\, \Ic^{d_1}_1\big]
 + \mathbb{E}\big[Y_{n}^{(d_{1},0_{2})} - Y_{n}^{(0_{1},0_{2})} \,\big|\, \Ic^{d_1}_1\big]
 + \mathbb{E}\big[Y_{n}^{(0_{1},0_{2})} \,\big|\, \Ic^{d_1}_1\big] \\
&= \mathbb{E}\big[\mathrm{Blip}_{2}(D_{2}) \,\big|\, \Ic^{d_1}_1\big] + \mathrm{Blip}_{1}(d_{1}) + \mathbb{E}\big[Y_{n}^{(0_{1},0_{2})}\big].
\end{align*}
The second equality applies Assumption~\ref{assumption:LTV_seq_exog} at $t=2$, given $(D_1,D_2)=(d_1,a)$ for each realized $a$ and then averaging over $D_2$, and at $t=1$, given $D_1 = d_1$; Assumption~\ref{assumption:LTV_factor_model} identifies the resulting $\gamma_{n,2,2}(a)$ and $\gamma_{n,2,1}(d_1)$ with $\mathrm{Blip}_{2}(a)$ and $\mathrm{Blip}_{1}(d_{1})$ of \eqref{eq:illustration}; and condition (ii) of \eqref{eq:LTV_donor} at $\bar{D}^1_j = d_1$ removes the conditioning in the last term. Rearranging,
\begin{equation*}
\mathrm{Blip}_{1}(d_{1})
= \mathbb{E}\big[Y_{n} - \mathrm{Blip}_{2}(D_{2}) \,\big|\, \Ic^{d_1}_1\big]
 - \mathbb{E}\big[Y_{n}^{(0_{1},0_{2})}\big],
\end{equation*}
which is ``observed'' for Type 3 donors: $Y_n$ and $D_2$ are data, $\mathrm{Blip}_{2}(\cdot)$ comes from Step 2 and the baseline from Step 1. A linear combination then identifies $\mathrm{Blip}_{1}(d_{1})$ for every unit, and \eqref{eq:illustration} identifies $\mathbb{E}[Y_{n}^{(d_{1},d_{2})}]$ for all units and all $(d_{1},d_{2})\in[A]^2$.

We now present the formal identification results for general $T$.

\LTVnum\begin{theorem}\label{thm:newLTV}
Let Assumptions~\ref{assumption:SUTVA} and \ref{assumption:LTV_factor_model}--\ref{assumption:LTV_seq_exog} hold.
Then, for any unit $n \in [N]$ and action sequence $\bar{d}^T \in [A]^T$, the expected counterfactual outcome can be expressed as:
\begin{align}
\Ex[Y_{n,T}^{(\bar{d}^T)}\mid \LFc]
=~& \sum_{t =1}^T \gamma_{n, T, t}(d_t) + b_{n,T} \mid \LFc, \tag{identification} \label{eq:LTV_identification}
\end{align}
where quantities on the right-hand side are identified as follows:

\noindent (i) We identify the baseline outcomes
\begin{align}
&\forall \ j \in \Ic^0_T~:~   b_{j,T} \mid \LFc  = \Ex[Y_{j, T} \mid \LFc, \ j \in \Ic^0_T], \tag{observed control}\label{eq:base_observed} \\
&\forall \ i \notin \Ic^0_T~:~  b_{i,T}\mid \LFc  =  \sum_{j \in \Ic^0_T}  \beta_j^{i,\Ic^0_T} \ b_{j,T} \mid \LFc, \Ic^0_T. \tag{synthetic control}\label{eq:base_synthetic}
\end{align}
(ii) We identify the blip effect at time $T$ for all $d \in [A]$:
\begin{align}
\forall \ j \in \Ic^d_T~:~& \gamma_{j, T, T}(d) \mid \LFc  =~ \Ex[Y_{j, T} \mid \LFc, \ j \in \Ic^d_T] -  b_{j, T} \mid \LFc, \tag{``observed'' blip at time $T$}\label{eq:base_blip_observed} \\
\forall \ i \notin \Ic^d_T~:~&  \gamma_{i, T, T}(d) \mid \LFc  =~  \sum_{j \in \Ic^d_T} \beta_j^{i, \Ic^d_T}
 \gamma_{j, T, T}(d) \mid \LFc, \Ic^d_T. \tag{synthetic blip at time $T$}\label{eq:base_blip_synthetic}
\end{align}
(iii) We identify the blip effect $\forall \ t < T, \ d \in [A]$:
\begin{align}
\forall \ j \in \Ic^d_t&~:~
\gamma_{j, T, t}(d) \mid \LFc  =~ \Ex\Big[Y_{j, T} - \sum_{\ell=t+1}^T  \gamma_{j,T, \ell}(D_{j, \ell}) \,\Big|\, \LFc, \Ic^d_t\Big] -  b_{j, T} \mid \LFc, \tag{``observed'' blip at time $t$}\label{eq:recursive_blip_observed} \\
\forall \ i \notin \Ic^d_t&~:~
\gamma_{i, T, t}(d) \mid \LFc  =~ \sum_{j \in \Ic^d_t} \beta_j^{i, \Ic^d_t}
\gamma_{j, T, t}(d) \mid \LFc, \Ic^d_t  \tag{synthetic blip at time $t$}\label{eq:recursive_blip_synthetic}.
\end{align}
%
\end{theorem}

\noindent{\bf Interpretation of identification result.}
\eqref{eq:LTV_identification} states that our target causal parameter of interest can be written as an additive function of $b_{n, T}$ and $\gamma_{n,T,t}(d_t)$ for $t \in [T]$ and $d_t \in [A]$.
Theorem~\ref{thm:newLTV} establishes that these various quantities can be expressed as functions of observed outcomes $\{Y_{j, T}\}_{j\in [N]}$, or more precisely, of the expected observed outcome conditional on the latent factors. This suffices because the latter can be estimated via principal component regression below. 

{\em Identifying baseline outcomes.}
For units $j \in \Ic^0_T$, \eqref{eq:base_observed} states that their baseline outcome $b_{j, T}$ is simply their expected observed outcome at time step $T$, i.e., $Y_{j,T}$.
For units $i \notin \Ic^0_T$, \eqref{eq:base_synthetic} states that we can identify $b_{i, T}$ by appropriately re-weighting the baseline outcomes $b_{j, T}$ of the units $j \in \Ic_T^0$ (identified via \eqref{eq:base_observed}).

{\em Identifying blip effects at time $T$.}
For any given $d \in [A]$:
For units $j \in \Ic^d_T$, \eqref{eq:base_blip_observed} states that their blip effect $\gamma_{j,T,T}(d)$ is equal to their observed outcome $Y_{j,T}$ minus the baseline outcome $b_{j, T}$ (identified via \eqref{eq:base_synthetic}).
For units $i \notin \Ic^d_T$, \eqref{eq:base_blip_synthetic} states that we can identify $\gamma_{i,T,T}(d)$ by appropriately re-weighting the blip effects $\gamma_{j,T,T}(d)$ of units $j \in \Ic^d_T$ (identified via \eqref{eq:base_blip_observed}).

{\em Identifying blip effects at time $t < T$.}
Suppose by induction $\gamma_{n, T, \ell}(d)$ is identified for every $\ell \in [t + 1, T]$, $n \in [N]$, $d \in [A]$, i.e., can be expressed in terms of observed outcomes.
Then for any given $d \in [A]$:
For units $j \in \Ic^d_t$, \eqref{eq:recursive_blip_observed} states that their blip effect $\gamma_{j,T,t}(d)$ is equal to their observed outcome $Y_{j,T}$ minus the baseline outcome $b_{j, T}$ (identified via \eqref{eq:base_synthetic}) minus the sum of blip effects $\sum_{\ell=t+1}^T \gamma_{j,T, \ell}(D_{j, \ell})$ (identified via the inductive hypothesis).
For units $i \notin \Ic^d_t$, \eqref{eq:recursive_blip_synthetic} states that we can identify $\gamma_{i,T,t}(d)$ by appropriately re-weighting the blip effects $\gamma_{j,T,t}(d)$ of units  $j \in \Ic^d_t$ (identified via \eqref{eq:recursive_blip_observed}).

The analogous result for the LTI model is the following.

\addtocounter{theorem}{-1}\LTInum\begin{theorem}\label{thm:newLTI}
Let Assumptions~\ref{assumption:SUTVA} and \ref{assumption:LTI_factor_model}--\ref{assumption:LTI_seq_exog} hold.
Then, for any unit $n \in [N]$ and action sequence $\bar{d}^T \in [A]^T$, the expected counterfactual outcome can be expressed as:
\begin{align}
\Ex[Y_{n,T}^{(\bar{d}^T)}\mid \LFc]
=~& \sum_{t =1}^T \gamma_{n, T - t}(d_t) + b_{n, T} \mid \LFc, \tag{identification} \label{eq:LTI_identification}
\end{align}
where quantities on the right-hand side are identified as follows:

\noindent (i) We identify the baseline outcomes for all $t \in [T]$:
\begin{align}
&\forall \ j \in \Ic^{0}_t~:~   b_{j,t} \mid \LFc  = \Ex[Y_{j, t} \mid \LFc, \ j \in \Ic^{0}_t], \tag{observed control}\label{eq:LTI_base_observed} \\
&\forall \ i \notin \Ic^{0}_t~:~  b_{i,t}\mid \LFc  =  \sum_{j \in \Ic^{0}_t}  \beta_j^{i,\Ic^{0}_t} \ b_{j,t} \mid \LFc, \Ic^{0}_t. \tag{synthetic control}\label{eq:LTI_base_synthetic}
\end{align}
(ii) We identify the blip effect with $0$ lag for all $d \in [A]$:
\begin{align}
\forall \ j \in \Ic^d~:~& \gamma_{j, 0}(d) \mid \LFc  =~ \Ex[Y_{j, t^*_j} \mid \LFc, \ j \in \Ic^d] -  b_{j, t^*_j} \mid \LFc, \tag{``observed'' lag $0$ blip}\label{eq:LTI_base_blip_observed} \\
\forall \ i \notin \Ic^d~:~&  \gamma_{i, 0}(d) \mid \LFc  =~  \sum_{j \in \Ic^d} \beta_j^{i, \Ic^d}
 \gamma_{j, 0}(d) \mid \LFc, \Ic^d. \tag{synthetic lag $0$ blip}\label{eq:LTI_base_blip_synthetic}
\end{align}
(iii) We identify the blip effect $\forall \ t \in [T - 1], \ d \in [A]$:\footnote{The restriction $t^*_j + t \le T$ ensures that only in-sample outcomes are used; the lag-$t$ blip is then identified from donors with $t^*_j \le T - t$, which is not restrictive under the fixed-lag structure of Assumption~\ref{assumption:outcome-fixed-lag-dep} below, where only lags $\ell \le q$ are needed.
}
\begin{align}
\forall \ j \in \Ic^d |_{t^*_j + t \le T}&~:~
\gamma_{j, t}(d) \mid \LFc  =~ \Ex\Big[Y_{j, t^*_j + t} - \sum^{t - 1}_{\ell  = 0}  \gamma_{j, \ell}(D_{j, t^*_j + t - \ell}) \,\Big|\, \LFc, \Ic^d\Big] -  b_{j, t^*_j + t} \mid \LFc, \tag{``observed'' lag $t$ blip}\label{eq:LTI_recursive_blip_observed} \\
\forall \ i \notin \Ic^d&~:~
\gamma_{i, t}(d) \mid \LFc  =~ \sum_{j \in \Ic^d} \beta_j^{i, \Ic^d}
\gamma_{j, t}(d) \mid \LFc, \Ic^d.  \tag{synthetic lag $t$ blip}\label{eq:LTI_recursive_blip_synthetic}
\end{align}
%
\end{theorem}
The interpretation parallels that of Theorem~\ref{thm:newLTV}, with three differences.
Blips are indexed by lag rather than by calendar time; the donor sets $\Ic^d$ are indexed by the action alone; and the recursion for the lag-$t$ blip uses donors' outcomes at their (unit-specific) time $t^*_j + t$, so the lag-$0$ blip is identified from outcomes at the first treatment time and higher-lag blips recursively from later outcomes.

\subsection{Discussion: Complexity and Adaptivity}\label{sec:LTV_complexity}\label{sec:LTI_complexity}

{\bf Donor sample complexity.}
To estimate $\Ex[Y^{(\bar{d}^T)}_{n, T} ~|~ \LFc]$ for all units $n \in [N]$ and any action sequence $\bar{d}^T \in [A]^T$, the LTV identification strategy requires a sufficiently large subset of donor units $\Ic^d_t$ for every $d \in [A]$ and $t \in [T]$.
However, the additional structure imposed by the time-varying factor model in Assumption~\ref{assumption:LTV_factor_model} leads to a decrease in sample complexity from $A^T$ to $A \times T$, compared with the general factor model in Assumption~\ref{assumption:general_latent_factor_model}. We view this as a useful trade-off between sample complexity and model flexibility.
The LTI identification strategy requires only a sufficiently large $\Ic^d$ for every $d \in [A]$, plus the control donor sets $\Ic^0_t$; since $\Ic^{0}_T \subseteq \Ic^{0}_t$ for all $t \in [T-1]$ (a unit under control through time $T$ is under control through any earlier $t$), it suffices that $\Ic^{0}_T$ is sufficiently large.
As a result, the total donor sample complexity needs to scale only on the order of $A + 1$: the time-invariant structure reduces the complexity from $A \times T$ to $A + 1$.

\begin{remark}[Relaxing additivity]\label{rem:hybrid}
The additive structure in Assumption~\ref{assumption:LTV_factor_model} can be relaxed to a hybrid structure that allows for flexible interaction within each fixed window of size $h$ (i.e., among the current and $h-1$ lagged treatments), while maintaining additivity across the windows: the latent factor $w_{\bar{d}^t}$ in Assumption~\ref{assumption:general_latent_factor_model} satisfies
\begin{align*}
    w_{\bar{d}^t} = [w_{\tilde d_1}, \dots, w_{\tilde d_t}],\qquad \tilde d_{\ell} = (d_{\ell-h+1},\dots,d_{\ell}),
\end{align*}
where we simply treat $d_{0}, d_{-1}, \dots, d_{-h+2}$ as control actions. Then the sample complexity can be bounded by $A^{h} \times T$. This remark further highlights the modeling-sampling trade-off.
\end{remark}

\begin{figure}
	\centering
	\begin{subfigure}[t]{0.48\linewidth}
	\centering
	\includegraphics[width=\linewidth]{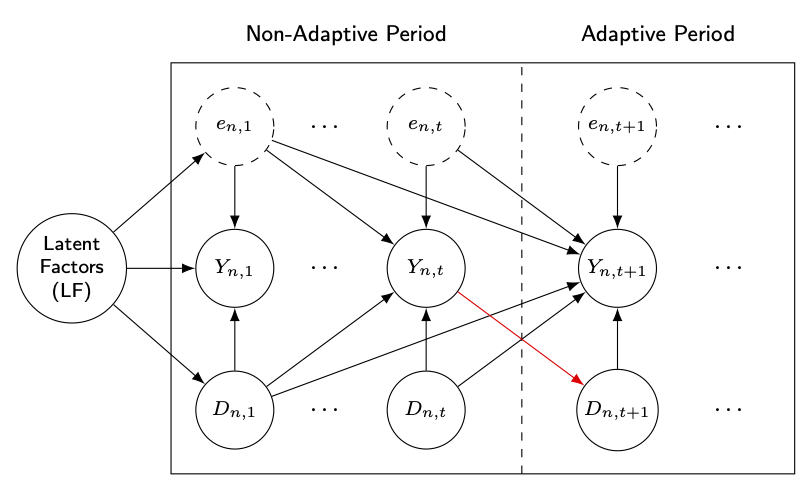}
	\caption{LTV donor set $\Ic^d_t$}
	\label{fig:LTV_Dag}
	\end{subfigure}\hspace{8pt}
	\begin{subfigure}[t]{0.41\linewidth}
	\centering
	\includegraphics[width=\linewidth]{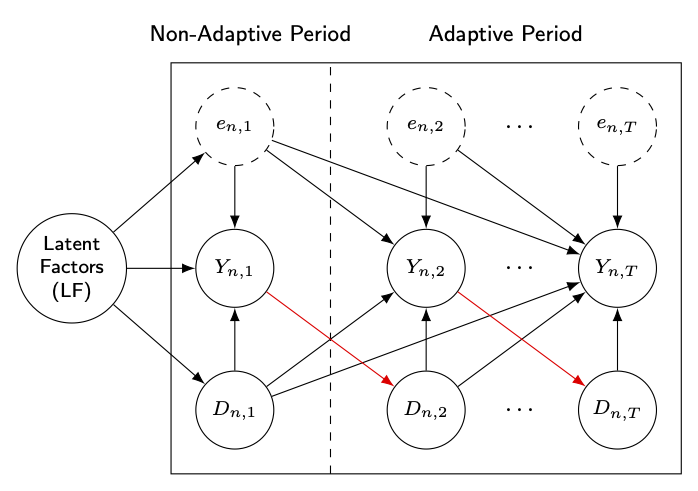}
	\caption{LTI donor set $\tilde\Ic^d$}
	\label{fig:LTI_Dag}
	\end{subfigure}
	\caption{
	    DAGs consistent with the exogeneity conditions implied by the donor set definitions.
	    (a) For $\Ic^d_t$, from time step $t + 1$ the action sequence $(D_{n, t + 1}, \dots, D_{n, T})$ can be adaptive, i.e., dependent on the observed outcomes $\{Y_{n, s}\}_{s \in [T]}$ (depicted by the red arrow).
	    (b) For $\tilde\Ic^d$, from time step $2$ the action sequence can be adaptive.
	}
	\label{fig:Dags}
\end{figure}

\noindent{\bf Donor exogeneity conditions.}
For $j \in \Ic^d_t$, we only require that $\Ex[\varepsilon^{(\bar{0}^T)}_{j, T} \mid \LFc, \bar{D}^t_{j}] = 0$: the actions picked for these donor units are only required to be non-adaptive until time step $t$, as opposed to being non-adaptive for the entire time period $T$, which was required for the SI identification strategy in Section~\ref{sec:identification_SI}.
The LTI strategy weakens this further: for $j \in \Ic^d$, non-adaptivity is required only until the unit's own first treatment time $t^*_j$.
As a special case, if we restrict ourselves to units $\tilde{\Ic}^d \coloneqq \{j \in \Ic^d: t^*_j = 1\}$, then the entire action sequence from time step $2$ can be adaptive.
Figure~\ref{fig:Dags} displays DAGs consistent with the exogeneity conditions implied by the definition of $\tilde{\Ic}^d$; these allow the feedback path (shown as the red arrow), which the SI strategy must rule out.

\section{Estimation: The \texttt{SBE-PCR} Algorithm}\label{subsection:SBE-Varying-Estimator}\label{subsection:SBE-Invariant-Estimator}

Here we introduce the algorithm that yields our estimator, which combines the \emph{synthetic blip effects} identification strategy of Section~\ref{sec:LTV_identification_strategy} with principal component regression (\texttt{SBE-PCR}), and then establish its consistency. We focus on the LTV model; the estimation algorithm and the consistency result for the LTI model are contained in the Appendix.

\subsection{The Algorithm}

The \texttt{SBE-PCR} algorithm uses additional covariates to estimate the weights for the synthetic baselines and blips, and hence the expected observed outcomes given the latent factors. For example, when evaluating the effects of a sequence of advertising campaigns on purchases, the covariates may consist of each customer's characteristics and past purchases before the campaigns start. These variables are assumed to be noisy measurements of the same latent factors that determine the subsequent potential outcomes, and they provide the information needed to estimate the synthetic weights.

\COMnum\begin{assumption}[Additional Covariates]\label{assumption:additional-covariates}
    For each unit $n \in [N]$, we assume access to covariates $X_n \in \mathbb{R}^p$ such that each element satisfies
    \begin{equation}
    X_{n,k} = \langle v_{n, T}, \rho_k \rangle + \varepsilon_{n,k},
    \end{equation}
    where $v_{n, T}$ is the unit latent factor defined in Assumption~\ref{assumption:general_latent_factor_model}, $\rho_k \in \Rb^{m(T)}$ is the covariate-specific latent factor, and $\varepsilon_{n,k}$ is mean-zero noise independent of everything else.
\end{assumption}
Denote $X = [X_1, \dots, X_N] \in \mathbb{R}^{p \times N}$, where $X_n$ is unit $n$'s pre-treatment time-invariant covariates. We maintain this design for simplicity; more general time-varying covariates (including pre-treatment outcomes) are discussed in SA~\ref{subsection:notes-on-ltv-setting}. We impose the leave-one-out analog of Assumption~\ref{assumption:LTV_well_supported_factors} to develop an algorithm with consistent estimators.

\LTVnum\begin{assumption}[Well-supported factors]\label{control-factors-support}
        For any donor set, i.e., any $t \in [T]$, $d \in [A]$, and unit $n \in \Ic_t^d$, there exist weights $\phi^{n, \Ic_t^d} \in \Rb^{|\Ic_t^d| - 1}$ such that $v_{n,T} = \sum_{k \in \Ic_t^d \setminus n} \phi_k^{n, \mathcal{I}_t^d} v_{k,T}$.
\end{assumption}

These assumptions allow us to detail the algorithm for estimating weights using principal component regression (PCR). Specifically in the LTV setting, for each $d \in [A]$, $t \in [T]$, and unit $n \in \Ic_t^d$, we estimate weights to express the response vector $X_n \in \Rb^p$ as a linear combination of the covariates from all other donor units in $\Ic_t^d$. The corresponding matrix of covariates is $X_{\Ic_t^d \setminus n} = X_{:,\Ic^d_t\setminus n} \in \Rb^{p \times |\Ic_t^d \setminus n|}$, which only chooses the relevant donor columns to have the leave-one-out structure. Similarly, we express $X_n$ for $n \notin \Ic_t^d$ as a linear combination of the covariates from donors in $\Ic_t^d$. Specifically, we apply PCR by regressing $X_n \in \Rb^p$ on the rank-$k_{\Ic_t^d \setminus n}$ approximation of $X_{\Ic_t^d \setminus n}$ with $k_{\Ic_t^d \setminus n} = \text{rank}(\EE[X_{\Ic_t^d \setminus n}|\LFc])$, i.e., conducting PCR with parameter $k_{\Ic_t^d \setminus n} \le \min \{p,|\Ic_t^d \setminus n|\}$.\footnote{Henceforth, $\LFc$ denotes the collection of the original latent factors together with the covariate factors $\{\rho_k\}_{k\in[p]}$.}

Denote the Singular Value Decomposition (SVD) of $X_{\Ic_t^d \setminus n}$ as
\[X_{\Ic_t^d \setminus n} = \sum_{l\geq 1}\hat{\sigma}_l\hat{u}_l\hat{v}_l^{\top},\]
where $\hat{u}_l \in \Rb^p$ and $\hat{v}_l \in \Rb^{|\Ic_t^d \setminus n|}$ are the left and right singular vectors arranged in decreasing order of corresponding singular values $\hat{\sigma}_l$.\footnote{Notice that, depending on whether unit $n$ is in the donor set $\Ic_t^d$, our covariate matrix size varies. This is intentional in order to unify notation between units in donor sets and those not in donor sets, since $\Ic_t^d \setminus n = \Ic_t^d$ if $n \notin \Ic_t^d$.} At this point, if $n \in \Ic_t^d$ we set
$$\hat{\phi}^{n, \Ic_t^d} = \left(\sum_{l =1}^{k_{\Ic_t^d \setminus n}}(1/\hat{\sigma}_l)\hat{v}_l \hat{u}_l^{\top}\right)X_n \in \Rb^{|\Ic_t^d| - 1},$$
and if $n \notin \Ic_t^d$ we set
$$\hat{\beta}^{n, \Ic_t^d} = \left(\sum_{l =1}^{k_{\Ic_t^d}}(1/\hat{\sigma}_l)\hat{v}_l \hat{u}_l^{\top}\right)X_n \in \Rb^{|\Ic_t^d|}.$$
The distinction between using $\beta$ and $\phi$ is to emphasize the difference in dimension (whether the unit is part of the donor set---interpolation---or not---extrapolation).
The use of PCR is natural here: under Assumption~\ref{assumption:additional-covariates}, the donor covariate matrix is a noisy, high-dimensional measurement of a low-rank latent-factor structure, so estimating weights from raw covariates is an errors-in-variables problem. In such settings, PCR consistently identifies the minimum-$\ell_2$ linear representation and generalizes out of sample when the target factor lies in the donor span (guaranteed for non-donor units by Assumption~\ref{assumption:LTV_well_supported_factors} and for donor units by Assumption~\ref{control-factors-support}), while remaining robust to noisy, sparse, and mixed-valued covariates \citep{agarwal2020principal,agarwal2020robustness,SI}.

Given our weight estimation procedure above, we are ready for our \texttt{SBE-PCR} algorithm:

\noindent{\bf Step 1: Estimate baseline outcomes.}
\begin{align*}
	\forall \ j \in \Ic^0_T~:~&\hat{b}_{j,T} = \sum_{k \in \Ic^0_T \setminus j}\hat{\phi}_k^{j, \Ic_T^0}Y_{k, T},\\
	\forall \ i \notin \Ic^0_T~:~&\hat{b}_{i,T} = \sum_{j \in \Ic^0_T} \hat{\beta}_{j}^{i,\Ic^0_T} \hat{b}_{j,T}.
\end{align*}

\noindent{\bf Step 2: Estimate blip effects at time $T$.} For $d \in [A]$:
\begin{align*}
	\forall \ j \in \Ic^d_T~:~&\hat{\gamma}_{j, T, T}(d) = \sum_{k \in \Ic^d_T \setminus j}\hat{\phi}_k^{j, \Ic_T^d}Y_{k, T} - \hat{b}_{j,T},\\
	\forall \ i \notin \Ic^d_T~:~&\hat{\gamma}_{i, T, T}(d) = \sum_{j \in \Ic^d_T} \hat{\beta}_{j}^{i,\Ic^d_T} \hat{\gamma}_{j, T, T}(d).
\end{align*}

\noindent{\bf Step 3: Recursively estimate blip effects for time $t < T$.} For $d \in [A]$ and $t \in \{T -1, \dots, 1\}$, recursively estimate as follows:
\begin{align*}
	\forall \ j \in \Ic^d_t~:~&\hat{\gamma}_{j, T, t}(d) = \sum_{k \in \Ic^d_t \setminus j}\hat{\phi}_k^{j, \Ic_t^d}\left(Y_{k, T} - \hat{b}_{k,T} - \sum_{\ell=t+1}^T \hat{\gamma}_{k,T,\ell}(D_{k,\ell})\right),\\
	\forall \ i \notin \Ic^d_t~:~&\hat{\gamma}_{i, T, t}(d) = \sum_{j \in \Ic^d_t} \hat{\beta}_{j}^{i,\Ic^d_t} \hat{\gamma}_{j, T, t}(d).
\end{align*}

\noindent{\bf Step 4: Estimate target causal parameter.} For $n \in [N]$ and $\bar{d}^T \in [A]^T$, estimate the causal parameter as follows:
\begin{align}\label{eq:causal_estimator}
\widehat{\Ex}[Y_{n, T}^{(\bar{d}^T)}] =~& \sum_{t =1}^T \hat{\gamma}_{n, T, t}(d_t) + \hat{b}_{n,T} .
\end{align}

We use the leave-one-out synthetic baseline in the first line of Step~1 (donor units $j \in \Ic^0_T$) rather than the raw observed outcome $Y_{j,T}$ because the estimand is the conditional mean baseline $b_{j,T}=\Ex[Y_{j,T}\mid \LFc,\ j \in \Ic^0_T]$, whereas $Y_{j,T}$ is only a single noisy realization. The PCR-based synthetic baseline pools information across control units that share the same latent-factor span, yielding a de-noised estimate.

\subsection{Consistency of  the \texttt{SBE-PCR} Estimator}\label{subsec:consistency-LTV}\label{subsec:consistency-LTI}

We state additional assumptions required to establish consistency.

\COMnum\begin{assumption}[Sub-Gaussian Noise and Bounded Conditional Means]
\label{assumption:sub-gaussian-noise}
For some constants $\sigma, C > 0$ and all $s \in [T]$, $\bar{d}^s \in [A]^s$:
(i) conditional on $\LFc$ and the treatment histories $\{\bar{D}^{s}_n\}_{n \in [N]}$, the outcome errors $\{\varepsilon_{n,s}^{(\bar{d}^s)}\}_{n \in [N]}$ and the covariate errors $\{\varepsilon_{n,k}\}_{n \in [N],\, k \in [p]}$ are mutually independent (across all $n$ and $k$) sub-Gaussian random variables, each with conditional variance at most $\sigma^2$ and conditional $\psi_2$ norm at most $C\sigma$; and
(ii)\label{assumption:bounded-expected-potential-outcomes}
$\Ex[Y_{n,s}^{(\bar{d}^s)} \mid \LFc]$ and $\Ex[X_{n,k} \mid \LFc]$ lie in $[-1,1]$ for all $n \in [N]$ and $k \in [p]$.
\end{assumption}

\LTVnum\begin{assumption}[Well-Balanced Singular Values and Row-Space Inclusion]\label{assumption:balanced-spectra}
For all $d \in [A]$ and $t \in [T]$:
(i) $\|\Ex[X_{\Ic^d_t}|\LFc]\|_F^2 \geq c'p|\Ic^d_t|$, where $X_{\Ic^d_t} \in \Rb^{p \times |\Ic^d_t|}$ is the relevant data matrix of observed covariates, and $\kappa^{-1} \geq c$, where $\kappa$ is the condition number of $\Ex[X_{\Ic^d_t}|\LFc]$, for constants $c, c' > 0$; and
(ii)\label{assumption:row-space} for $s \in \{t-1, t\}$, there exists a common vector $\xi^{(d,t,s)} \in \Rb^p$ such that, for every $j \in \Ic^d_t$,
$$
\Ex\!\left[Y_{j,T}^{(\bar{D}_j^{\,s},\, \underline{0}^{\,s+1})} \,\middle|\, \LFc, j \in \Ic^d_t\right]
= \sum_{i=1}^p \xi^{(d,t,s)}_i \cdot \Ex\!\left[(X_{\Ic_t^d})_{ij} \,\middle|\, \LFc, j \in \Ic^d_t\right],
$$
where $(\bar{D}_j^{\,s},\, \underline{0}^{\,s+1})$ denotes the donor's realized actions through period $s$ followed by the control sequence thereafter.\footnote{Both conditions (i) and (ii) are also assumed for the leave-one-out donor sets $\Ic_t^d \setminus n$.}
\end{assumption}

Assumption~\ref{assumption:sub-gaussian-noise} and part (i) of Assumption~\ref{assumption:balanced-spectra} are standard and closely follow the conditions presented in \cite[Section 4.3]{SI}. The row-space inclusion in part (ii) facilitates consistency by ensuring that the test data lie within the subspace spanned by the training data---specifically, within its row space---thereby enabling generalization of \texttt{SBE-PCR}; it is not a restrictive condition and is standard within the literature. SA~\ref{subsection:notes-on-ltv-setting} lists a sufficient condition for it.

We establish that the \texttt{SBE-PCR} estimator is consistent: the estimation error decays as the donor-set cardinalities and the covariate dimension $p$ grow. However, the error of the time-$t$ blip estimate compounds geometrically in the recursion depth, so the error for the target parameter scales as $k^T$, where $k$ is the (maximal) rank of the expected donor covariate matrices. This is benign for fixed $T$ (which we allow) but explosive as $T$ grows. It motivates the following restriction, under which outcomes depend only on the current and $q$ most recent treatments, so that the recursion depth is $q$ rather than $T$:

\COMnum\begin{assumption}[Fixed lag dependence]\label{assumption:outcome-fixed-lag-dep}
    Let the setup of Assumption~\ref{assumption:LTV_factor_model} hold. We further assume the counterfactual potential outcomes depend only on the current and $q$ most recent blips, for a constant $q$; namely, for all units $n \in [N]$ and $t \in [q+1, T]$ we have $\psi_n^{t, t-q-i} = 0$ for all $i \in [t-q-1]$. Notably, this implies that for any $n \in [N]$ and $\bar{d}^T \in [A]^T$ we have
    $$\EE[Y_{n,T}^{(\bar{d}^T)}|\LFc] = \sum_{\ell = T- q }^T\langle\psi_n^{T,\ell}, w_{d_{\ell}}\rangle.$$
    As a special case,\label{assumption:outcome-fixed-lag-dep-invariant} in the LTI model (Assumption~\ref{assumption:LTI_factor_model}) the same restriction reads $\psi_n^{q+i} = 0$ for all $i \in [T-q-1]$, so that $\EE[Y_{n,T}^{(\bar{d}^T)}|\LFc] = \sum_{\ell = T- q}^T\langle\psi_n^{T-\ell}, w_{d_{\ell}}\rangle$.
\end{assumption}

\LTVnum\begin{theorem}\label{thm:consistency-time-varying-fixed-lags}
    Let Assumptions~\ref{assumption:SUTVA}, \ref{assumption:LTV_factor_model}--\ref{assumption:LTV_seq_exog}, \ref{assumption:additional-covariates}, \ref{control-factors-support}, \ref{assumption:sub-gaussian-noise}, \ref{assumption:balanced-spectra}, and \ref{assumption:outcome-fixed-lag-dep} hold. Consider the \texttt{SBE-PCR} estimator in Section~\ref{subsection:SBE-Varying-Estimator} modified to only estimate the baseline, terminal blip, and previous $q$ blips, and suppose $k = \max_{d \in [0,A], t\in [T]}\text{rank}(\EE[X_{\Ic_t^d}|\LFc])$. Then, conditional on $\LFc$ and the donor-defining treatment histories $\{\bar{D}^{t}_{j} : j \in \Ic^d_t\}$ of all donor sets, we have for any $n \in [N]$ and $\bar{d}^T \in [A]^T$:
$$
\widehat{\Ex}[Y_{n, T}^{(\bar{d}^T)}] - \EE[Y_{n, T}^{(\bar{d}^T)}\mid \LFc]=O_p\left(\sqrt{\log(p\pi_{\Ic})}\left(\frac{k^{q+7/4}}{p^{1/4}}
+
k^{q+3}\max\left\{\frac{\sqrt{\pi_{\Ic}}}{p^{3/2}}, \frac{1}{\sqrt{\alpha_{\Ic}-1}}, \frac{1}{\sqrt{p}}\right\}\right)\right),
$$
where $\pi_{\Ic} = \max\Cc$ and $\alpha_{\Ic} = \min\Cc$ with $\Cc = \{|\Ic_T^0|, (|\Ic_{t}^{d_{t}}|)_{t \in [T-q,T]} ,(|\Ic_{t}^{D_{n,t}}|)_{n \in [N], t \in [T-q+1,T]}\}$.
\end{theorem}

Theorem~\ref{thm:consistency-time-varying-fixed-lags} concludes that, upon modifying the \texttt{SBE-PCR} estimator to account for the system depending on only a \textit{constant} $q$ lags, we have a consistent estimate of the causal estimand. More precisely, for fixed $k$, the estimation error decays as donor set cardinalities and the number of covariates $p$ grow, provided $p/\pi_{\mathcal{I}}^{1/3} \rightarrow\infty$. Crucially, the theorem allows $T \to \infty$ as well, which justifies the growing number of covariates by including time-varying covariates, i.e., $p$ can now depend on $T$ asymptotically (see SA~\ref{subsection:notes-on-ltv-setting}); in the empirical application of Section~\ref{sec:application}, we include time-varying covariates in estimation. The result establishes pointwise consistency, i.e., there is \emph{no} average across units in the statement. The proof is in SA~\ref{subsection:proof-of-varying-consistency-fixed-lags}; the analogous result for the LTI model is Theorem~\ref{thm:consistency-time-invariant-fixed-lags} in the Appendix.

\begin{remark}[Uniform guarantees]\label{rem:uniform}
The pointwise guarantees above can be upgraded to uniform---over
units and feasible treatment schedules---and hence average
consistency, provided sufficiently strong tail bounds make a
union bound over the relevant index set vanish.
For a fixed schedule and each fixed $\varepsilon>0$, an
$o(1/N)$ bound on the probability of an absolute error exceeding
$\varepsilon$, uniform over units, suffices; see
SA~\ref{app:uniform-consistency}.
The implications for policy learning and the corresponding conditions are discussed in
Remark~\ref{rem:plugin_EWM} and SA~\ref{sec:appendix_EWM}.
\end{remark}



\COMnum
\section{Application: Export Financial Support}\label{sec:application}

The goal of this section is to showcase the usefulness of our approach in understanding individual dynamic treatment effects and developing optimal targeting rules in a real-world panel data application. We first introduce the background on financial credit support for exporting firms and the data (Sections~\ref{subsec:backgrounds}--\ref{subsec:data}) and describe diagnostic exercises for the synthetic blip estimator (Section~\ref{subsec:model_validation}). We report the synthetic blip estimates of support impacts (Section~\ref{subsec:effects_financial_support}). We then investigate the extent of possible improvement of support allocation for each firm in the data (Section~\ref{subsubsec:optimal_firms_in_data}) and develop an optimal targeting rule for allocating support to new firms based on firm characteristics (Section~\ref{subsubsec:optimal_new_firms}).

\subsection{Background}\label{subsec:backgrounds}

Exporting is inherently risky, requiring firms to secure upfront working capital, offer extended payment terms, and protect themselves against non-payment or foreign market shocks. When trade finance dried up during the Great Recession, the resulting contraction disproportionately hit firms reliant on weak banks or operating in finance-dependent sectors \citep{Amiti2011, Chor2012, Paravisini2015}. These vulnerabilities matter particularly in economies that rely heavily on international trade. The Korean economy is a salient example of this, as exports accounted for 45--58\% of Gross Domestic Product (GDP) between 2006 and 2015, making the economy highly sensitive to fluctuations in global trade and financial conditions. In this environment, ECAs play a critical role by using public funds to provide insurance and loans that enable firms to sustain and expand their export activities.

Korea has two independent ECAs: the Korea Trade Insurance Corporation (K-SURE), which specializes in export insurance, and the Export-Import Bank of Korea (EXIM), which provides export loans. Firms seeking support apply to the relevant agency, which evaluates the application and selects firms based on observable firm characteristics---K-SURE assesses the creditworthiness of exporters and their foreign buyers, while EXIM evaluates financial stability and contract documents. Because these criteria depend not only on firm characteristics but also on performance after earlier support, the agencies' support decisions can be adaptive---an aspect our framework accommodates (Section~\ref{sec:LTV_complexity}). This selection on firm characteristics connects to our later analyses of heterogeneity and optimal treatment allocation. Moreover, the two agencies select firms independently, leaving room for improvement through coordination, which we investigate as well.

Using the firm-level data described below, we estimate the dynamic effects of the two treatments and then ask how treatments, timing, and allocation could be improved.

\subsection{Data and Variables}\label{subsec:data}

Our empirical analysis relies on a novel Korean firm-level panel dataset for 2006--2015 that links three sources of firm-level data: (i) the Survey of Business Activities (SBA) from Statistics Korea,\footnote{This annual survey provides detailed information on inputs, outputs, and trade activities of all incorporated firms with at least 50 employees and paid-in capital of at least 300 million Korean won.} which provides detailed firm characteristics; (ii) export insurance data from K-SURE; and (iii) export loan data from EXIM. Combining these sources allows us to track which firms received support, the form and timing of support, and their subsequent performance.

We construct a balanced panel and further restrict the sample to firms that received no export credit support during the first five years (2006--2010), so that all firms share a common, fully untreated initial period. This leaves 2,052 unique firms. The treatment group consists of 167 firms that received at least one form of support in the later period (2011--2015), and the control group consists of the remaining 1,885 firms, which were never supported during the sample period.

\begin{table}[!htbp]
\centering
\caption{Pre-treatment Summary Statistics: Estimation Sample}
\label{tab:summary_stats_pretreatment}
\footnotesize
\begin{tabular}{lcccc}
\toprule
Variable & Supported & Never-supported & $t$-statistic & $p$-value \\
\midrule
Export share & 0.207 & 0.183 & -1.272 & 0.203 \\
$\ln$(Sales) & 10.606 & 10.574 & -0.374 & 0.708 \\
Workers & 210.743 & 178.946 & -1.149 & 0.250 \\
$\ln$(Capital) & 9.317 & 9.225 & -0.946 & 0.344 \\
$\ln$(Value added) & 9.021 & 9.015 & -0.083 & 0.933 \\
TFP & 239.834 & 229.238 & -0.209 & 0.834 \\
$\ln$(Wage bill) & 8.507 & 8.534 & 0.385 & 0.700 \\
$\ln$(R\&D) & 5.340 & 4.566 & -3.479 & 0.001 \\
Debt-asset ratio & 0.553 & 0.525 & -1.434 & 0.151 \\
Liquidity ratio & 0.085 & 0.112 & 1.196 & 0.231 \\
FDI indicator & 0.382 & 0.299 & -2.123 & 0.034 \\
Parent affiliation & 0.092 & 0.228 & 3.827 & 0.000 \\
Establishment year & 1987 & 1985 & -1.145 & 0.252 \\
Industry capital intensity & 0.267 & 0.291 & 0.599 & 0.548 \\
Industry wage indicator & 0.450 & 0.458 & 0.195 & 0.844 \\
\bottomrule
\end{tabular}

\vspace{0.5em}
\begin{minipage}{\linewidth}
\footnotesize \textit{Note:} Pre-treatment averages over 2006--2010 for the estimation sample of 661 firms. \emph{Supported} firms receive at least one form of export credit support during 2011--2015 and none during 2006--2010 (162 firms); \emph{never-supported} firms receive none throughout (499 firms). The last two columns report two-sample $t$-tests for equality of means.
\end{minipage}
\end{table}

The outcome of interest is the export value of firm $n$ in year $t$, $Y_{n,t}$, reported in millions of US dollars.\footnote{Estimation is carried out in Korean won; reported figures are converted at 1,100 won per dollar, the average market exchange rate over 2006--2015 (Bank of Korea).} The vector of firm-level covariates, $X_{n,t}$, comprises the eleven time-varying and two time-invariant firm characteristics listed in Table~\ref{tab:summary_stats_pretreatment}, capturing firm performance, scale, financial constraints, and innovation capacity.\footnote{Total factor productivity (TFP) is measured as value added divided by $K^{1/3}L^{2/3}$, where $K$ denotes tangible capital stock and $L$ the number of workers. Sales, capital stock, value added, wage bill, and R\&D expenditure are expressed in natural logarithms, with the underlying unit being one million won.} At the industry level, we control for $Z_{m}$, a vector of two indicators for whether industry $m$ has above-average capital intensity and above-average wage per worker.\footnote{Capital intensity is defined as tangible capital stock per worker, and wage per worker is measured as the total wage bill divided by the number of workers.} Before any sample restriction, the two groups differ on ten of the fifteen characteristics: supported firms are larger and more export-intensive, and they carry more debt and less liquidity (SA Table~\ref{tab:summary_stats_full}).

These pre-treatment differences motivate restricting the candidate comparison pool before estimation. As \citet{abadie2021using} and \citet{abadie2021penalized} emphasize, comparison units with pre-treatment characteristics far from those of the treated units can degrade counterfactual accuracy, especially when the candidate pool is large. Although the \texttt{SBE-PCR} estimator does not solely rely on convex combinations of donor outcomes, a similar concern can arise. We therefore restrict the pool by matching never-supported firms to each supported firm based on pre-treatment characteristics measured over 2006--2010. We match on two-digit industry, export share, and sales.\footnote{Within two-digit industries, donor pools are formed using each treated firm's third-nearest-neighbor distance in pre-treatment characteristics, with a 10-percent tolerance around that distance.} Industry captures sector-specific export dynamics, while export share and sales capture two key dimensions of firm heterogeneity, namely export intensity and firm size. We further refine the resulting pool by dropping firms with the poorest fit---those with the largest ratios of post-treatment to pre-treatment mean squared prediction error (MSPE), as described in Section~\ref{subsec:model_validation}.\footnote{We drop five supported and ten never-supported firms with the largest post-treatment to pre-treatment MSPE ratios.} The final estimation sample includes 661 firms, with 162 supported firms and 499 never-supported firms. Table~\ref{tab:summary_stats_pretreatment} reports pre-treatment averages for this estimation sample: the two groups are now statistically indistinguishable on twelve of the fifteen characteristics, the exceptions being R\&D intensity, FDI status, and parent affiliation.

In each year $t\in \{T_0 +1,\dots,T_0+T\}$, firm $n$ receives treatment $D_n\in[0,A]=\{0,1,2,3\}$: $D_n=1$ if firm $n$ receives insurance, $D_n=2$ if loans, $D_n=3$ if both, and $D_n=0$ if none. In this application, $T_0 = 5$ and $t_n^{*} = T_0+1 = 6$ for all $n$ (i.e., no firm receives treatment through period $T_0$), and $T=5$; recall that $\bar{d}^t=(d_{1},\dots,d_{t})$ for $t \in \{1,\dots,T\}$, which here concerns the periods after $T_0$. The dimension of covariates in the PCR (analogous to the number of pre-treatment periods in the standard SC) is $p = 11\times 5 + 2 + 2 + 5 = 64$, with each term corresponding to time-varying firm characteristics over the pre-treatment years 2006--2010, time-invariant firm characteristics, time-invariant industry-level characteristics, and pre-treatment outcomes, respectively. We assume the LTV latent factor model (Section~\ref{sec:time_varying_systems}) and the number of lags to be $q=1$ in Assumption~\ref{assumption:outcome-fixed-lag-dep}. Using the \texttt{SBE-PCR} algorithm, we estimate $\Ex[Y_{n,T_0+t}^{(\bar{d}^{t})}\mid\LFc]$ for given $\bar{d}^t \in [0,A]^t$ and for each firm $n$ and $t\in [T]$ (i.e., the last five periods) in the data.\footnote{Unlike in the theoretical sections, the expression $\Ex[Y_{n,T_0+t}^{(\bar{d}^{t})}\mid\LFc]$ is explicit about the no-treatment periods. We can still estimate $\Ex[Y_{n,T_0+t}^{(\bar{d}^{t})}\mid\LFc]$ at $t = 1$ even when $q \ge 1$, because lagged treatments can be set to zero, as no firm is treated through year $T_0$.} These quantities are the crucial ingredient for all analyses below: they are used to calculate average counterfactual outcomes and average treatment effects, and to conduct policy learning.

\subsection{Diagnostic Exercises}\label{subsec:model_validation}
Before turning to the main results, we report two diagnostics for the credibility of the \texttt{SBE-PCR} counterfactual estimates: within-donor-set fit and pre-treatment fit.

\begin{figure}
    \centering
    \subfloat[Baseline]{\includegraphics[width=0.48\textwidth]{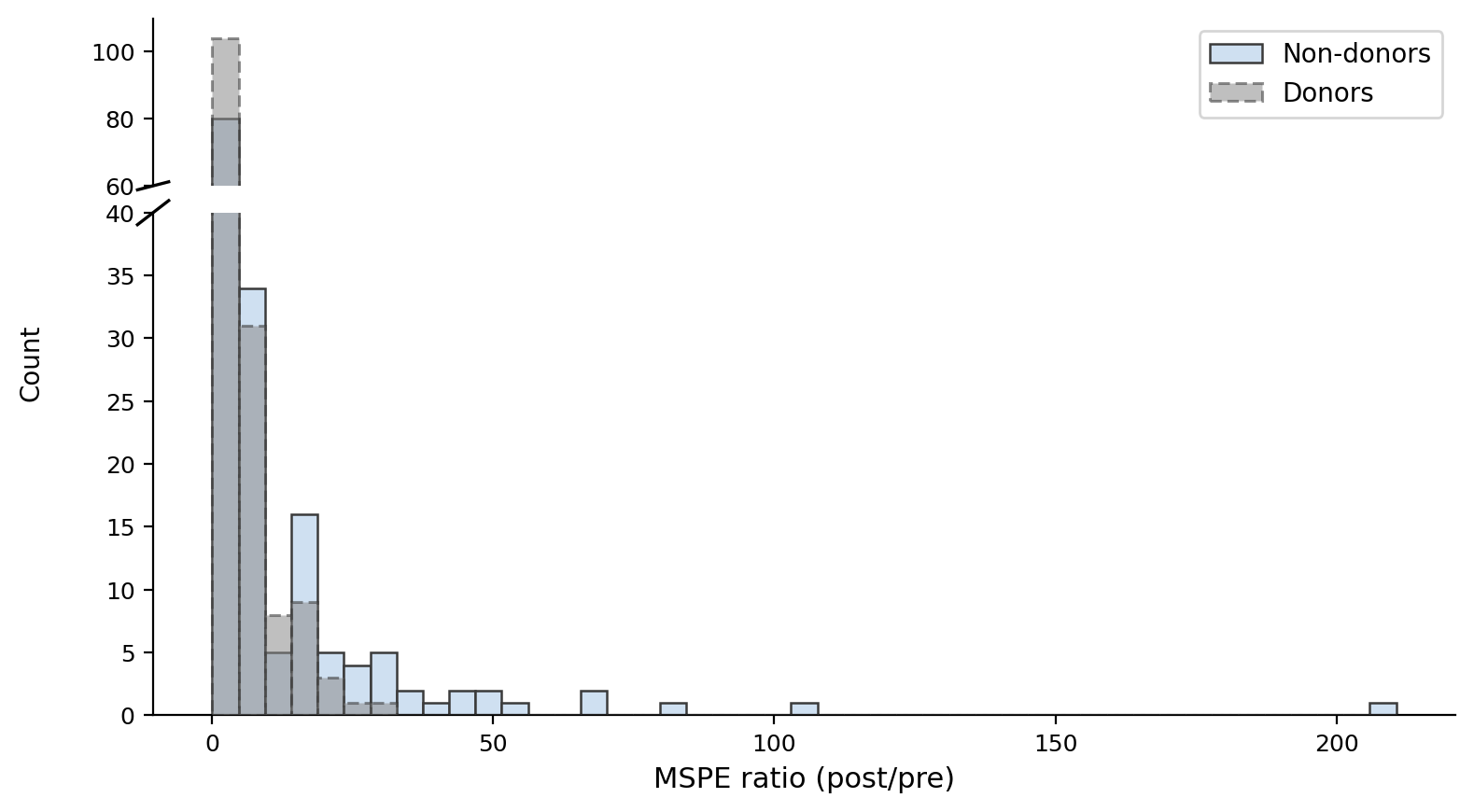}}\hspace{0.02\textwidth}
    \subfloat[Pooled]{\includegraphics[width=0.48\textwidth]{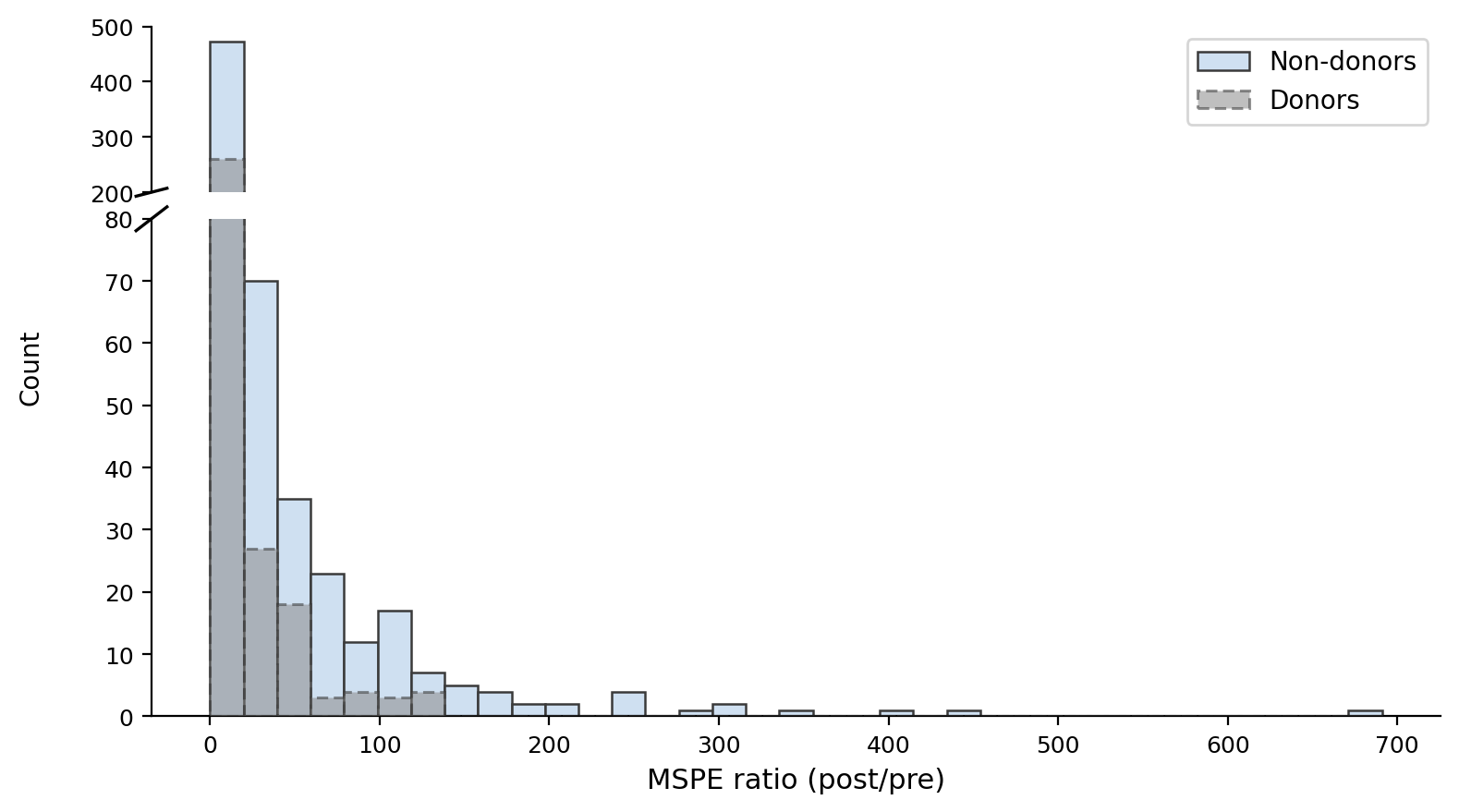}}
    \caption{Histogram of MSPE Ratios (Post/Pre)}
    \vspace{0.5em}
    \noindent\begin{minipage}[t]{1\columnwidth}%
    \small Notes: Each panel shows the distribution for donors and non-donors. Panel~(a) uses the baseline donor set, which estimates the no-support counterfactual; Panel~(b) pools the baseline set and all other donor sets used in the recursive estimation. All bars beyond 30 in Panel~(a) and beyond 140 in Panel~(b) correspond to non-donors.%
    \end{minipage}
    \label{fig:histogram}
\end{figure}

We extend the diagnostic approach of \citet{abadie2} to the sequential procedure of \texttt{SBE-PCR} that involves multiple target periods (each of which requires a separate \texttt{SBE-PCR} run) and potentially many donor sets. First, for each donor set, we compute the ratios of post-treatment to pre-treatment MSPE for firms within the donor set (i.e., donors) and those outside (i.e., non-donors); the MSPE ratio comparison of these two groups can be viewed as a per-donor-set placebo exercise. We then pool the ratios across donor sets for donors and non-donors. Absent systematic post-treatment divergence in within-donor fit, the ratios should be less dispersed and less shifted toward large values for donors than non-donors. Figure~\ref{fig:histogram} plots the distribution of the MSPE ratio separately for the two groups.\footnote{For this diagnostic exercise, donors and non-donors are randomly subsampled to equal numbers within each donor set (40 per set; 167 for the baseline set) so that the comparison is not driven by group size. In each panel, the vertical axis is broken so that the outsized first bin does not compress the others.} Panel~(a) uses only the baseline donor set used to estimate the no-support counterfactual for the terminal outcome. Panel~(b) pools observations across all donor sets used in the recursive estimation procedure, including the baseline set.\footnote{There are $3 \times 5 = 15$ treatment--period donor sets
$\Ic^{d}_{t}$, for treatments $d \in \{1,2,3\}$ and periods
$t \in \{1,\dots,5\}$, together with the baseline (never-supported) set,
giving 16 in total. No firm in our sample first receives both treatments
(i.e., $d = 3$) in period 4 or 5, so $\Ic^{3}_{4}$ and $\Ic^{3}_{5}$ are
empty and are dropped; Panel~(b) therefore pools 14 sets. This does not affect the policy-evaluation exercises, but it restricts the policy-learning exercise, as discussed in Section~\ref{subsec:optimal_targeting}.} In both panels, the MSPE-ratio distribution for non-donors is more dispersed and more shifted toward large values than for units in the donor sets.

A second diagnostic examines whether the estimated counterfactual outcomes track observed outcome trajectories during the pre-treatment period. Figure~\ref{fig:dynamic_treatment_effect}(a) below reports the average difference between observed outcomes and estimated counterfactual outcomes over time.\footnote{Note that the pre-treatment fit can be calculated for each set of \texttt{SBE-PCR} weights corresponding to each target period $t \ge T_0 + 1$ and firm $n$. The reported average is taken over all such pairs $(n, t)$. Thus, this average should be compared to the post-treatment fit (currently averaged across $n$ in the figure) after it is averaged across $(n, t)$ pairs.} The pre-treatment residuals remain centered around zero, contrasting with the larger estimated treatment effects reported for the post-treatment period.

\subsection{Dynamic Effects of Financial Support}\label{subsec:effects_financial_support}

\begin{figure}
\centering
\subfloat[Dynamic treatment effects]{\includegraphics[width=0.38\textwidth]{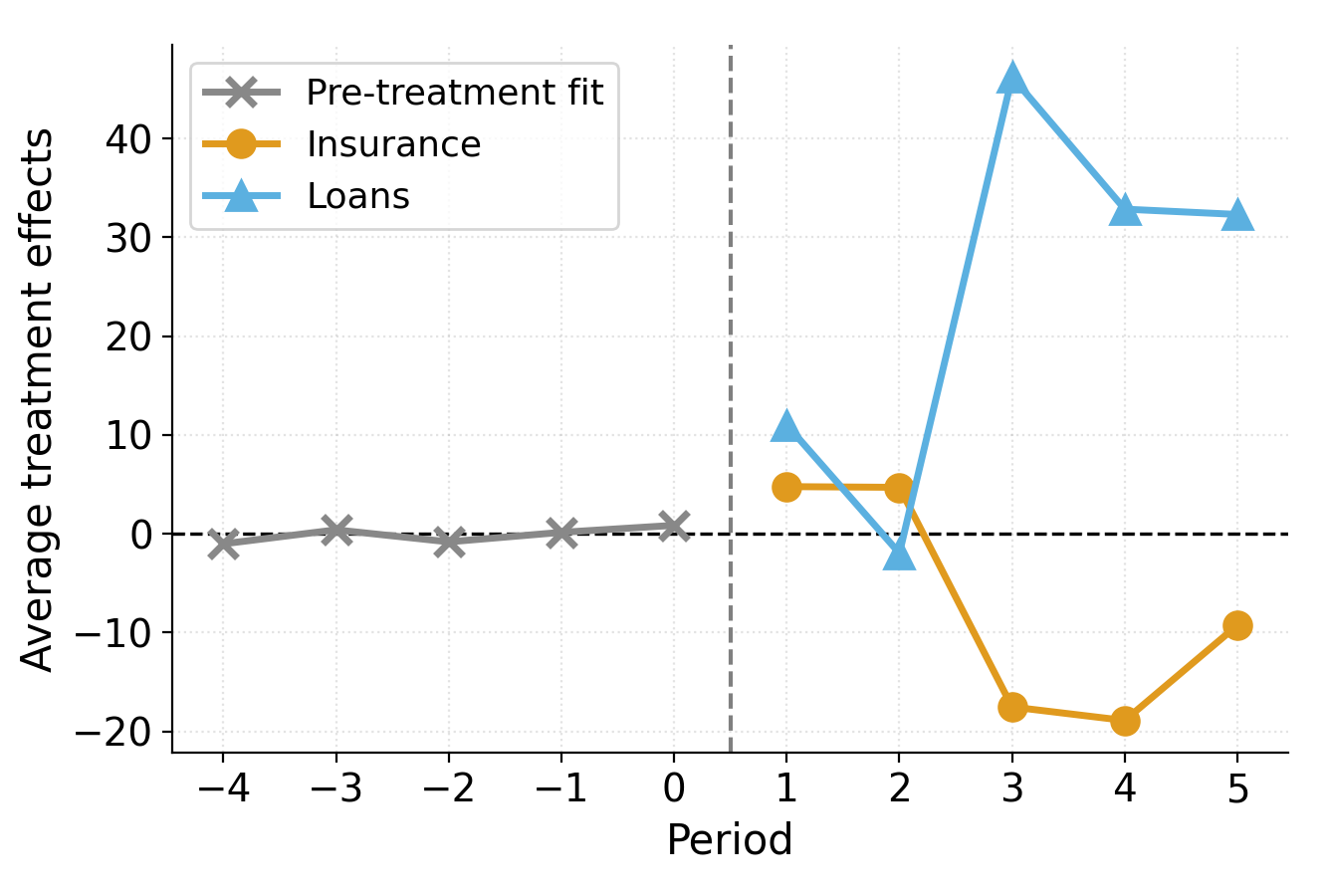}}\hspace{0.06\textwidth}
\subfloat[Heterogeneity\label{fig:dynamic_treatment_effect_hetero}]{\includegraphics[width=0.38\textwidth]{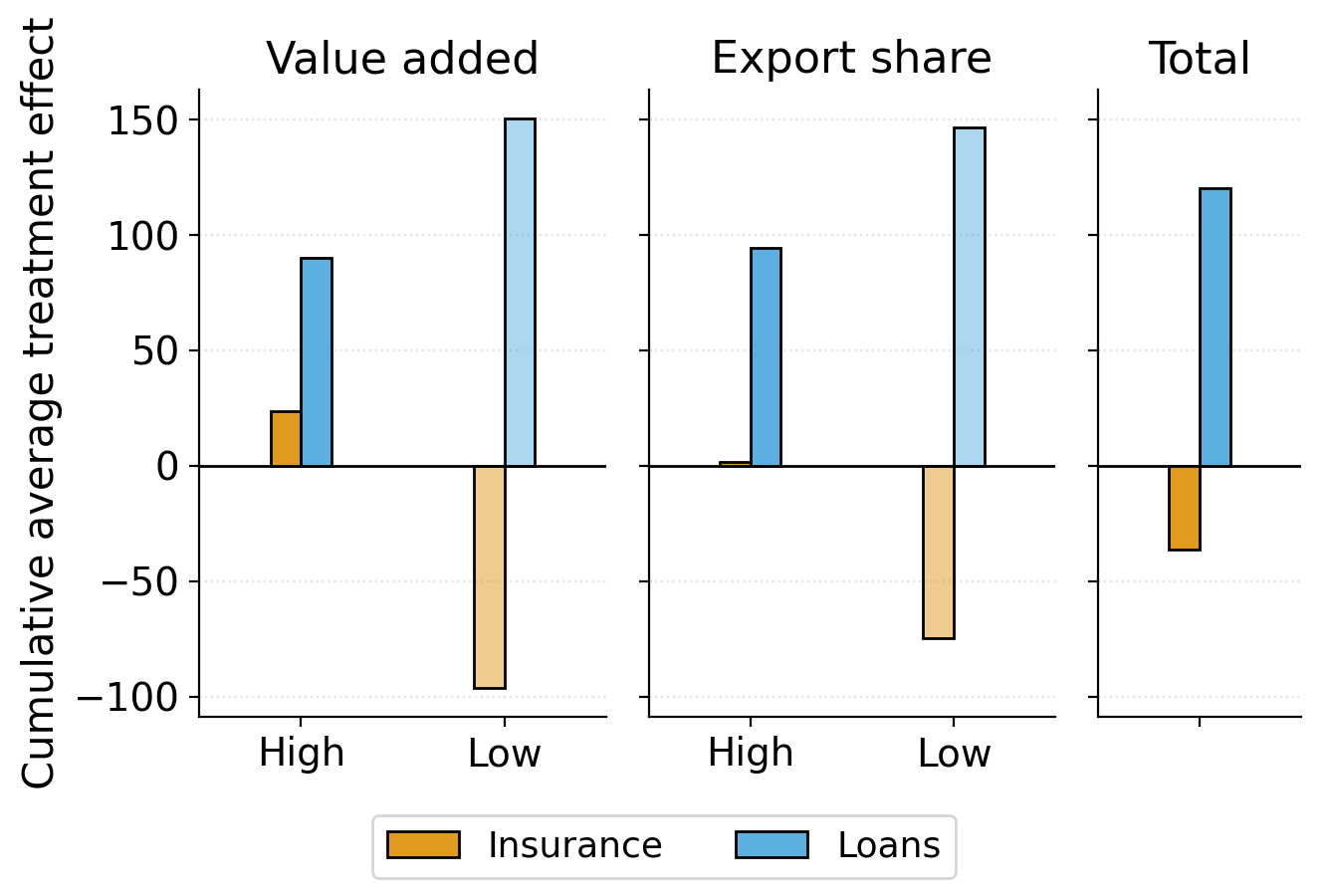}}
\caption{Dynamic Treatment Effects on Export Values}
\vspace{0.5em}
\begin{minipage}[t]{\linewidth}%
\small Notes: Panel~(a) depicts the trajectory of average treatment effects across firms, with units in millions of USD. The cumulative average treatment effect amounts to $-36.3$ for insurance and 120.4 for loans. Panel~(b) depicts cumulative treatment effects by firm group, where firms are split at the sample median of pre-treatment value added and export share, with the rightmost column reporting the full sample.%
\end{minipage}
\label{fig:dynamic_treatment_effect}
\end{figure}

To understand time-varying effects of the financial support, we report the trajectories of average dynamic treatment effects. We consider two exercises---sustained support from each instrument, and alternative schedules that vary the timing of a fixed total---each reported on average across firms and then by firm group. We begin with sustained support, where \emph{the average is taken across firms}. Each support sequence is defined as $\bar{d}^T=(d,d,d,d,d)$ for $d=1$ or $2$ (again, $1$ being insurance support and $2$ being loans support). Figure~\ref{fig:dynamic_treatment_effect}(a) reports both the pre-treatment fit of the synthetic counterfactual and the post-treatment effects of the two treatment types.

The two treatments exhibit sharply contrasting dynamic patterns. Insurance displays modest positive effects initially, followed by sizable negative effects in later periods, resulting in a cumulative average treatment effect of $-36.3$ million USD. This pattern should be interpreted with caution, as it is consistent with two distinct explanations that we discuss in turn. First, the negative estimate may arise from comparing publicly with privately insured firms rather than from the effect of insurance itself. Because export transactions are rarely undertaken without some form of coverage, firms without K-SURE support are unlikely to be truly uninsured; many rely on private insurance or other arrangements we do not observe. Since public insurance is designed to cover transactions that the private market will not, publicly insured firms plausibly carry higher underlying risk. Second, the negative estimate may also reflect a behavioral response to insurance (i.e., moral hazard): by shifting transaction risk off the firm, insurance can lead firms to enter riskier markets or take on more uncertain projects, which can lower realized performance even when insurance expands the set of feasible transactions.

Loans, by contrast, generate positive effects that grow over time, yielding a cumulative average treatment effect of 120.4 million USD. This pattern is consistent with a working-capital channel: export production requires firms to finance inputs---materials and labor---and to fulfill contracts before export revenues are realized, and loans directly relax this constraint \citep{antras2015, Paravisini2015, niepmann2017}. By covering these upfront costs, loans enable firms to carry out production and, in turn, expand their export sales.

We next ask whether the two explanations above are consistent with how effects vary across firms. We split firms along two dimensions: value added (i.e., a firm's own production scale) and export sales share. For each dimension, we average over the pre-treatment period and split firms at the sample median.

Figure~\ref{fig:dynamic_treatment_effect}(b) reports cumulative treatment effects by firm characteristics. For insurance, firms with high value added or high export share exhibit small positive effects, whereas firms with low value added or low export share experience substantially negative effects. For loans, effects are positive across all groups, but larger for firms with weaker pre-treatment characteristics. These patterns are consistent with the explanations above. For insurance, the negative effects are largest among smaller firms, which also receive the most complete coverage from K-SURE.\footnote{For short-term export insurance covering general exports and consignment processing trade, K-SURE reports compensation ratios of 100\% for small, 97.5\% for medium, and 95\% for large firms. However, the applicable coverage may vary depending on country-specific underwriting policies.} For loans, the larger gains among low value added or low export share firms are consistent with binding working-capital constraints: for these firms, additional liquidity can translate more directly into production and export sales.

\begin{figure}
    \centering
    \subfloat[Insurance]{\includegraphics[width=0.38\textwidth]{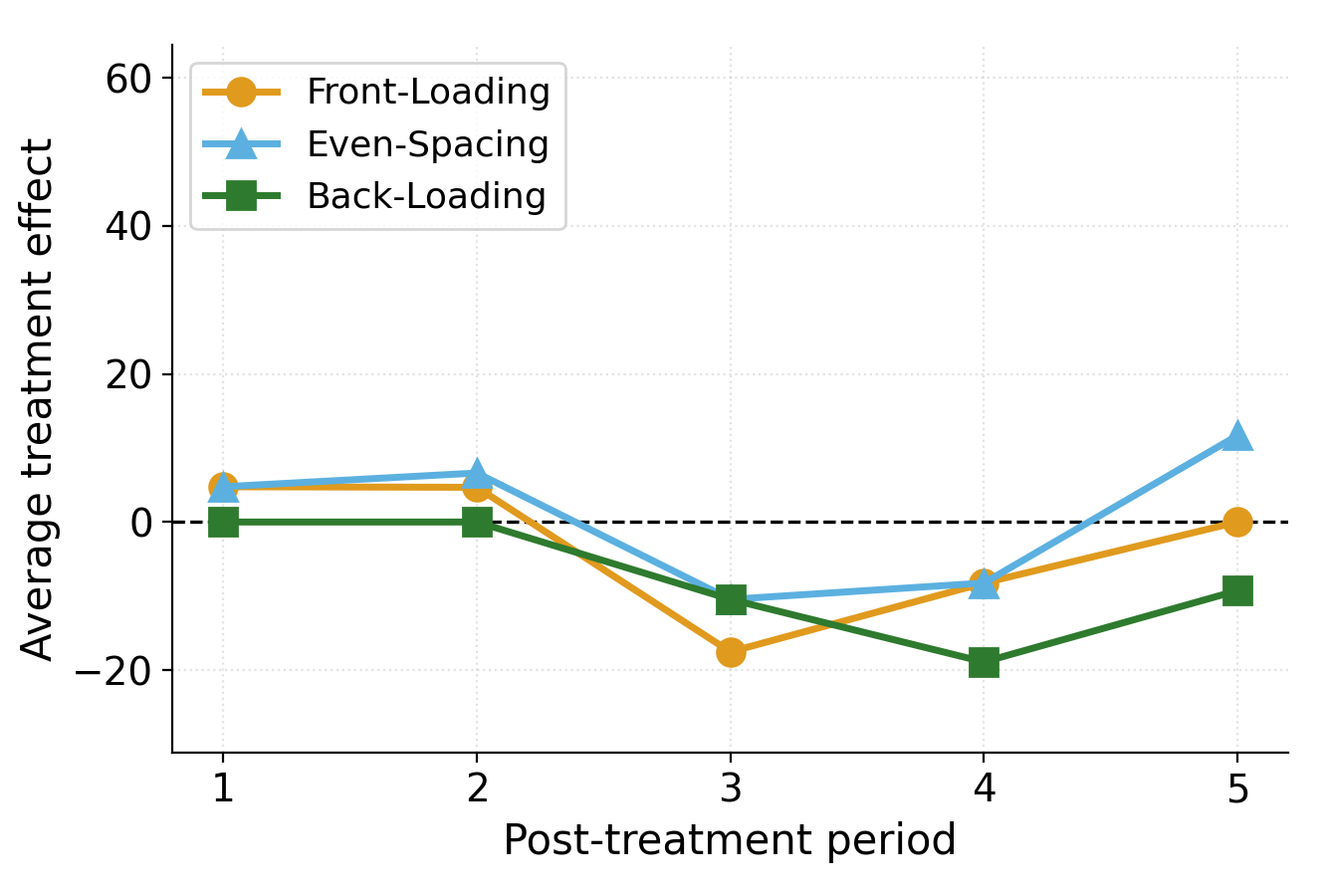}}\hspace{0.06\textwidth}
    \subfloat[Insurance: heterogeneity]{\includegraphics[width=0.38\textwidth]{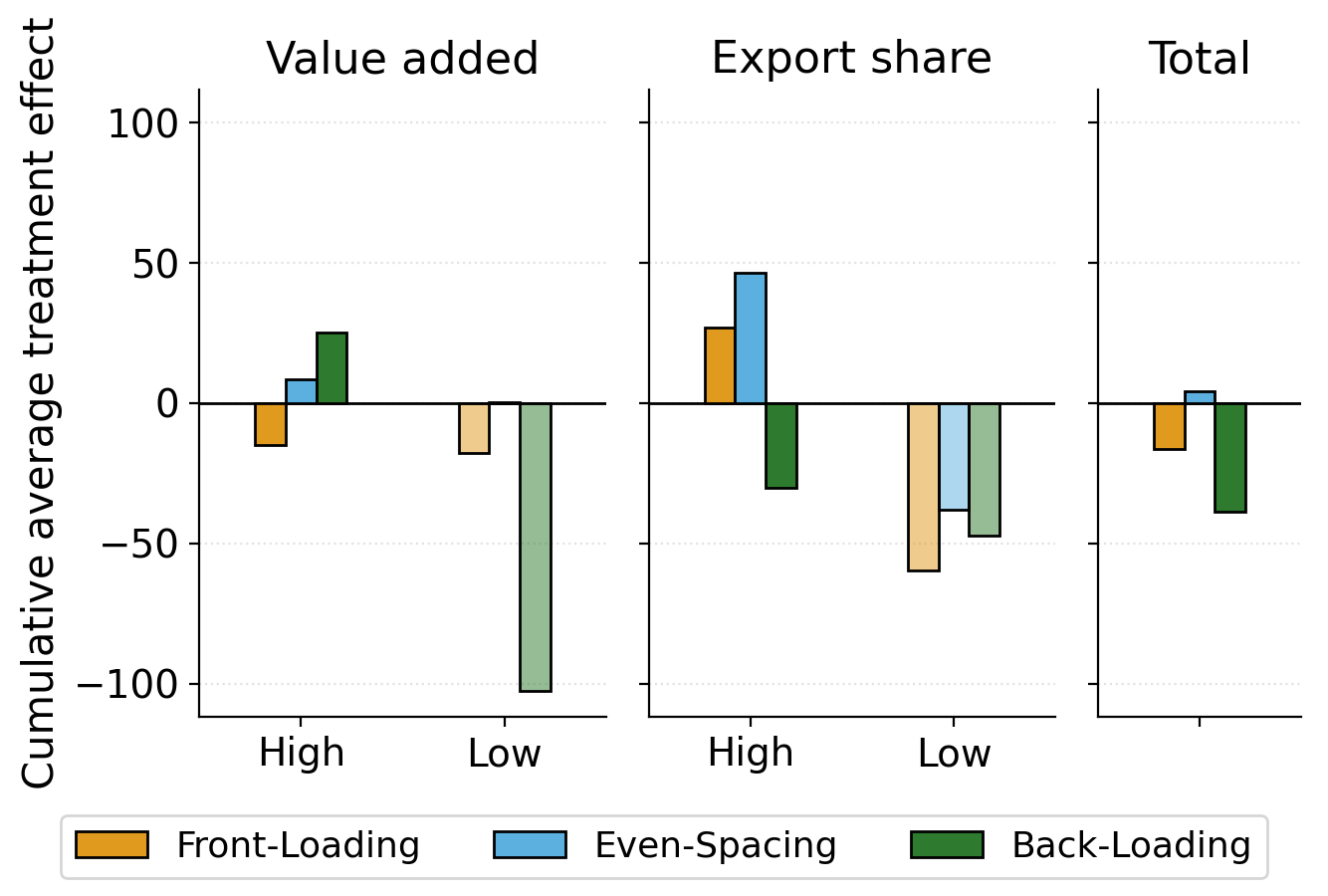}}\\
    \subfloat[Loans]{\includegraphics[width=0.38\textwidth]{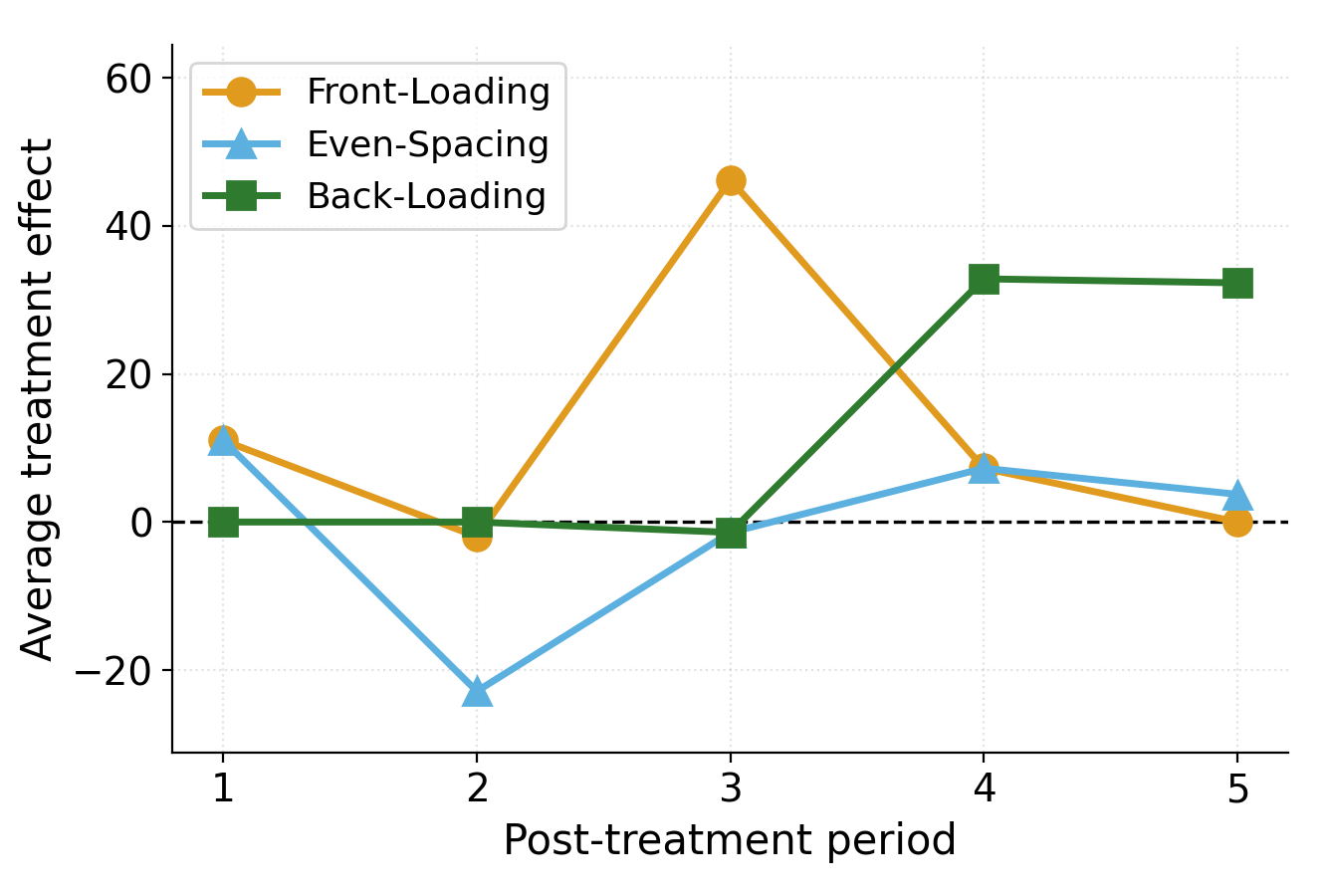}}\hspace{0.06\textwidth}
    \subfloat[Loans: heterogeneity]{\includegraphics[width=0.38\textwidth]{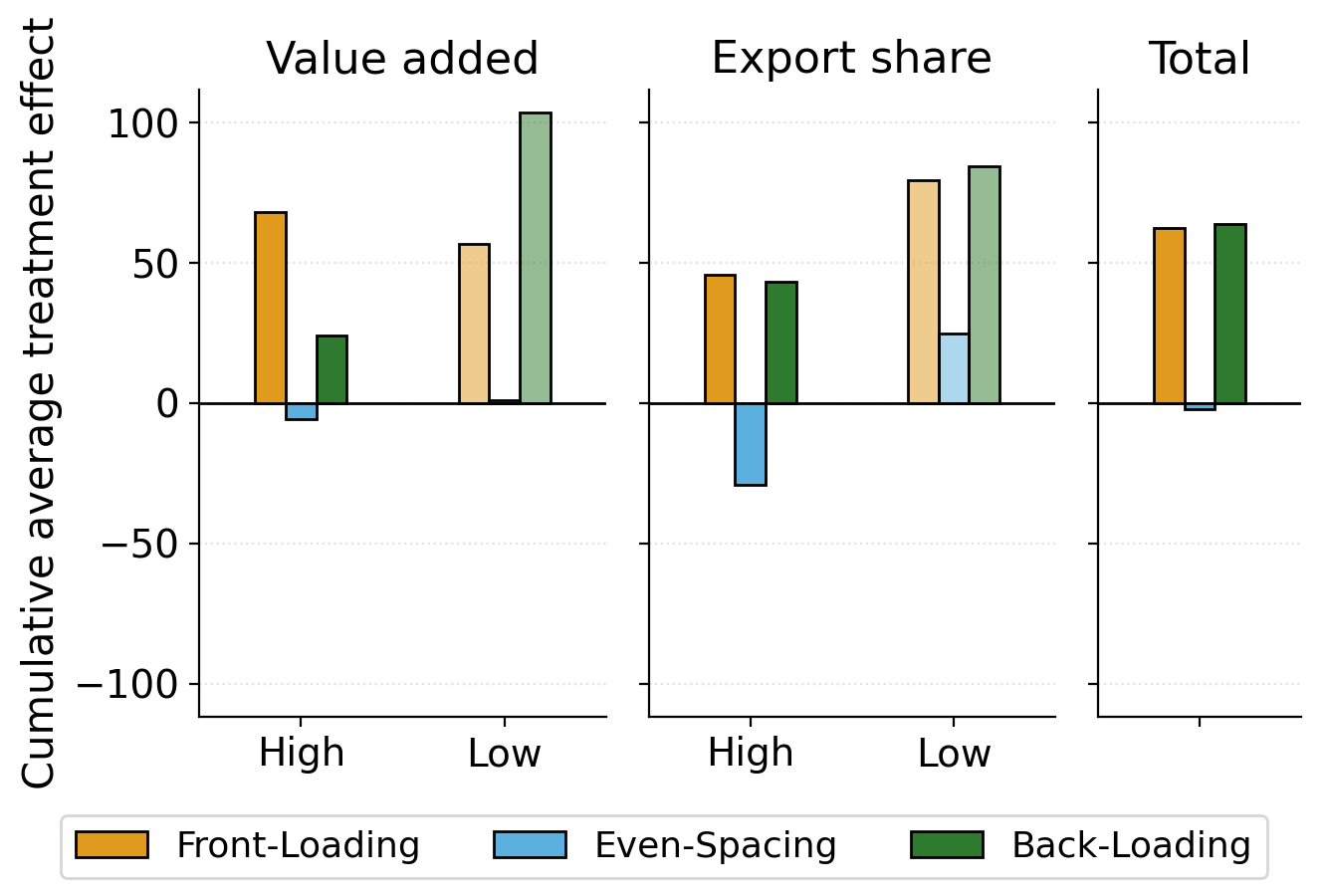}}
    \caption{Treatment Effects Under Front-, Even- and Back-Loading Treatment Schedules}
    \vspace{0.5em}
    \noindent\begin{minipage}[t]{1\columnwidth}%
    \small Notes: Panels~(a) and (c) depict the trajectory of average treatment effects under front-, even- and back-loading hypothetical treatment schedules, for insurance and loans, respectively. The outcome is export value, with units in millions of USD. Cumulative average treatment effects are $-16.3$, 4.4 and $-38.6$ for insurance and 62.5, $-2.3$ and 63.7 for loans, under front-, even- and back-loading, respectively. Panels~(b) and (d) report cumulative average treatment effects by firm group, where firms are split at the sample median of pre-treatment value added and export share, with the rightmost columns reporting the full sample.%
    \end{minipage}
    \label{fig:front_even_back}
\end{figure}

We next turn to a sequence-specific analysis. We attempt to understand how timing (but not the total amount of support) affects the trajectory of treatment effects by considering the following: (i) a front-loading treatment $\bar{d}^T=(d,d,d,0,0)$ for either $d=1$ or $2$, which concentrates support at the start; (ii) an even-loading treatment $(d,0,d,0,d)$, which spreads it evenly; and (iii) a back-loading treatment $(0,0,d,d,d)$, which defers it to the end.\footnote{The analysis with $\bar{d}^T=(d,d,0,0,0)$, $(0,d,0,d,0)$, and $(0,0,0,d,d)$ produces a similar result, especially for insurance as support.} Figures~\ref{fig:front_even_back}(a) and (c) plot the trajectories of average treatment effects under the three schedules for insurance and loans, respectively, and the rightmost columns in panels~(b) and (d) report the corresponding cumulative average treatment effects. For insurance, front-loading and back-loading generate negative cumulative effects, while even-loading is the only schedule associated with a small positive effect. For loans, both front-loading and back-loading generate substantial cumulative gains, while even-loading produces a small negative effect.

For insurance, concentrating support in consecutive periods performs worse than spreading it out, and even-loading is the only schedule with a positive cumulative effect. For loans, by contrast, concentrated schedules perform better: consecutive support sustains working capital across successive production cycles, whereas alternating support leaves gaps in which firms must self-finance. Overall, the results suggest that both timing and spacing matter for export support policies, and that the optimal schedule differs substantially across treatment types.\footnote{Under $q=1$, support affects outcomes in the year of provision and the following year. This restriction partially influences the estimates in Figure~\ref{fig:front_even_back}, i.e., support provided two or more years earlier has no effect by assumption. Assessing their sensitivity to larger $q$ would be a natural robustness exercise.}

Figures~\ref{fig:front_even_back}(b) and (d) report cumulative effects under the three schedules by firm group. For insurance, estimates for firms with weaker pre-treatment characteristics are negative under both concentrated schedules and closest to zero under even-loading, while estimates for stronger firms are positive under at least one schedule. For loans, effects are positive for every group under both concentrated schedules and consistently larger for low export share firms, while even-loading is negative for the two high groups; the largest gains are for low value added firms under back-loading.

\subsection{Optimal Targeting of Financial Support}\label{subsec:optimal_targeting}

In providing financial support, each ECA has its own rules for selecting export firms. It would be interesting to investigate (i) whether better (statistical) selection rules could have been used for each support program compared to the observed selections, (ii) whether coordination among agencies in the selection process, rather than independent selection, would have led to additional gains, and (iii) what selection rule could be implemented for new and not-yet-treated firms. Questions (i) and (ii) relate to \emph{retrospective policy learning}, while (iii) corresponds to \emph{prospective policy learning}.

\subsubsection{Retrospective Policy Learning}\label{subsubsec:optimal_firms_in_data}

For each firm $n$ in the data, we consider the optimal treatment schedule $\bar{d}^{T*}(n) \in \mathcal{D}$ that maximizes the aggregate outcome:
\begin{equation}\label{eq:opt_alloc}
\bar{d}^{T*}(n) \in \argmax_{\bar{d}^{T}\in \mathcal{D}}\sum_{t=1}^{T}\Ex\left[Y_{n,T_{0}+t}^{(\bar{d}^t)}\mid\LFc\right],
\end{equation}
where $\Ex\left[Y_{n,T_{0}+t}^{(\bar{d}^t)}\mid\LFc\right]$ is estimated using the \texttt{SBE-PCR} algorithm. Here the set of possible schedules $\mathcal{D}$ can be restricted for institutional reasons or due to budget constraints. An example of the latter would be $\mathcal{D}=\{\bar{d}^{T}:\sum_{t=1}^{T}c(d_t) \le B\}$, where $c(\cdot)$ is the per-period cost of a treatment, with $c(0)=0$, $c(1)=c_{I}$ (insurance), $c(2)=c_{L}$ (loans), and $c(3)=c_{I}+c_{L}$ (both), and $B$ is the budget. An example of the former is the independent selection process of each ECA, in which case K-SURE is equipped with $\mathcal{D}=\{\bar{d}^{T}:d_t \in \{0,1\}\}$ and EXIM is equipped with $\mathcal{D}=\{\bar{d}^{T}:d_t \in \{0,2\}\}$; this example is investigated below.

\begin{figure}
\centering
\subfloat[Optimal vs. observed allocations]{\includegraphics[width=0.38\textwidth]{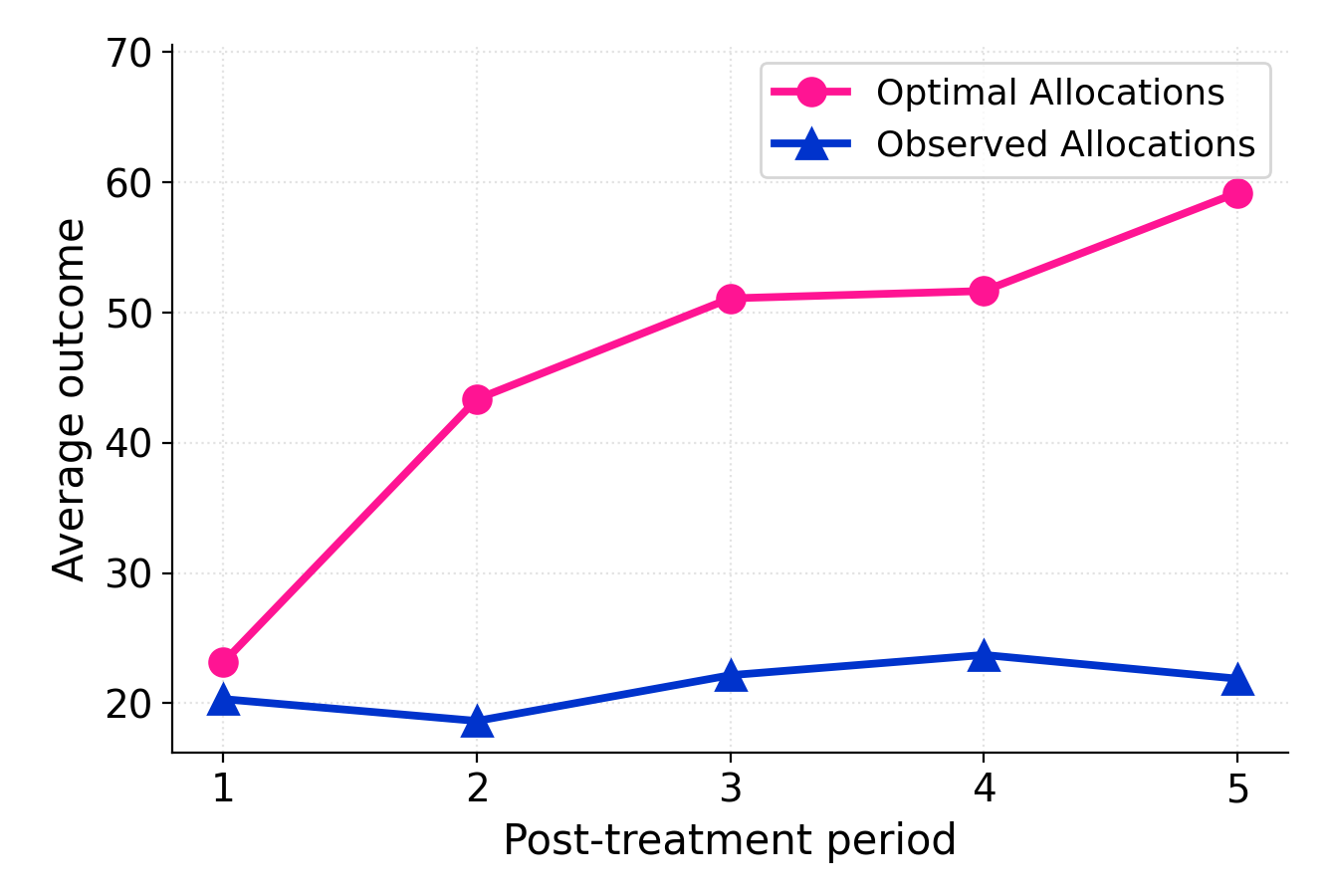}}\hspace{0.06\textwidth}
\subfloat[Heterogeneity\label{fig:optimal_observed_hetero}]{\includegraphics[width=0.38\textwidth]{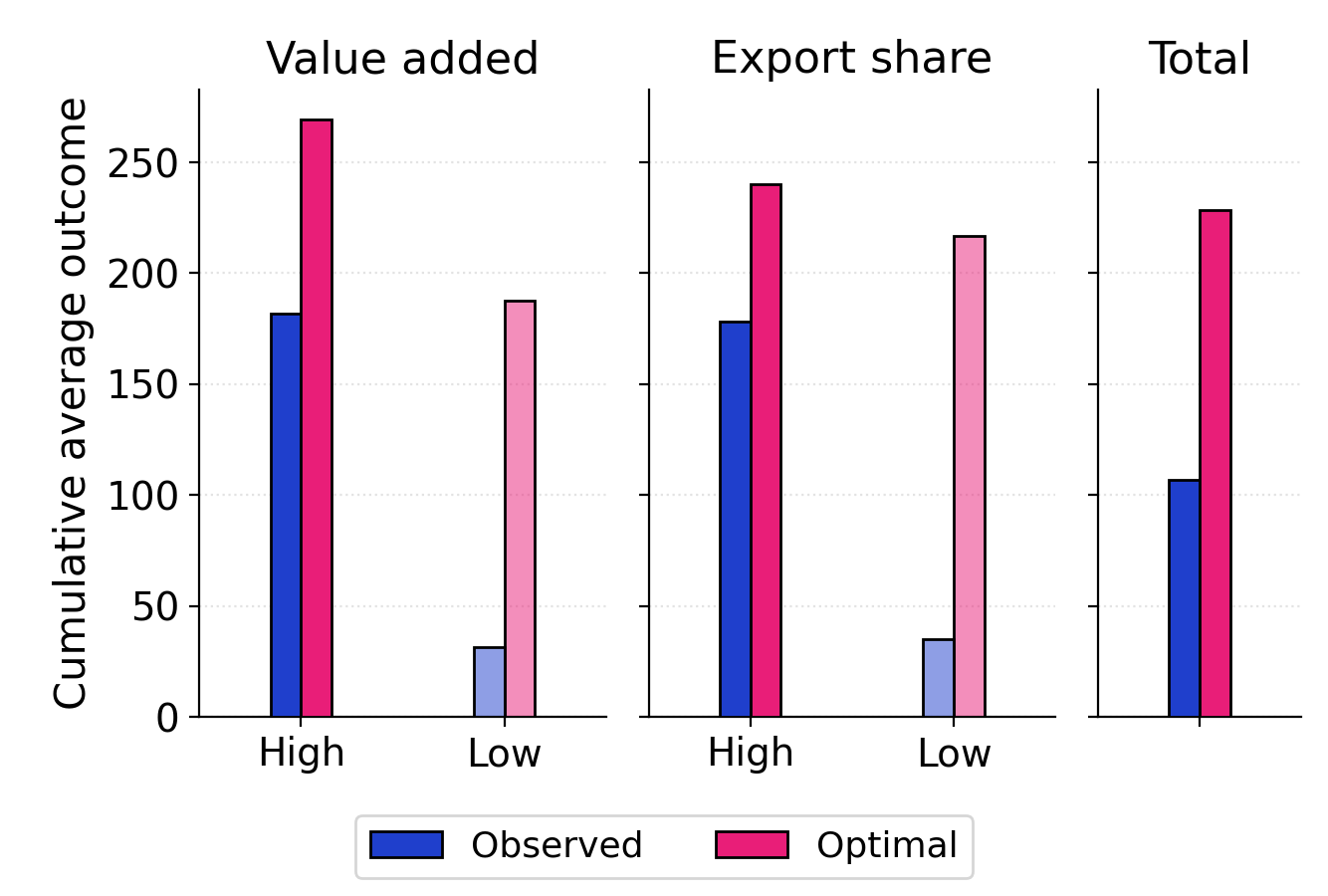}}
\caption{Potential Export Values Under Optimal vs. Observed Treatment Schedules}
\vspace{0.5em}
\noindent\begin{minipage}[t]{1\columnwidth}%
\small Notes: Panel~(a) depicts the trajectory of counterfactual export values under optimal versus observed treatment schedules, with units in millions of USD. The cumulative average outcome amounts to 106.7 under the observed treatment schedules, compared with 228.5 under the optimal schedules. Panel~(b) reports cumulative outcomes by firm group, where firms are split at the sample median of pre-treatment value added and export share, with the rightmost column reporting the full sample.%
\end{minipage}
\label{fig:optimal_observed}
\end{figure}

Using this policy learning framework, we calculate the optimal counterfactual
allocation \emph{subject to each firm's budget not exceeding its observed level}, treating one firm-year of insurance and one of loans as equally costly.\footnote{Receiving both treatments in the same year counts as two units. We base the relative cost on the margin each agency forgoes: EXIM's net interest margin averaged 0.43 percent over 2011--2015 against 1.93 percent for private banks (Korea's Financial Supervisory Service); K-SURE's premiums net of claims and expenses averaged $-0.06$ percent of insured turnover against 0.04 percent for the three largest private trade credit insurers (Euler Hermes, Atradius, and Coface). These correspond to about USD 41K and 36K per supported firm-year, a relative cost of 1.1 for a loan, rounded to one.} Because never-treated firms have an observed budget of zero, this constraint effectively restricts the exercise to firms treated at least once; policy learning for never-treated firms is considered below. Figure
\ref{fig:optimal_observed}(a) reports the average counterfactual trajectories across firms under
the observed and optimal treatment schedules. Relative to the observed
allocations, the optimal (cost-constrained) paths yield systematically
higher outcomes in every post-intervention period. Across the five 
post-intervention periods, the optimal allocation raises cumulative outcomes from 106.7 million USD under the observed allocation to 228.5, by approximately 114\%, with gains increasing over time.\footnote{Because the reported gains evaluate treatment schedules that were themselves selected by maximizing estimated values, they are subject to a winner's-curse upward bias: optimizing over noisy estimates tends to select schedules whose values are overestimated. See \citet{andrews2024inference} for inference methods that account for this selection; the in-sample gains reported here (and the subgroup gains below) should be read as optimistic estimates of the achievable improvement. Another caveat to the retrospective exercise is that the two (out of fifteen) blip effects corresponding to the empty donor sets $\Ic^{3}_{4}$ and $\Ic^{3}_{5}$ are not identified; allocations using them are hence excluded from the policy class. Since this only restricts the feasible set, the reported optimal value is, if anything, understated.} Moreover, these gains are achieved with lower resource use: the total cost of the optimal allocation is 335 supports, compared with 346 under the observed allocation. These conclusions are unchanged under alternative assumptions about the cost of a loan relative to insurance.\footnote{Varying the insurance premium over the 0.1 to 0.5 percent range quoted in this market and the imputation exponent between 0.8 and 1.2 places the implied relative cost of a loan between 0.4 and 2.5. Setting this relative cost to 0.5 and to 2 raises cumulative outcomes by 112 and 104\%, and in both cases the optimal allocation again costs less than the observed one.} Overall, the findings indicate that the current targeting rules employed by the ECAs have
substantial scope for improvement in terms of sequencing and timing of support.

We next examine heterogeneity in the gains from the optimal allocation. Figure~\ref{fig:optimal_observed}(b) reports cumulative outcomes under the observed and optimal allocations by firm group. Relative to the observed allocation, the optimal allocation improves outcomes across all groups, with particularly large gains for firms with low value added or low export share. For these firms, cumulative outcomes increase by roughly 500--520\%, compared with around 35--50\% for firms with high value added or high export share.

\begin{figure}
\centering
\subfloat[Joint vs. independent allocations]{\includegraphics[width=0.38\textwidth]{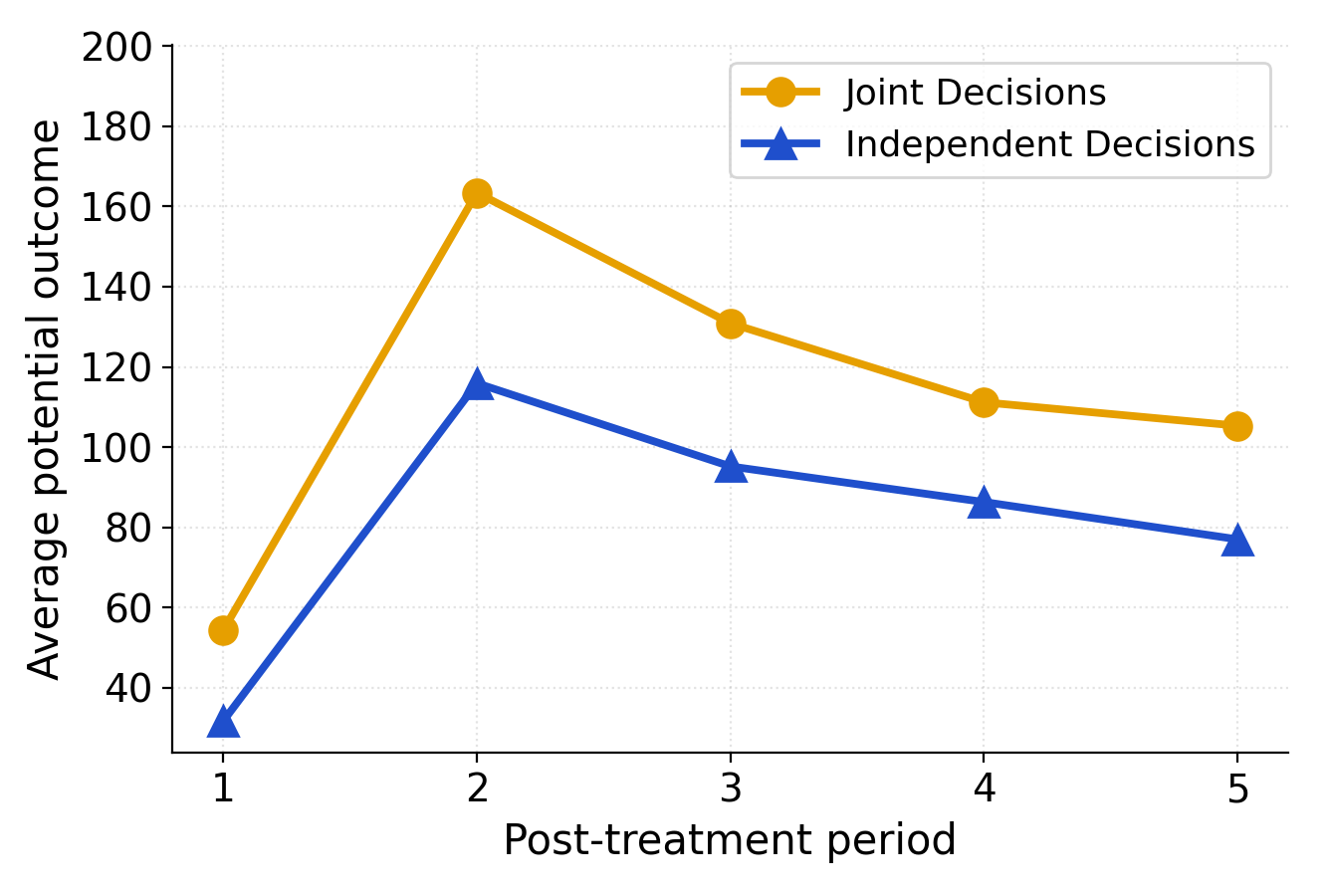}}\hspace{0.06\textwidth}
\subfloat[Heterogeneity\label{fig:joint_independent_hetero}]{\includegraphics[width=0.38\textwidth]{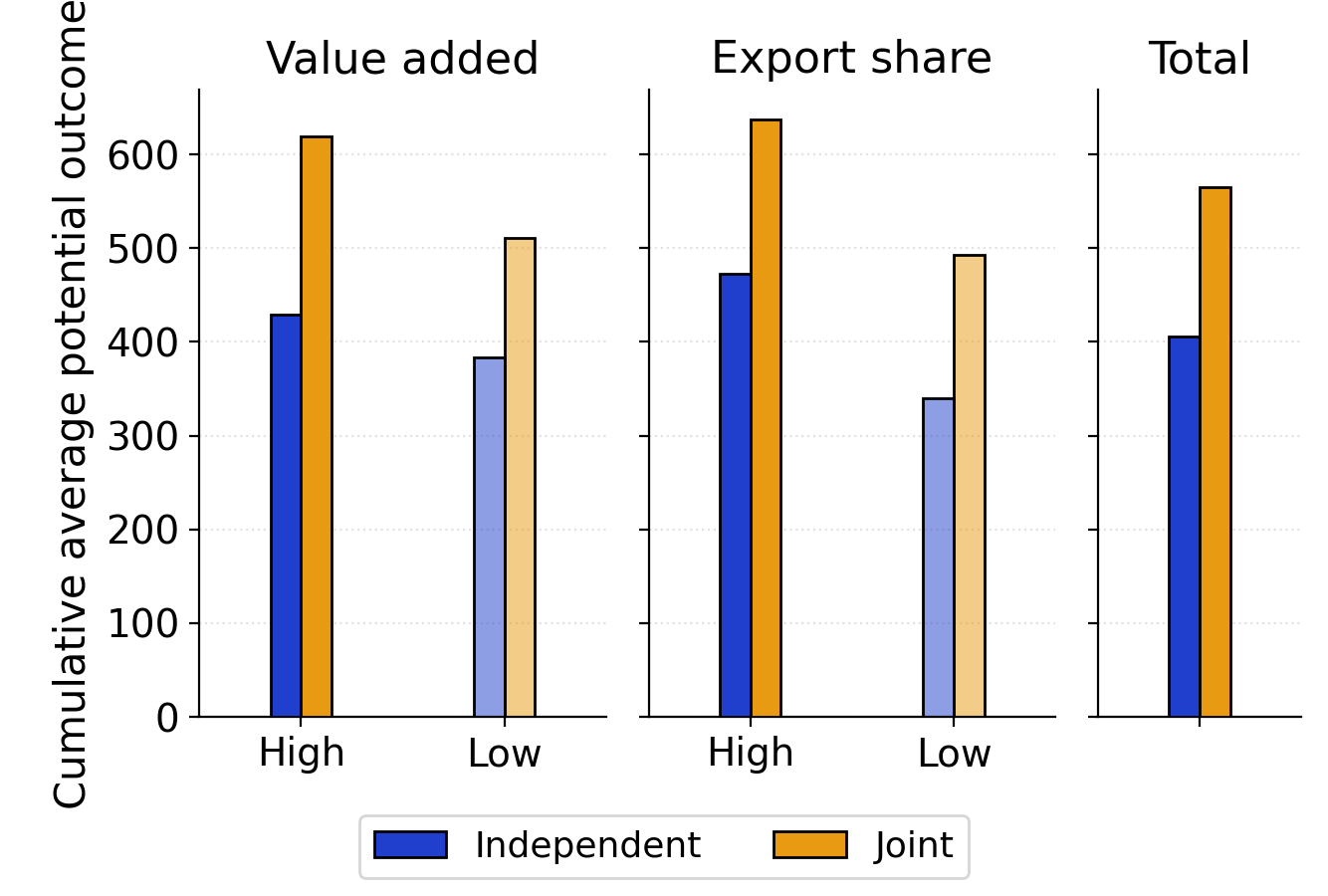}}
\caption{Potential Export Values Under Treatment Schedules Jointly vs. Independently Optimized}
\vspace{0.5em}
\noindent\begin{minipage}[t]{1\columnwidth}%
\small Notes: Panel~(a) depicts the trajectory of counterfactual export values under treatment schedules jointly versus independently optimized by the agencies, with units in millions of USD. The cumulative average outcome amounts to 565.2 under joint optimization and 406.1 under independent optimization. Panel~(b) reports cumulative outcomes by firm group, where firms are split at the sample median of pre-treatment value added and export share, with the rightmost column reporting the full sample.%
\end{minipage}
\label{fig:joint_independent}
\end{figure}

We next compare outcomes under independent allocation, in which the agencies act separately, and joint allocation, in which decisions are coordinated as if by a single agency. We impose no budget constraint, so that the total budget spent in each case can also be compared. As illustrated in Figure~\ref{fig:joint_independent}(a), average potential outcomes are higher under joint allocation in every period, by approximately 39\%, at \emph{lower} total cost: 3,176 supports under joint allocation versus 3,650 under independent allocation.\footnote{Note that unlike in Figure \ref{fig:optimal_observed}, no budget constraint is imposed here; joint allocation remains more efficient under the alternative relative costs considered above.} These results indicate that coordination among agencies reduces misallocation across treatments and leverages treatment complementarities. Figure~\ref{fig:joint_independent}(b) reports the corresponding outcomes by group. Joint allocation yields higher outcomes than independent allocation in every group, and the gains are similar in magnitude across groups.

\subsubsection{Prospective Policy Learning}\label{subsubsec:optimal_new_firms}

Policymakers may want to estimate optimal targeting rules for new firms that are \emph{not} observed in the data as they arrive after the data period or for not-yet-treated (i.e., never-supported) firms observed in the data. To that end, we consider allocating support based on firms' observed covariates. We focus on a one-period targeting problem: support is assigned once, in the period after the last observed period $\bar{T} := T_0 + T$; SA~\ref{sec:appendix_EWM} considers multi-period schedules. Let $\delta:\widetilde{\mathcal{X}}\rightarrow\tilde{\mathcal{D}}$ be a targeting rule, where $\tilde{\mathcal{D}}\subseteq[0,A]$ is the feasible set of actions and $\widetilde{\mathcal{X}}$ is the support of the vector of policy features $\widetilde{X}_n$ of firm $n$ (i.e., firm characteristics observed before the policy period, specified below), which may differ from the covariates $X_n$ used by \texttt{SBE-PCR}. Let $\Pi$ be the (possibly restricted) class of targeting rules. Then define an optimal targeting rule as
\begin{align}\label{eq:opt_alloc_x}
\delta^{*} & \in\argmax_{\delta\in \Pi}\Ex\left[Y_{n,\bar{T}+1}^{(\bar{0}^{T},\,\delta(\widetilde{X}_n))}\right].
\end{align}
Since the no-support value $\Ex[Y_{n,\bar{T}+1}^{(\bar{0}^{T+1})}]$ is common to all rules, \eqref{eq:opt_alloc_x} equivalently maximizes the expected \emph{gain} relative to no support, which is the form analyzed in SA~\ref{sec:appendix_EWM}. The historical analogue of the expected gain in the data would be $\tau^H_n(d) \coloneqq \Ex[Y_{n,\bar{T}}^{(\bar{0}^{T-1},\,d)} - Y_{n,\bar{T}}^{(\bar{0}^{T})} \mid \LFc]$ for $d \in \tilde{\mathcal{D}}$, which is the blip effect of $d$ at the terminal period and is therefore identified by Theorem~\ref{thm:newLTV} and estimated by \texttt{SBE-PCR}. In SA~\ref{subsec:EWM_setup} we show that, under an i.i.d.\ sampling scheme with no covariate shift and temporal stability of conditional treatment gains (e.g., between the two consecutive periods as in this application), the expected gain of any rule $\delta$ in the prospective period equals $\Ex[\tau^H_n(\delta(\widetilde{X}^H_n))]$, where $\widetilde{X}^H_n$ denotes the policy features observed before period $\bar{T}$. Hence the optimal rule $\delta^{*}$ is identified (Lemma~\ref{lem:prospective-policy-identification}). The estimated gains $\hat{\tau}^H_n(\cdot)$ then serve as pseudo-outcomes for a prediction problem with predictors $\widetilde{X}^H_n$ to fit the decision tree, and then this fitted rule is deployed using the policy features $\widetilde{X}^P_n$ observed at the end of the sample, $\bar{T}$.\footnote{SA Figure~\ref{fig:prospective-policy-timeline} illustrates the timeline of the prospective policy learning.}

Based on this framework, we implement a tree-based policy learning algorithm that yields interpretable decision rules. For the policy features $\widetilde{X}_n$, we consider thirteen firm characteristics: averages over the $5$ years preceding the policy period of export share, sales, employment, tangible capital, value added, TFP, total wage bill, R\&D expenditure, debt-to-asset ratio, liquidity ratio, an indicator for FDI status, parent-company affiliation, and firm age. Based on $\widetilde{X}_n$, the algorithm selects the most predictive variables and thresholds,
partitioning firms into subgroups with distinct optimal actions. Each leaf in a decision tree corresponds to one recommended action, providing a transparent mapping from firm characteristics to the type of support. Since considering all possible targeting rules in $\Pi$ is computationally and practically demanding, we restrict the class in the analysis through the tree structure.

\begin{figure}
\centering
\includegraphics[width=0.495\textwidth]{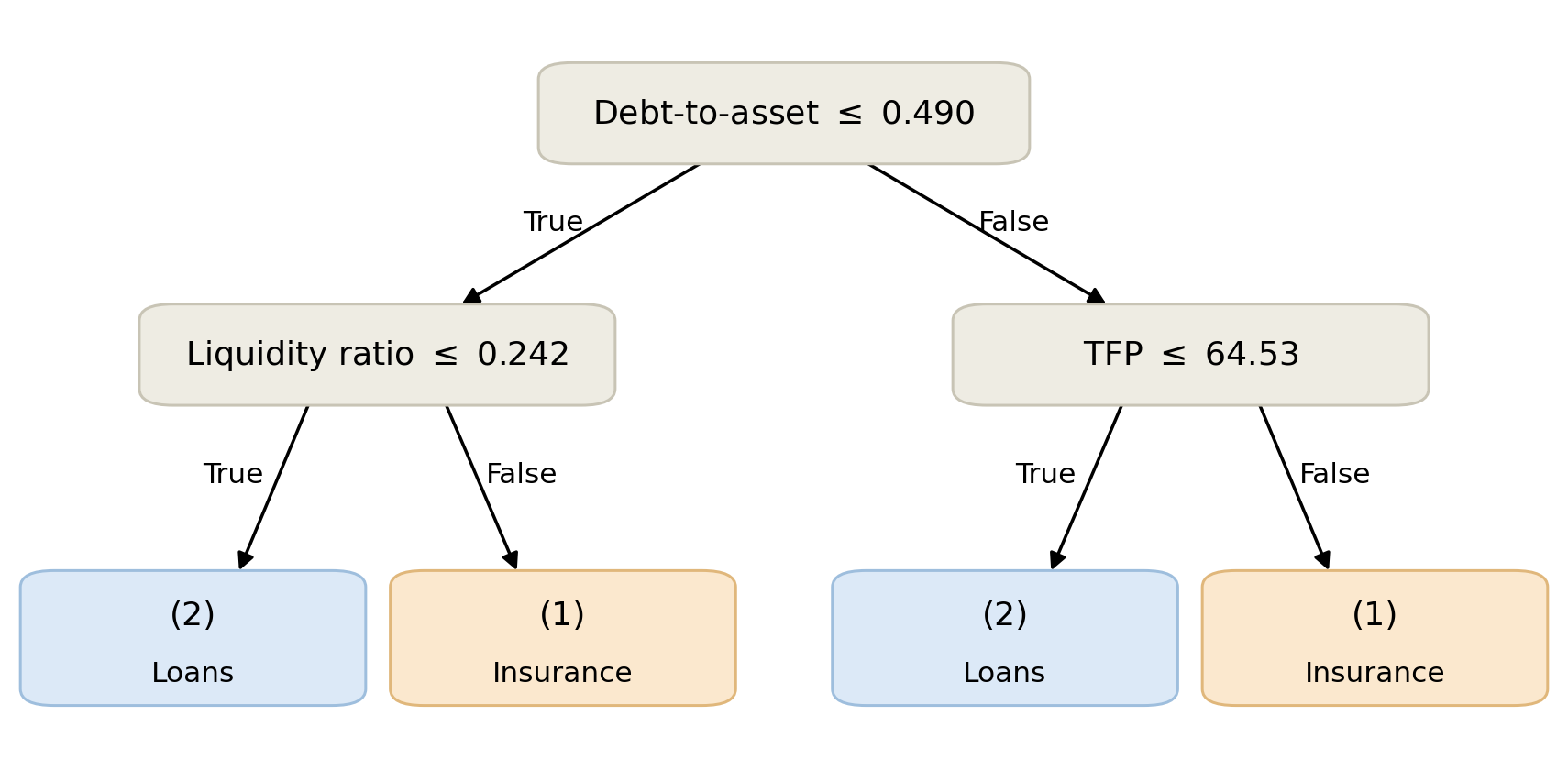}
\caption{Decision Tree for Prospective Policy Learning}
\vspace{0.5em}
\label{fig:decision_trees}
\end{figure}

In the application, $q = 1$ (Section~\ref{subsec:data}), so the support given in the policy period also affects the outcome of the following period. We include this continuation effect in the objective by adding to \eqref{eq:opt_alloc_x} the outcome of the following period under no further support, and define $\tau^H_n(\cdot)$ accordingly (see Remark~\ref{rem:continuation} on continuation effects in SA~\ref{subsec:EWM_setup}). The historical analogue then assigns support in 2014 to firms untreated through 2013 and evaluates its continuation in 2015, with policy features averaged over 2009--2013; then the fitted rule is deployed using features averaged over 2011--2015. We take $\tilde{\mathcal{D}}=\{0,1,2\}$, i.e., no support, insurance, or loans.\footnote{We abstain from learning for the joint action, whose blip effect is
not identified: the corresponding donor set is empty, as no firm in the sample
moves directly from control to the joint action.} Figure \ref{fig:decision_trees} shows the optimal policy tree. The tree splits first on the debt-to-asset ratio, and then on the liquidity ratio among less leveraged firms and on productivity among more leveraged ones. Leverage decides which of the two is used, and the second split decides the support. Within each leverage group, firms that are more liquid or more productive are assigned insurance, while the rest receive loans. This is consistent with a difference in which constraint binds: stronger firms need cover against non-payment in order to expand, whereas weaker ones need working capital.\footnote{SA~\ref{sec:appendix_five_period} reports a five-period version of this exercise, in which the rule chooses not only which support to give but also when to give it and how to spread it.}

\begin{remark}[Regret bounds]\label{rem:plugin_EWM}
Our prospective policy-learning procedure for
\eqref{eq:opt_alloc_x} follows the empirical welfare maximization
framework \citep{kitagawa2018should,athey2021policy}. Our welfare criterion uses plug-in estimates obtained by \texttt{SBE-PCR}.
The retrospective exercise \eqref{eq:opt_alloc} is the historical analogue over the full observation window in which a schedule is chosen separately for each observed firm, so it involves no policy class.
By standard plug-in arguments, prospective regret is bounded by an empirical-process
term plus twice the uniform plug-in estimation error;
retrospective regret is bounded by twice the corresponding
uniform plug-in estimation error alone.
For fixed $|\tilde{\mathcal{D}}|$, the empirical-process term is
$O_p(\sqrt{d_{\mathrm{Nat}}\log N/N})$, where
$d_{\mathrm{Nat}}$ is the finite Natarajan dimension of the
policy class (e.g., depth-bounded trees).
The uniform plug-in condition would follow from tail-bound versions of the convergence rates in Theorems~\ref{thm:consistency-time-varying-fixed-lags} and \ref{thm:consistency-time-invariant-fixed-lags}.
See SA~\ref{sec:appendix_EWM} for formal statements and proofs.
\end{remark}

\section{Conclusion}\label{sec:conclusion}
In this work, we formulate a causal framework for dynamic treatment effects under unobserved confounding using panel data. 
We propose a latent factor model, which admits linear time-varying and time-invariant dynamic triangular systems as special cases.
Depending on the structure placed on this factor model, we quantify the trade-off between the sample complexity and the level of adaptivity allowed in the intervention policy, for estimating mean counterfactual outcomes. The estimated counterfactual outcomes are useful in estimating the impact of a particular treatment schedule relative to another, as well as the optimal rules for allocating treatment schedules. We showcase this usefulness in the context of government financial support. We hope this work spurs further research connecting the growing fields of SC and panel data methods with dynamic treatment models studied in econometrics, and potentially sequential learning methods such as reinforcement learning studied in computer science.

\section*{Appendix: Estimation and Consistency for the LTI Model}

This appendix collects the LTI counterparts of the estimation algorithm and the consistency result of Section~\ref{subsection:SBE-Varying-Estimator}; all proofs are in SA~\ref{app:LTI_proofs}.

\noindent{\bf Algorithm.} In the LTI setting the \texttt{SBE-PCR} algorithm is identical in structure, with the donor sets $\Ic \in \{\Ic^d : d \in [A]\} \cup \{\Ic^0_t : t \in [T]\}$ in place of $\Ic^d_t$ and $\Ic^0_T$: the PCR weights $\hat{\phi}^{n, \Ic}$ ($n \in \Ic$) and $\hat{\beta}^{n, \Ic}$ ($n \notin \Ic$) are constructed exactly as in Section~\ref{subsection:SBE-Varying-Estimator}; baselines $\hat{b}_{j, t}$ are estimated for every $t \in [T]$; the lag-$0$ blip $\hat{\gamma}_{j, 0}(d)$ is estimated from outcomes at each donor's first treatment time $t^*_k$; and the lag-$\ell$ blips are estimated recursively from outcomes at time $t^*_k + \ell$. The full listing is given in SA~\ref{app:lti-algorithm}. The weights rely on the leave-one-out analog of Assumption~\ref{assumption:LTI_well_supported_factors}:

\setcounter{assumption}{6}\LTInum\begin{assumption}[Well-supported factors]\label{control-factors-support-LTI}
For any $\Ic \in \{\Ic^d : d \in [A]\} \cup \{\Ic^0_t : t \in [T]\}$ and $n \in \Ic$, there exist weights $\phi^{n, \Ic}\in \Rb^{|\Ic| - 1}$ such that $v_{n, T} = \sum_{k \in \Ic \setminus n} \phi^{n, \Ic}_k v_{k, T}$.
\end{assumption}

\noindent{\bf Consistency.} The analog of Assumption~\ref{assumption:balanced-spectra} is imposed on the LTI donor sets.

\setcounter{assumption}{8}\LTInum\begin{assumption}[Well-Balanced Singular Values and Row-Space Inclusion]\label{assumption:lti-balanced-spectra}
(i) Part (i) of Assumption~\ref{assumption:balanced-spectra} holds with $X_{\Ic^d_t}$ replaced by $X_{\Ic}$ for each LTI donor set $\Ic$; and
(ii)\label{assumption:lti-row-space}
for every $t \in [T]$, there exists a common vector $\xi^{(0,t)} \in \Rb^p$ such that, for every $j \in \Ic^0_t$,
\[
\Ex[Y_{j,t} \mid \LFc, j \in \Ic^0_t]
= \sum_{i=1}^p \xi^{(0,t)}_i \cdot \Ex[(X_{\Ic^0_t})_{ij} \mid \LFc, j \in \Ic^0_t];
\]
and for every $d \in [A]$, $t \in [0,T-1]$, and $\ell \in \{t, t+1\}$, there exists a common vector $\xi^{(d,t,\ell)} \in \Rb^p$ such that, for every $j \in \Ic^d$,
\[
\Ex\!\left[
Y_{j,t_j^*+t}^{\left(\bar D_j^{\,t_j^*+t-\ell},\, \overbar{\tilde{0}}^{\,\ell}\right)}
\bigm| \LFc, j \in \Ic^d
\right]
= \sum_{i=1}^p \xi_i^{(d,t,\ell)} \cdot \Ex[(X_{\Ic^d})_{ij} \mid \LFc, j \in \Ic^d].
\]
Here $\overbar{\tilde{0}}^{\,\ell}$ denotes a control sequence of length $\ell$, with $\overbar{\tilde{0}}^{\,0}$ the empty history.\footnote{As in Assumption~\ref{assumption:balanced-spectra}, both conditions (i) and (ii) are also assumed for the leave-one-out donor sets $\Ic \setminus n$, for each LTI donor set $\Ic$.} 
\end{assumption}

\setcounter{theorem}{1}\LTInum\begin{theorem}\label{thm:consistency-time-invariant-fixed-lags}
    Let Assumptions~\ref{assumption:SUTVA}, \ref{assumption:LTI_factor_model}--\ref{assumption:LTI_seq_exog}, \ref{assumption:additional-covariates}, \ref{control-factors-support-LTI}, \ref{assumption:sub-gaussian-noise}, \ref{assumption:lti-balanced-spectra}, and \ref{assumption:outcome-fixed-lag-dep-invariant} hold. Consider the LTI variant of the \texttt{SBE-PCR} estimator (SA~\ref{app:lti-algorithm}) modified to only estimate the baseline, terminal blip, and previous $q$ blips, and suppose $k = \max_{\Ic \in \{\Ic^d\} \cup \{\Ic^0_t\}}\text{rank}(\EE[X_{\Ic}|\LFc])$. Then, conditional on $\LFc$ and the donor-defining treatment histories $\{\bar{D}^{t^*_j}_{j} : j \in \Ic^d\}$ and $\{\bar{D}^{t}_{j} : j \in \Ic^0_t\}$ of all donor sets, we have for any $n \in [N]$ and $\bar{d}^T \in [A]^T$:
$$
\widehat{\Ex}[Y_{n, T}^{(\bar{d}^T)}] - \EE[Y_{n, T}^{(\bar{d}^T)}\mid \LFc]=O_p\left(\sqrt{\log(p\pi_{\Ic})}\left(\frac{k^{q+9/4}}{p^{1/4}}
+
k^{q+7/2}\max\left\{\frac{\sqrt{\pi_{\Ic}}}{p^{3/2}}, \frac{1}{\sqrt{\alpha_{\Ic}-1}}, \frac{1}{\sqrt{p}}\right\}\right)\right),
$$
where $\pi_{\Ic} = \max\Cc$ and $\alpha_{\Ic} = \min\Cc$ with $\Cc = \{|\Ic_T^0|, |\Ic^0_1|,(|\Ic^{d_{t}}|)_{t \in [T-q, T]} ,(|\Ic^{D_{n,t_n^* + t}}|)_{n \in [N], t \in [q]}\}$.
\end{theorem}

Theorem~\ref{thm:consistency-time-invariant-fixed-lags} is the LTI counterpart of Theorem~\ref{thm:consistency-time-varying-fixed-lags} and has the same qualitative implications; its proof is in SA~\ref{subsection:proof-of-invariant-consistency-fixed-lags}. Since the control donor sets are nested, $\Ic^0_T \subseteq \Ic^0_{T-1} \subseteq \cdots \subseteq \Ic^0_1$, their cardinalities are weakly decreasing in $t$; if at least one unit initiates treatment in each period, the gap between the largest and smallest cardinalities in $\Cc$ is at least of order $T$.
\COMnum

\bibliography{main}
\bibliographystyle{apalike}

\clearpage

\vspace*{3em}
\begin{center}
{\Large Supplemental Appendix to\\[0.5em]
``Synthetic Blips: Generalizing Synthetic Controls for Dynamic Treatment Effects''}\\[2em]
{\large Anish Agarwal \qquad Sukjin Han \qquad Dwaipayan Saha}\\[0.5em]
{\large Vasilis Syrgkanis \qquad Haeyeon Yoon}\\[1.5em]
\date{September 10, 2026}
\end{center}
\vspace{2.5em}


\appendix

\counterwithin{equation}{section}
\counterwithin{theorem}{section}
\counterwithin{lemma}{section}
\counterwithin{proposition}{section}
\counterwithin{corollary}{section}
\counterwithin{definition}{section}
\counterwithin{assumption}{section}
\counterwithin{remark}{section}
\counterwithin{figure}{section}
\counterwithin{table}{section}
\def\theHequation{\thesection.\arabic{equation}}
\def\theHtheorem{\thesection.\arabic{theorem}}
\def\theHlemma{\thesection.\arabic{lemma}}
\def\theHproposition{\thesection.\arabic{proposition}}
\def\theHcorollary{\thesection.\arabic{corollary}}
\def\theHdefinition{\thesection.\arabic{definition}}
\def\theHassumption{\thesection.\arabic{assumption}}
\def\theHremark{\thesection.\arabic{remark}}
\def\theHfigure{\thesection.\arabic{figure}}
\def\theHtable{\thesection.\arabic{table}}

\tocentrieson

\clearpage
\makeatletter
\@starttoc{toc}
\makeatother
\clearpage

\section{Empirical Welfare Maximization}\label{sec:appendix_EWM}

This section of the SA formalizes the plug-in empirical welfare maximization interpretation
\citep{kitagawa2018should,athey2021policy} of the policy-learning exercises in
the main text, stated informally in Remark~\ref{rem:plugin_EWM} of
Section~\ref{subsubsec:optimal_new_firms}.

\subsection{Setup and Learning Algorithm}\label{subsec:EWM_setup}

Let $X_n\in\cX$ be the covariate vector used by \texttt{SBE-PCR}
(Assumption~\ref{assumption:additional-covariates}).  For policy learning, fix
a policy-feature history length $r$, and let
$\widetilde X_n^H,\widetilde X_n^P\in\widetilde{\cX}\subseteq\Rb^{\widetilde p}$
denote the same vector of policy features constructed immediately before the
historical and prospective policy windows, respectively.  Each contains the
same time-invariant characteristics and the same summaries of the preceding
$r$ periods of time-varying covariates.  The policy features may overlap with
$X_n$, but need not coincide with it or satisfy
Assumption~\ref{assumption:additional-covariates}.

Let $\bar T\coloneqq T_0+T$ denote the last observed period.  For a prospective
policy horizon $S\leq T$, take the historical policy window to be the most
recent $S$ periods in the data, $[\bar T-S+1,\bar T]$, and the prospective
policy window to be the following $S$ periods.  Therefore,
$\widetilde X_n^H$ uses the $[\bar T-S-r+1,\bar T-S]$ covariates and
$\widetilde X_n^P$ uses the $[\bar T-r+1,\bar T]$ covariates, for some
chosen $r\leq\bar T-S$.  A leading example is $S=1$ (i.e., a one-period policy problem based on the closest
observed one-period analogue); if additionally $r=5$, $\widetilde X_n^H$ uses the 2010--14 covariates and
$\widetilde X_n^P$ uses the 2011--15 covariates. See Figure~\ref{fig:prospective-policy-timeline} below for the timeline.

Let $\tilde{\mathcal{D}}\subseteq[0,A]^S$ be the feasible set of treatment schedules.  For
$\bar d^S\in\tilde{\mathcal{D}}$, define the historical and prospective cumulative
potential-outcome gains, relative to no support:
\begin{align}
\tau_{n,\mathrm{PO}}^H(\bar d^S)
&\coloneqq
\sum_{s=1}^S
\left\{
Y_{n,\bar T-S+s}^{(\bar 0^{T-S},\bar d^s)}
-Y_{n,\bar T-S+s}^{(\bar 0^{T-S+s})}
\right\},
\label{eq:historical-gain}\\
\tau_{n,\mathrm{PO}}^P(\bar d^S)
&\coloneqq
\sum_{s=1}^S
\left\{
Y_{n,\bar T+s}^{(\bar 0^T,\bar d^s)}
-Y_{n,\bar T+s}^{(\bar 0^{T+s})}
\right\}.
\label{eq:prospective-gain}
\end{align}
As in the main text, the initial $T_0$ untreated periods are suppressed in the
treatment-history superscripts.  Define the historical causal effect parameter by 
\[
\tau_n^H(\bar d^S)
\coloneqq
\Ex\!\left[
\tau_{n,\mathrm{PO}}^H(\bar d^S)
\,\middle|\,\LFc
\right].
\] 
Under the conditions of Section~\ref{sec:LTV_identification_strategy},
$\tau_n^H(\cdot)$ is identified and can be estimated from the observed panel using
$X_n$ via the \texttt{SBE-PCR}
algorithm.

A policy is a measurable mapping
$\pi:\widetilde{\cX}\rightarrow\tilde{\mathcal{D}}$, and $\PiC$ denotes the candidate policy
class.  For $j\in\{H,P\}$, define its gain value by
\[
V^j(\pi)
\coloneqq
\Ex\!\left[\tau_{n,\mathrm{PO}}^j\bigl(\pi(\widetilde X_n^j)\bigr)\right].
\]
Maximizing gain value is equivalent to maximizing welfare in levels because
the corresponding no-support value is common to all policies.  For
prospective policy learning, the objective is to maximize $V^P(\pi)$ over
$\pi\in\PiC$.  Identification of $\tau_n^H(\cdot)$ alone does not necessarily
identify $V^P(\cdot)$.  We therefore impose additional assumptions.

\begin{assumption}[Policy-learning sample]\label{ass:sampling}
The historical pairs
$\{(\widetilde X_n^H,\tau_n^H(\cdot))\}_{n=1}^N$ are i.i.d.\ draws from a
superpopulation.  Moreover,
$\widetilde X_n^P\ \overset{d}{=}\ \widetilde X_n^H$.
\end{assumption}

The distributional equality is a no-covariate-shift condition.  It does not
impose equality between historical and prospective gains and can be weakened
by covariate reweighting.  Next, we assume temporal stability only on the
conditional cumulative gains relevant for the policy problem.

\begin{assumption}[Temporal stability of conditional treatment gains]
\label{ass:TT-contrast}
For every $\bar d^S\in\tilde{\mathcal{D}}$ and almost every
$x\in\widetilde{\cX}$,
\begin{equation}
\Ex\!\left[
\tau_{n,\mathrm{PO}}^P(\bar d^S)
\,\middle|\,\widetilde X_n^P=x
\right]
=
\Ex\!\left[
\tau_{n,\mathrm{PO}}^H(\bar d^S)
\,\middle|\,\widetilde X_n^H=x
\right].
\label{eq:TT-contrast}
\end{equation}
\end{assumption}

Assumption~\ref{ass:TT-contrast} is typically most plausible for $S=1$,
because it then transports only the next-period gain from the closest
observed analogue.  Larger $S$ requires stability of cumulative gains for
longer treatment schedules.
The assumption is weaker than an analogous assumption in terms of levels
(rather than gains) but is sufficient for policy choice.

\begin{lemma}[Identification in prospective policy learning]
\label{lem:prospective-policy-identification}
Suppose Assumptions~\ref{ass:sampling} and~\ref{ass:TT-contrast} hold.  For
every $\bar d^S\in\tilde{\mathcal{D}}$, $s\in[S]$, suppose that $\Ex\!\left[
  \varepsilon_{n,\bar T-S+s}^{(\bar 0^{T-S},\bar d^s)}
  -
  \varepsilon_{n,\bar T-S+s}^{(\bar 0^{T-S+s})}
  \,\middle|\,
  \LFc,\widetilde X_n^H
\right]
=0$. Then, under the identifying conditions of
Section~\ref{sec:LTV_identification_strategy}, for every $\pi\in\PiC$,
\begin{equation}
V^P(\pi)
=V^H(\pi)
=\Ex\!\left[
\tau_n^H\bigl(\pi(\widetilde X_n^H)\bigr)
\right],
\label{eq:prospective-value-identification}
\end{equation}
and this common value is identified.
\end{lemma}

\begin{proof}
By iterated expectations, Assumption~\ref{ass:TT-contrast}, and the equality
of the marginal policy-covariate distributions in
Assumption~\ref{ass:sampling},
\begin{align*}
V^P(\pi)
&=
\int
\sum_{\bar d^S\in\tilde{\mathcal{D}}}
1\{\pi(x)=\bar d^S\}
\Ex\!\left[
\tau_{n,\mathrm{PO}}^P(\bar d^S)
\,\middle|\,\widetilde X_n^P=x
\right]
dF_{\widetilde X^P}(x)\\
&=
\int
\sum_{\bar d^S\in\tilde{\mathcal{D}}}
1\{\pi(x)=\bar d^S\}
\Ex\!\left[
\tau_{n,\mathrm{PO}}^H(\bar d^S)
\,\middle|\,\widetilde X_n^H=x
\right]
dF_{\widetilde X^H}(x)=V^H(\pi).
\end{align*}
The conditional mean-zero restriction and the latent-factor representation
imply, for every $\bar d^S\in\tilde{\mathcal{D}}$, $\Ex\!\left[
\tau_{n,\mathrm{PO}}^H(\bar d^S)
\,\middle|\,\LFc,\widetilde X_n^H
\right]
=
\Ex\!\left[
\tau_{n,\mathrm{PO}}^H(\bar d^S)
\,\middle|\,\LFc
\right]
=\tau_n^H(\bar d^S)$. Therefore, by iterated expectations,
\begin{align*}
V^H(\pi)
&=
\Ex\!\left[
\sum_{\bar d^S\in\tilde{\mathcal{D}}}
1\{\pi(\widetilde X_n^H)=\bar d^S\}
\Ex\!\left[
\tau_{n,\mathrm{PO}}^H(\bar d^S)
\,\middle|\,
\LFc,\widetilde X_n^H
\right]
\right]
\\
&=
\Ex\!\left[
\sum_{\bar d^S\in\tilde{\mathcal{D}}}
1\{\pi(\widetilde X_n^H)=\bar d^S\}
\Ex\!\left[
\tau_{n,\mathrm{PO}}^H(\bar d^S)
\,\middle|\,
\LFc
\right]
\right]=
\Ex\!\left[
\tau_n^H\bigl(\pi(\widetilde X_n^H)\bigr)
\right].
\end{align*}
The identifying conditions of
Section~\ref{sec:LTV_identification_strategy} identify
$\tau_n^H(\bar d^S)$ for every $\bar d^S\in\tilde{\mathcal{D}}$;
$\widetilde X_n^H$ is observed; and $\pi$ is known.  Hence the last
expression, and therefore the common value in
\eqref{eq:prospective-value-identification}, is identified.  Equality for
every $\pi\in\PiC$ also implies equality of the historical and prospective
oracle-policy sets and of their policy-value gaps.
\end{proof}

Given Lemma~\ref{lem:prospective-policy-identification}, one can conduct policy learning using the following three
steps.

\noindent\textbf{Step 1}:
Using $X_n$, estimate $\widehat\tau_n^H(\cdot)$ by the \texttt{SBE-PCR}
algorithm.

\noindent\textbf{Step 2}:
Based on $(\widetilde X_n^H,\widehat\tau_n^H(\cdot))$, fit the decision tree
$\widehat\pi$.\footnote{We avoid using $\widetilde X_n^P$ to fit the tree
because it is observed at the end of the historical policy window and may
contain covariates affected by support received during 2011--2015; it would
also condition the historical gain on a later firm state rather than a
pre-policy state.}

\noindent\textbf{Step 3}:
Using $\widetilde X_n^P$, deploy the tree
$\widehat\pi(\widetilde X_n^P)$.

Step 2 maximizes the plug-in
analogue of the last expression in
\eqref{eq:prospective-value-identification}, which equals $V^P(\pi)$, and
Step 3 deploys the resulting policy using $\widetilde X_n^P$. Figure~\ref{fig:prospective-policy-timeline} depicts the entire timeline. Here are further details on implementation:

\begin{figure}[t]
\centering
\begin{tikzpicture}[
  x=1cm,y=1cm,>=stealth,
  every node/.style={font=\footnotesize},
  smalllabel/.style={font=\scriptsize,align=center},
  covariate/.style={draw,fill=black!8},
  window/.style={draw,fill=black!20}
]
\def\xstart{0}
\def\xhiststart{2.8}
\def\xTzero{3.8}
\def\xprospectstart{5.5}
\def\xhistorical{8.0}
\def\xobserved{10.7}
\def\xfuture{13.4}

\draw[<->] (\xstart,0.95)--(\xTzero,0.95)
  node[midway,smalllabel,inner sep=1pt] {$T_0$ observed periods};
\draw[<->] (\xTzero,0.95)--(\xobserved,0.95)
  node[midway,smalllabel,inner sep=1pt] {$T$ observed periods};
\draw[<->] (\xobserved,0.95)--(\xfuture,0.95)
  node[midway,smalllabel,inner sep=1pt] {$S$ prospective periods};

\draw[thick,->] (\xstart,0.45)--(13.65,0.45);
\foreach \x/\lab in {
  \xstart/$1$,
  \xTzero/$T_0$,
  \xhistorical/$\bar T-S$,
  \xobserved/$\bar T$,
  \xfuture/$\bar T+S$
}{
  \draw (\x,0.36)--(\x,0.54);
  \node[smalllabel,anchor=north] at (\x,0.31) {\lab};
}
\draw[densely dotted,gray]
  (\xhistorical,0.36)--(\xhistorical,-3.00);
\draw[densely dotted,gray]
  (\xobserved,0.36)--(\xobserved,-5.05);

\draw[covariate]
  (\xstart,-0.45) rectangle (\xTzero,-1.20);
\node[smalllabel] at (1.9,-0.825)
  {$X_n$: covariates used by\\\texttt{SBE-PCR}};
\draw[window]
  (\xTzero,-0.45) rectangle (\xobserved,-1.20);
\node[smalllabel] at (7.25,-0.825)
  {observed treatment and outcome panel};
\draw
  (\xstart,-1.32)--(\xstart,-1.45)
  --(\xobserved,-1.45)--(\xobserved,-1.32);
\node[smalllabel,anchor=north] at (5.35,-1.50)
  {\texttt{SBE-PCR} estimation:
   $(X_n,\text{observed panel})
   \mapsto\widehat\tau_n^H(\cdot)$};

\draw[covariate]
  (\xhiststart,-2.20) rectangle (\xhistorical,-3.00);
\node[smalllabel] at (5.4,-2.60)
  {$\widetilde X_n^H$: policy features\\
   $r$ periods ending before\\
   the historical policy window};
\draw[window]
  (\xhistorical,-2.20) rectangle (\xobserved,-3.00);
\node[smalllabel] at (9.35,-2.60)
  {historical policy/gain\\
   window: $S$ periods};
\draw
  (\xhiststart,-3.12)--(\xhiststart,-3.25)
  --(\xobserved,-3.25)--(\xobserved,-3.12);
\node[smalllabel,anchor=north] at (6.75,-3.30)
  {Tree fitting:
   $(\widetilde X_n^H,\widehat\tau_n^H(\cdot))
   \mapsto\widehat\pi$};

\draw[thin,->]
  (\xobserved,-2.60)
  to[out=0,in=90]
  node[pos=.55,smalllabel,anchor=west,inner sep=1pt]
       {Temporal stability}
  (11.35,-4.00);

\draw[covariate]
  (\xprospectstart,-4.00) rectangle (\xobserved,-4.80);
\node[smalllabel] at (8.1,-4.40)
  {$\widetilde X_n^P$: policy features\\
   $r$ periods ending before deployment};
\draw[window]
  (\xobserved,-4.00) rectangle (\xfuture,-4.80);
\node[smalllabel] at (12.05,-4.40)
  {prospective policy\\
   window: $S$ periods};
\draw
  (\xprospectstart,-4.92)--(\xprospectstart,-5.05)
  --(\xfuture,-5.05)--(\xfuture,-4.92);
\node[smalllabel,anchor=north] at (9.45,-5.10)
  {Deployment:
   $\widetilde X_n^P
   \mapsto\widehat\pi(\widetilde X_n^P)$};
\end{tikzpicture}
\caption{Estimation, prospective policy-learning, and deployment windows}
\label{fig:prospective-policy-timeline}
\end{figure}

\begin{remark}[Policy-learning sample and deployment]
When $S<T$, the clean baseline restricts the tree-learning sample to firms
untreated through period $\bar T-S$, so that
$\widetilde X_n^H$ is observed before treatment and matches the all-control
history in \eqref{eq:historical-gain}.\footnote{Under the additive LTV model,
one may instead include previously treated firms by evaluating gains after
each firm's realized history before the learning window, because the terms
generated by that common history cancel from the treated-versus-control
contrast.  This again relies on potentially treatment-affected covariates and
requires appropriate overlap and temporal stability for the intended target
population. 
}
The deployment vector $\widetilde X_n^P$ is directly available for a firm
observed and not-yet-treated through $\bar T$.  A firm first observed only
after that date requires policy covariates constructed from information
available at entry or from a shorter history.
\end{remark}

\begin{remark}[Continuation effects]\label{rem:continuation}
Equations~\eqref{eq:historical-gain}--\eqref{eq:prospective-gain} sum
outcomes only over the $S$ policy periods.  If welfare instead includes the
full $q$-period continuation of the final action, extend $\bar d^S$ by setting
$d_s$ to the control (no-support) action for $s\in[S+1,S+q]$, shift the historical action window to
$[\bar T-S-q+1,\bar T-q]$, construct $\widetilde X_n^H$ from
$[\bar T-S-q-r+1,\bar T-S-q]$, replace $\bar T-S$ and $T-S$ in
\eqref{eq:historical-gain} by $\bar T-S-q$ and $T-S-q$, respectively, and
extend both gain sums through $S+q$.
The prospective policy-feature window remains $[\bar T-r+1,\bar T]$.  In the
application (Section \ref{subsubsec:optimal_new_firms}), with $S=q=1$, $r=5$ gives 2009--13 for
$\widetilde X_n^H$ and 2011--15 for $\widetilde X_n^P$. The historical analogue then
assigns the action in 2014 and observes its continuation in 2015; the
prospective policy assigns the action in 2016 and evaluates its continuation
in 2017. Our robustness
exercise (SA~\ref{sec:appendix_five_period}) considers $S=5$, where $\widetilde X_n^H$ uses the
2006--10 covariates and $\widetilde X_n^P$ uses the 2011--15 covariates.
\end{remark}

\subsection{Plug-In Estimator and Regret Bounds}

To simplify notation, throughout this subsection and the remainder of
SA~\ref{sec:appendix_EWM}, write
\[
V(\pi)\coloneqq V^H(\pi),\qquad
\tau_n(\bar d^S)\coloneqq\tau_n^H(\bar d^S),\qquad
\widetilde X_n\coloneqq\widetilde X_n^H.
\]
For $\pi\in\PiC$, define its regret by
\[
R(\pi)
\coloneqq
\sup_{\pi'\in\PiC}V(\pi')-V(\pi).
\]
Under Assumptions~\ref{ass:sampling} and~\ref{ass:TT-contrast},
Lemma~\ref{lem:prospective-policy-identification} implies $R(\pi)
=
\sup_{\pi'\in\PiC}V^P(\pi')-V^P(\pi)$, so the analysis below applies to the prospective regret under the additional Assumption~\ref{ass:TT-contrast}.

For each $n$ and $\bar d^S\in\tilde{\mathcal{D}}$, let
$\widehat\tau_n(\bar d^S):=\widehat\tau_n^H(\bar d^S)$ denote the plug-in
estimate of $\tau_n(\bar d^S)$ produced by the \texttt{SBE-PCR} algorithm
(or a cross-fitted version).  Define the plug-in empirical value and the
learned policy by
\[
\widehat V(\pi)
\coloneqq
\frac1N\sum_{n=1}^N
\widehat\tau_n\bigl(\pi(\widetilde X_n)\bigr),
\qquad
\widehat V(\widehat\pi)
\ \ge\
\sup_{\pi\in\PiC}\widehat V(\pi)-\eta_N,
\]
where $\eta_N\geq0$ is an optimization tolerance.  Define the (random)
\emph{uniform plug-in error}
\[
r_N
\coloneqq
\max_{1\leq n\leq N}\ \max_{\bar d^S\in\tilde{\mathcal{D}}}
\left|\widehat\tau_n(\bar d^S)-\tau_n(\bar d^S)\right|.
\]
The \texttt{SBE-PCR} results enter the policy-learning analysis only through
the estimation error $r_N$ in $\widehat\tau_n(\cdot)$.

\begin{assumption}[Bounded gains]\label{ass:bounded}
There exists $M<\infty$ such that
$|\tau_n(\bar d^S)|\leq M$ almost surely for all
$\bar d^S\in\tilde{\mathcal{D}}$.
\end{assumption}

\noindent Under
Assumption~\ref{assumption:bounded-expected-potential-outcomes}(ii) (applied
to each of the $S$ outcome dates), each per-period conditional mean lies in
$[-1,1]$, so Assumption~\ref{ass:bounded} holds with $M=2S$.

\begin{assumption}[Uniform plug-in accuracy]\label{ass:uniform_plugin}
$r_N=o_p(1)$.
\end{assumption}

\noindent Sufficient conditions for Assumption~\ref{ass:uniform_plugin} are
developed in SA~\ref{sec:plugin}.

\begin{assumption}[Approximate maximization]\label{ass:approxmax}
$\eta_N=o_p(1)$.
\end{assumption}

\noindent In the application the schedule sets are small
($|\tilde{\mathcal{D}}|\in\{3,9,25\}$) and the tree search is exact \citep{zhou2023offline}, so
$\eta_N=0$.

\begin{assumption}[Finite Natarajan dimension]\label{ass:natarajan}
$\Nat(\PiC)=d_{\Nat}<\infty$.\footnote{The Natarajan dimension extends the VC
dimension to multiclass prediction; see \citet{natarajan1989learning} or
\citet[Ch.~29]{shalev2014understanding}.  When $|\tilde{\mathcal{D}}|=2$ it coincides with the
VC dimension.}
\end{assumption}

Define the oracle empirical value and the empirical-process deviation
\[
V_N(\pi)
\coloneqq
\frac1N\sum_{n=1}^N
\tau_n\bigl(\pi(\widetilde X_n)\bigr),
\qquad
\Delta_N
\coloneqq
\sup_{\pi\in\PiC}\bigl|V_N(\pi)-V(\pi)\bigr|.
\]

\begin{lemma}[Regret decomposition]\label{lem:decomp}
For any $\widehat\pi\in\PiC$ with
$\widehat V(\widehat\pi)
\geq\sup_{\pi\in\PiC}\widehat V(\pi)-\eta_N$,
\[
R(\widehat\pi)
\ \leq\
2\Delta_N+2r_N+\eta_N.
\]
\end{lemma}

\begin{proof}
For every $\pi\in\PiC$,
\[
\left|\widehat V(\pi)-V_N(\pi)\right|
\leq
\frac1N\sum_n
\left|
\widehat\tau_n\bigl(\pi(\widetilde X_n)\bigr)
-\tau_n\bigl(\pi(\widetilde X_n)\bigr)
\right|
\leq r_N.
\]
Hence
\begin{align*}
R(\widehat\pi)
&=\bigl(V(\pi^*)-V_N(\pi^*)\bigr)
+\bigl(V_N(\pi^*)-\widehat V(\pi^*)\bigr)
+\bigl(\widehat V(\pi^*)-\widehat V(\widehat\pi)\bigr)\\
&\qquad
+\bigl(\widehat V(\widehat\pi)-V_N(\widehat\pi)\bigr)
+\bigl(V_N(\widehat\pi)-V(\widehat\pi)\bigr)\leq
\Delta_N+r_N+\eta_N+r_N+\Delta_N,
\end{align*}
where $\pi^*$ maximizes $V$ over $\PiC$, assuming such a policy exists.
\qedhere
\end{proof}

\begin{theorem}[Natarajan-type regret bound]\label{thm:natarajan}
Suppose Assumptions~\ref{ass:sampling}, \ref{ass:bounded}, and
\ref{ass:natarajan} and the conditions of Lemma~\ref{lem:prospective-policy-identification} hold.  There is a universal constant $C>0$ such that for
any $\delta\in(0,1)$, with probability at least $1-\delta$,
\[
\Delta_N
\ \leq\
C M\,\sqrt{
\frac{
d_{\Nat}\bigl(\log N+\log|\tilde{\mathcal{D}}|\bigr)+\log(1/\delta)
}{N}
},
\]
and hence, by Lemma~\ref{lem:decomp}, with probability at least $1-\delta$,
\[
R(\widehat\pi)
\ \leq\
C M\,\sqrt{
\frac{
d_{\Nat}\bigl(\log N+\log|\tilde{\mathcal{D}}|\bigr)+\log(1/\delta)
}{N}
}
+2r_N+\eta_N.
\]
In particular, if Assumptions~\ref{ass:uniform_plugin} and
\ref{ass:approxmax} also hold, then
\[
R(\widehat\pi)
\ \leq\
O_p\!\left(
\sqrt{\frac{d_{\Nat}\bigl(\log N+\log|\tilde{\mathcal{D}}|\bigr)}{N}}
\right)
+2r_N+\eta_N
\ \xrightarrow{\ p\ }\ 0
\]
provided $d_{\Nat}\bigl(\log N+\log|\tilde{\mathcal{D}}|\bigr)/N\to0$.
\end{theorem}

\begin{proof}
Write $Z_n\coloneqq(\widetilde X_n,\tau_n(\cdot))$ and $f_\pi(Z_n)\coloneqq\tau_n(\pi(\widetilde X_n))$, so that $V_N(\pi)$ is the empirical mean of $f_\pi$ over the i.i.d.\ sample $Z_{1:N}$ (Assumption~\ref{ass:sampling}), $V(\pi)=\Ex[f_\pi(Z_1)]$, and $|f_\pi|\le M$ (Assumption~\ref{ass:bounded}). By McDiarmid's inequality, $\Delta_N\le\Ex[\Delta_N]+M\sqrt{2\log(1/\delta)/N}$ with probability at least $1-\delta$. By symmetrization and Massart's finite-class lemma \citep[Lemma 26.8]{shalev2014understanding}, $\Ex[\Delta_N]\le 2M\sqrt{2\log(2\Gamma_{\PiC}(N))/N}$, where $\Gamma_{\PiC}(N)$ is the number of distinct label vectors $(\pi(x_1),\dots,\pi(x_N))$ realizable by $\PiC$, and the Natarajan lemma \citep[Lemma 29.4]{shalev2014understanding} gives $\Gamma_{\PiC}(N)\le N^{d_{\Nat}}|\tilde{\mathcal{D}}|^{2d_{\Nat}}$. Combining the three bounds (using $\sqrt a+\sqrt b\le\sqrt{2(a+b)}$ and $\log 2\le\log N$) yields the first display with $C=7$; Lemma~\ref{lem:decomp} doubles the $\Delta_N$ term, so both displays hold with $C=14$, and the $O_p$ statement follows from Assumptions~\ref{ass:uniform_plugin} and~\ref{ass:approxmax}.
\end{proof}

\begin{remark}[The $\log|\tilde{\mathcal{D}}|$ term]\label{rem:logD}
With $\tilde{\mathcal{D}}\subseteq[0,A]^S$ one has
$\log|\tilde{\mathcal{D}}|\leq S\log(A+1)$, which is \emph{not} negligible when $S$
grows---the regime covered by the fixed-lag consistency theorems.  Absorbing
it into the constant is justified only for fixed $|\tilde{\mathcal{D}}|$; for the restricted
schedule sets of the application ($|\tilde{\mathcal{D}}|\leq25$) the term can be ignored.
\end{remark}

\begin{remark}[Decision trees]\label{rem:trees}
For depth-$L$ axis-aligned decision trees on $\Rb^{\widetilde p}$ with leaf
labels in $\tilde{\mathcal{D}}$, $\Nat(\PiC)
\ \leq\
O\!\bigl(L\,2^L\log(\widetilde p\,|\tilde{\mathcal{D}}|)\bigr)
$ \citep[Thm.~2.2]{jin2023natarajan}.  This is exponential in depth and only
\emph{logarithmic} in the number of policy features $\widetilde p$ and the
number of schedules $|\tilde{\mathcal{D}}|$.  For fixed $L$,
Theorem~\ref{thm:natarajan} then gives a regret rate of order
$\sqrt{d_{\Nat}\log N/N}$ with
$d_{\Nat}\lesssim L2^L\log(\widetilde p|\tilde{\mathcal{D}}|)$.
\end{remark}

\begin{remark}[Retrospective learning]\label{rem:retrospective-application}
The retrospective exercise in \eqref{eq:opt_alloc} is obtained by setting
$S=T$ in the historical gain problem.  For $\bar d^T\in\tilde{\mathcal{D}}$, the relevant
gain relative to no support is
\[
\tau_n(\bar d^T)
\coloneqq
\sum_{t=1}^T
\Ex\!\left[
Y_{n,T_0+t}^{(\bar d^t)}
-
Y_{n,T_0+t}^{(\bar 0^t)}
\,\middle|\,\LFc
\right].
\]
Its plug-in counterpart selects $\widehat{\bar d}^{\,T}(n)
\in
\argmax_{\bar d^T\in\tilde{\mathcal{D}}}\widehat\tau_n(\bar d^T)$ and
\begin{align*}
\tau_n\bigl(\bar d^{T*}(n)\bigr)
-
\tau_n\bigl(\widehat{\bar d}^{\,T}(n)\bigr)&\leq
\left[
\tau_n\bigl(\bar d^{T*}(n)\bigr)
-
\widehat\tau_n\bigl(\bar d^{T*}(n)\bigr)
\right]
+
\left[
\widehat\tau_n\bigl(\widehat{\bar d}^{\,T}(n)\bigr)
-
\tau_n\bigl(\widehat{\bar d}^{\,T}(n)\bigr)
\right]\leq 2r_N
\end{align*}
simultaneously for all $n\leq N$.  Unlike the prospective exercise, the
retrospective exercise optimizes a schedule for each observed
firm.  It therefore involves neither a policy class nor an empirical-process
term; its gain loss is controlled solely by the uniform plug-in error.
\end{remark}

\subsection{Verifying Assumption~\ref{ass:uniform_plugin}}\label{sec:plugin}

Under the fixed-lag structure of
Assumption~\ref{assumption:outcome-fixed-lag-dep}, the baseline terms cancel
from the treatment gains.  Re-index the historical blips relative to the
beginning of the historical action window as $\gamma^H_{n,s,\ell}(d)\coloneqq\gamma_{n,T-S+s,\,T-S+\ell}(d)$ for $s\in\{\ell,\ldots,\min\{S,\ell+q\}\}$ and $\ell\in[S]$. Then
\begin{equation}\label{eq:fixedlag}
\tau_n(\bar d^S)
=
\sum_{\ell=1}^S\sum_{s=\ell}^{\min\{S,\ell+q\}}
\gamma^H_{n,s,\ell}(d_\ell),
\qquad
\widehat\tau_n(\bar d^S)
=
\sum_{\ell=1}^S\sum_{s=\ell}^{\min\{S,\ell+q\}}
\widehat\gamma^H_{n,s,\ell}(d_\ell).
\end{equation}
Define the primitive errors $e^H_{n,s,\ell}(d)
\coloneqq
\widehat\gamma^H_{n,s,\ell}(d)-\gamma^H_{n,s,\ell}(d)$ and, for each action time $\ell\in[S]$, the feasible non-control action set $\cA_\ell^+(\tilde{\mathcal{D}})
\coloneqq
\{d_\ell:\bar d^S\in\tilde{\mathcal{D}}\}\cap[A]$.

\begin{lemma}[Primitive union bound, fixed-lag form]
\label{lem:primitive_union}
Let \eqref{eq:fixedlag} hold and set $\kappa\coloneqq S(q+1)$.  For every
$\varepsilon>0$,
\[
\PP(r_N>\varepsilon)
\ \leq\
\sum_{n=1}^N\sum_{\ell=1}^S
\sum_{s=\ell}^{\min\{S,\ell+q\}}
\sum_{d\in\cA_\ell^+(\tilde{\mathcal{D}})}
\PP\!\left(
|e^H_{n,s,\ell}(d)|>\frac{\varepsilon}{\kappa}
\right).
\]
Consequently, if for some tail function $\psi_N(\cdot)$, $\sup_{\substack{
n\leq N,\ \ell\in[S],\\
s\in[\ell,\min\{S,\ell+q\}],\ d\in[A]
}}
\PP\bigl(|e^H_{n,s,\ell}(d)|>u\bigr)
\leq\psi_N(u)$, then
\[
\PP(r_N>\varepsilon)
\ \leq\
N S(q+1)A\,
\psi_N\!\left(\frac{\varepsilon}{S(q+1)}\right).
\]
In particular, even for $\tilde{\mathcal{D}}=[0,A]^S$ the union bound runs over at most
$NS(q+1)A$ primitive events (i.e., polynomial in $(N,S,q,A)$) rather than
over the $N|\tilde{\mathcal{D}}|=N(A+1)^S$ schedule values.
\end{lemma}

\begin{proof}
By \eqref{eq:fixedlag} and the triangle inequality, for every $n$ and
$\bar d^S\in\tilde{\mathcal{D}}$,
\[
\left|
\widehat\tau_n(\bar d^S)-\tau_n(\bar d^S)
\right|
\leq
\sum_{\ell=1}^S\sum_{s=\ell}^{\min\{S,\ell+q\}}
\left|e^H_{n,s,\ell}(d_\ell)\right|,
\]
a sum of at most $\kappa=S(q+1)$ nonzero primitive errors.  If all primitive
errors with indices in the feasible range are at most
$\varepsilon/\kappa$, then $r_N\leq\varepsilon$.  Hence
$\{r_N>\varepsilon\}$ is contained in the union of the events displayed in
the lemma.  The union bound gives the first inequality, and the uniform tail
bound together with $|\cA_\ell^+(\tilde{\mathcal{D}})|\leq A$ gives the second.
\end{proof}

For the continuation-aware criterion of Remark~\ref{rem:continuation}, the
same argument uses the origin $T-S-q$ in place of $T-S$ and lets
$s\in[\ell,\ell+q]$; thus the bound above remains valid.

\begin{remark}[Tail regimes]\label{rem:tails}
The consistency theorems of the main text deliver $O_p$ rates for each primitive error, whereas Lemma~\ref{lem:primitive_union} requires tail bounds; strengthening the former to the latter is a separate step. If the primitive errors are sub-Gaussian, $\psi_N(u)=C\exp(-u^2/B_N^2)$, then with probability at least $1-\delta$, $r_N\lesssim S(q+1)B_N\sqrt{\log(NS(q+1)A/\delta)}$; if they are only sub-exponential, $\psi_N(u)=C\exp(-u/B_N)$, the bound degrades to $S(q+1)B_N\log(NS(q+1)A/\delta)$. In either case Assumption~\ref{ass:uniform_plugin} holds once $S(q+1)B_N$ times the corresponding logarithmic factor vanishes, and the logarithm is over the polynomially many primitives of Lemma~\ref{lem:primitive_union}, never over the $(A+1)^S$ schedules.
\end{remark}

\section{Connection to Models for Dynamic Treatment Effects in Biostatistics}
\label{app:snmm_assm}

We relate our assumptions to the notation and assumptions of structural nested mean models (SNMMs) and marginal structural models (MSMs) in the biostatistics literature on dynamic treatment effects.
A typical assumption in this literature is sequential conditional exogeneity: for all $t\in[T]$ and some sequence of random state variables $S_{n,t}$,
\begin{align}\label{eq:blip_seq_exog}
\forall \ell \in [t], \ \forall \bar{d}^t \in [A]^t: Y_{n, t}^{(\bar{d}^t)} \independent \ D_{n, \ell} \mid \bar{S}^{\ell -1}_{n}, \bar{D}^{\ell - 1}_{n} = \bar{d}^{\,\ell - 1}, \LFc,
\end{align}
where $\bar{S}^{\ell -1}_{n}=(S_{n,0}, \ldots, S_{n,\ell-1})$.
Moreover, assume that the blip effects admit the factor model representation
\begin{align}\label{eq:blip_effect}
\Ex\left[Y_{n, t}^{(\bar{d}^\ell, \underline{0}^{\ell + 1})} - Y_{n, t}^{(\bar{d}^{\ell - 1}, \underline{0}^{\ell})} \mid \bar{S}^{\ell -1}_{n}, \bar{D}_n^{\ell} = \bar{d}^\ell, \LFc \right] = \ldot{\psi^{t, \ell}_{n}}{w_{d_\ell} - w_{0_\ell}}  \mid \LFc ,
\end{align}
so that the conditional mean of the blip effect is invariant to past states and actions.
Lastly, assume that the baseline potential outcome has a factor model representation:
\begin{align}\label{eq:blip_baseline_outcomes}
\Ex\left[ Y^{(\bar{0}^t)}_{n, t} \mid \LFc \right]
= \sum^t_{\ell = 1} \ldot{\psi^{t, \ell}_{n}}{w_{0_\ell}}  \mid \LFc.
\end{align}
\begin{proposition}\label{prop:snmm_connection}
Let \eqref{eq:blip_seq_exog}--\eqref{eq:blip_baseline_outcomes} hold.
Then Assumptions~\ref{assumption:LTV_factor_model} and \ref{assumption:LTV_seq_exog} hold.
\end{proposition}
The inductive proof below is in essence known in the literature: SNMMs that are past action and state independent also imply a marginal structural model, i.e., Assumption~\ref{assumption:LTV_factor_model} (see, e.g., Technical Point~21.4 of \citet{hernan2020causal}); we include it for completeness and in our notation.
Thus, one could instead impose \eqref{eq:blip_seq_exog}--\eqref{eq:blip_baseline_outcomes}, which are more in line with the dynamic treatment effect literature, and our identification argument would immediately apply.
Our assumptions are more permissive, however: the blip condition in Assumption~\ref{assumption:LTV_seq_exog} only requires that the blip effect is not modified by past actions, while potentially allowing modification conditional on past states that confound the treatment; the full SNMM specification above precludes such effect modifications.

\paragraph{Verification of Assumption~\ref{assumption:LTV_factor_model}.}
In what follows, all conditional expectations are also conditioned on the latent factors $\LFc$, which we omit.
By telescoping, $\Ex[Y_{n,t}^{(\bar{d}^t)} - Y_{n,t}^{(\bar{0}^t)}] = \sum_{\ell=1}^t \Ex[Y_{n,t}^{(\bar{d}^\ell, \underline{0}^{\ell+1})} - Y_{n,t}^{(\bar{d}^{\ell-1}, \underline{0}^{\ell})}]$, so it suffices to show that, for each $\ell \leq t$,
\begin{align}
Q_{n,t,\ell}
\coloneqq \Ex\left[Y_{n,t}^{(\bar{d}^\ell, \underline{0}^{\ell+1})} - Y_{n,t}^{(\bar{d}^{\ell-1}, \underline{0}^{\ell})}\right]
= \ldot{\psi_{n}^{t,\ell}}{w_{d_\ell} - w_{0_\ell}}. \label{eq:blip_key_representation}
\end{align}
We establish this via a nested mean argument, proving by induction over $r \leq \ell$ that
\begin{align}
Q_{n,t,\ell}
=~& \Ex\left[\Ex\left[\ldots \Ex\left[Y_{n,t}^{(\bar{d}^\ell, \underline{0}^{\ell+1})} - Y_{n,t}^{(\bar{d}^{\ell-1}, \underline{0}^{\ell})} \mid \bar{S}_n^{r-1}, \bar{D}_n^{r}
= \bar{d}^{r}\right]\ldots \mid S_{n,0}, D_{n,1}=d_1\right]\right]. \label{eq:blip_inducive_proof}
\end{align}
For the base case $r=1$, the tower law gives $Q_{n,t,\ell} = \Ex[\Ex[\,\cdot\mid S_{n,0}]]$, and \eqref{eq:blip_seq_exog} allows adding the conditioning event $D_{n,1}=d_1$.
For the inductive step, apply the tower law to the innermost expectation in \eqref{eq:blip_inducive_proof} to condition additionally on $\bar{S}_n^{r}$, and then use \eqref{eq:blip_seq_exog} to add the conditioning event $D_{n,r+1}=d_{r+1}$; this yields \eqref{eq:blip_inducive_proof} with $r$ replaced by $r+1$, concluding the induction.
At $r = \ell$, the innermost expectation in \eqref{eq:blip_inducive_proof} equals, by \eqref{eq:blip_effect}, $\gamma_{n,t,\ell}(d_\ell) \coloneqq \ldot{\psi_{n}^{t,\ell}}{w_{d_\ell} - w_{0_\ell}}$, which is independent of $\bar{S}^{\ell-1}_n$ and $\bar{D}^{\ell}_n$; hence all outer expectations collapse and \eqref{eq:blip_key_representation} follows.
Summing \eqref{eq:blip_key_representation} over $\ell \leq t$ and rearranging the telescoping identity gives $\Ex[ Y_{n,t}^{(\bar{d}^t)}]
= \Ex[Y_{n,t}^{(\bar{0}^t)}] + \sum_{\ell=1}^t \gamma_{n,t,\ell}(d_\ell)$, which combined with \eqref{eq:blip_baseline_outcomes} implies Assumption~\ref{assumption:LTV_factor_model} holds.

\paragraph{Verification of Assumption~\ref{assumption:LTV_seq_exog}.}
Assumption~\ref{assumption:LTV_seq_exog} is immediately implied by \eqref{eq:blip_effect} and the tower law of expectations: integrate $\bar{S}^{\ell-1}_{n}$ out of both sides of \eqref{eq:blip_effect}, applied at outcome date $t = T$ and $\ell = t$.

\section{Identification by Synthetic Interventions}\label{app:SI}

We provide an identification argument for dynamic treatments that builds upon the SI framework \citep{SI}, formalizing the discussion in Section~\ref{sec:identification_SI}.

\paragraph{Donor units.} We first define a collection of donor units.

\begin{definition}[SI donor units]\label{def:general_factor_model_donor_units}
For $\bar{d}^T \in [A]^T$,
\begin{align}
    \Ic^{\bar{d}^T} &\coloneqq \{j \in [N]:
    (i) \ \bar{D}^T_j = \bar{d}^T,
    \ (ii) \ \ \forall \ \bar{\delta}^T \in [A]^T, \ \    \Ex[\varepsilon^{(\bar{\delta}^T)}_{j ,T} \mid \bar{D}^T_j, \LFc] = 0 \}. \label{eq:general_factor_model_donor_units}
\end{align}
\end{definition}

The donor set $\Ic^{\bar{d}^T}$ thus contains units that receive exactly the sequence $\bar{d}^T$ and whose noise terms $\varepsilon^{(\bar{\delta}^T)}_{j ,T}$ are conditionally mean independent of the action sequence $\bar{D}^T_j$.
A sufficient condition for property (ii) is that $\forall \ \bar{\delta}^T \in [A]^T, \ Y^{(\bar{\delta}^T)}_{j ,T} \independent \bar{D}^T_j \mid \LFc$: the entire action sequence is chosen at $t = 0$ conditional on the latent factors, i.e., the policy for these units is {\em not adaptive} (cannot depend on observed outcomes $Y_{j, t}$ for $t \in [T]$).

\begin{assumption}[Well-supported factors]\label{assumption:ID1_well_supported_factors}
$\forall n\in [N], \bar{d}^T \in [A]^T$ suppose that $v_{n, T}$ satisfies a well-supported condition, i.e., there exist linear weights $\beta^{n,\Ic^{\bar{d}^T}} \in \Rb^{|\Ic^{\bar{d}^T}|}$ such that $v_{n, T} =\sum_{j \in \Ic^{\bar{d}^T}}
\beta_j^{n,\Ic^{\bar{d}^T}} v_{j, T}$.
\end{assumption}

Assumption~\ref{assumption:ID1_well_supported_factors} states that, for each $\bar{d}^T \in [A]^T$, the target unit's latent factor $v_{n, T}$ lies in the linear span of the donor units' factors $\{v_{j, T}\}_{j \in \Ic^{\bar{d}^T}}$; see the discussion under Assumption~\ref{assumption:LTV_well_supported_factors} in Section~\ref{sec:time_varying_systems}.
We then have identification: the target parameter \eqref{eq:target_causal_param} can be expressed as a function of observed outcomes by adapting the identification argument in \cite{SI}.

\begin{theorem}[SI Identification Strategy]\label{thm:newID1}
Let Assumptions~\ref{assumption:SUTVA}, \ref{assumption:general_latent_factor_model}, and \ref{assumption:ID1_well_supported_factors} hold.
Then, for all $n\in [N]$ and $\bar{d}^T \in [A]^T$, the mean counterfactual outcome can be expressed as:
\begin{align*}
\Ex[Y_{n,T}^{(\bar{d}^T)}\mid \LFc]
=~& \Ex\left[\sum_{j \in \Ic^{\bar{d}^T}} \beta_j^{n,\Ic^{\bar{d}^T}} Y_{j ,T}  \mid \LFc, \Ic^{\bar{d}^T}\right].
\end{align*}
\end{theorem}

\paragraph{Interpretation and limitations.}
Theorem~\ref{thm:newID1} establishes that the mean counterfactual outcome of unit $n$ under $\bar{d}^T$ is a linear re-weighting, with weights $\beta_j^{n,\Ic^{\bar{d}^T}}$, of the observed outcomes of donors that received exactly that sequence non-adaptively.
This strategy has two demanding requirements. First, it needs a sufficiently large donor set $\Ic^{\bar{d}^T}$ for {\em every} $\bar{d}^T \in [A]^T$, so the number of donor units must scale as $A^T$, which grows exponentially in $T$. Second, the actions chosen for the donor units cannot be adaptive.

\subsection{Proof of Theorem~\ref{thm:newID1}}
Since $v_{n, T}$ and $w_{\bar{d}^T}$ are deterministic conditional on the latent factors $\LFc$ (and thus are independent of $\Ic^{\bar{d}^T}$), Assumption~\ref{assumption:general_latent_factor_model} and Assumption~\ref{assumption:ID1_well_supported_factors} give
\begin{align*}
\Ex\left[Y_{n, T}^{(\bar{d}^T)} \mid \LFc \right]
= \ldot{v_{n, T}}{\ w_{\bar{d}^T}}\mid \LFc,\Ic^{\bar{d}^T}
= \sum_{j \in \Ic^{\bar{d}^T}} \beta^{n, \Ic^{\bar{d}^T}}_j\ldot{v_{j, T}}{\ w_{\bar{d}^T}}\mid \LFc,\Ic^{\bar{d}^T}.
\end{align*}
Moreover, by the conditional mean exogeneity of $\varepsilon_{j, T}^{(\bar{d}^T)}$ for donor units in Definition~\ref{def:general_factor_model_donor_units}, $\Ex[\varepsilon_{j, T}^{(\bar{d}^T)} \mid \LFc, \Ic^{\bar{d}^T}] = 0$ for all $j \in \Ic^{\bar{d}^T}$; hence each summand satisfies $\ldot{v_{j, T}}{\ w_{\bar{d}^T}}\mid \LFc,\Ic^{\bar{d}^T} = \Ex[Y_{j, T}^{(\bar{d}^T)} \mid \LFc, \Ic^{\bar{d}^T}] = \Ex[Y_{j, T} \mid \LFc, \Ic^{\bar{d}^T}]$, where the first equality also uses Assumption~\ref{assumption:general_latent_factor_model} and the second follows from Assumption~\ref{assumption:SUTVA} since units in $\Ic^{\bar{d}^T}$ received exactly the sequence $\bar{d}^T$.
This completes the proof.

\section{Proofs and Remarks for LTV Factor Models}\label{app:LTV_proofs}

\subsection{Details for Example~\ref{ex:dyn_triangular_model} (LTV Dynamic Triangular Models)}\label{app:LTV_example_details}
The factors and noise terms in representation \eqref{eq:LTV_factor_model_representation} are, for $\ell \in [t - 1]$,
\begin{align*}
    \psi^{t, \ell}_{n} &= \Big( \Big(\prod^t_{k = \ell + 1}\bB_{n, k} \Big) \bC_{n, \ell}\Big)^{\top} \theta_{n, t},
    &&\varepsilon_{n, t, \ell} = \Big(\Big(\prod^t_{k = \ell + 1}\bB_{n, k}\Big) \eta_{n, \ell} \Big)^{\top} \theta_{n, t},
    \\
    \psi^{t, t}_{n} &= \bC_{n, t}^{\top}\theta_{n, t} + \tilde{\theta}_{n, t},
    &&\varepsilon_{n, t, t} = \theta_{n, t}^{\top}\eta_{n, t} + \tilde{\eta}_{n, t}.
\end{align*}
In this example, the target parameter $\Ex[Y^{(\bar{d}^T)}_{n, T} \mid \LFc]$ in
\eqref{eq:target_causal_param} conditions on the latent quantities
$\{\psi^{t, \ell}_{n}, w_{d_\ell}\}$, which are functions of
$(\bB_{n, t}, \bC_{n, t}, \theta_{n, t}, \tilde{\theta}_{n, t})$, and averages over
the per-step independent mean-zero innovations.

\subsection{Proof of Theorem~\ref{thm:newLTV}}
For simplicity, we omit the conditioning on $\LFc$ in all derivations; all expectations are conditioned on $\LFc$.

\noindent{\bf 1. Verifying \eqref{eq:LTV_identification}.}
First, we verify \eqref{eq:LTV_identification}, which expresses the counterfactual outcome in terms of the blips and the baseline.
For all $n \in [N]$, using Assumption~\ref{assumption:LTV_factor_model},
\begin{align*}
\Ex[Y^{(\bar{d}^T)}_{n, T}]
&= \Ex[Y^{(\bar{d}^T)}_{n, T} - Y^{(\bar{0}^T)}_{n, T}] + \Ex[Y^{(\bar{0}^T)}_{n, T}]
= \sum^{T}_{t = 1} \gamma_{n, T, t}(d_t)  + b_{n, T},
\end{align*}
where the second equality substitutes the representation of Assumption~\ref{assumption:LTV_factor_model} into both expectations and applies the definitions of $\gamma_{n, T, t}(d_t)$ and $b_{n, T}$.

\noindent{\bf 2. Verifying \eqref{eq:base_observed} and \eqref{eq:base_synthetic}.}
We first show \eqref{eq:base_observed}. For $j \in \Ic^0_T$:
\begin{align}
b_{j, T}
&= \sum^T_{t = 1} \ldot{\psi^{T, t}_{j}}{w_{0_t}}
= \Ex\left[ \sum^T_{t = 1} \ldot{\psi^{T, t}_{j}}{w_{0_t}} + \varepsilon^{(\bar{0}^T)}_{j, T}  \mid j \in \Ic^0_T \right]
= \Ex\left[Y_{j,T} \mid j \in \Ic^0_T \right], \label{eq:LTV_base_1}
\end{align}
where the second equality in \eqref{eq:LTV_base_1} follows from Assumption~\ref{assumption:LTV_factor_model} together with the facts that $\ldot{\psi^{T, t}_{j}}{w_{0_t}}$ is deterministic conditional on $\LFc$ and that $\Ex[\varepsilon^{(\bar{0}^T)}_{j, T} \mid j \in \Ic^0_T] = \Ex[\varepsilon^{(\bar{0}^T)}_{j, T}]$ by the definition of $\Ic^0_T$; the last equality follows from Assumption~\ref{assumption:LTV_factor_model} (which gives $\Ex[Y^{(\bar{0}^T)}_{j,T} \mid j \in \Ic^0_T]$) and Assumption~\ref{assumption:SUTVA}.

Next we show \eqref{eq:base_synthetic}. For $i \notin \Ic^0_T$:
\begin{align}
b_{i, T}
=  \sum^{T}_{t = 1} \ldot{\psi^{T, t}_{i}}{w_{0_t}}
= \sum^{T}_{t = 1} \ldot{\psi^{T, t}_{i}}{w_{0_t}} \mid \Ic^0_T = \sum^{T}_{t = 1} \sum_{j \in \Ic^0_T} \beta_j^{i,\Ic^0_T} \ldot{\psi^{T, t}_{j}}{w_{0_t}} \mid \Ic^0_T
=  \sum_{j \in \Ic^0_T} \beta_j^{i,\Ic^0_T}   b_{j, T} \mid \Ic^0_T, \label{eq:LTV_base_synth_1}
\end{align}
where the second equality in \eqref{eq:LTV_base_synth_1} follows from the fact that $\ldot{\psi^{T, t}_{i}}{w_{0_t}}$ is deterministic conditional on $\LFc$, and the third from Assumption~\ref{assumption:LTV_well_supported_factors}.

\noindent{\bf 3. Verifying \eqref{eq:base_blip_observed} and \eqref{eq:base_blip_synthetic}.}
We first show \eqref{eq:base_blip_observed}. For all $d \in [A]$ and $j \in \Ic^d_T$, adding and subtracting $\sum^{T - 1}_{t = 1} \ldot{\psi_{j}^{T,t}}{w_{0_t}}$,
\begin{align}
\gamma_{j, T, T}(d)
&=  \ldot{\psi^{T, T}_{j}}{w_{d} - w_{0_T}} \nonumber\\
&= \Ex\left[\ldot{\psi_{j}^{T,T}}{w_{d}} + \varepsilon^{(\bar{0}^{T - 1}, d)}_{j, T} + \sum^{T - 1}_{t = 1} \ldot{\psi_{j}^{T,t}}{w_{0_t}}  \mid j \in \Ic^d_T \right] - b_{j, T} \label{eq:LTV_base_blip_2}
\\ &= \Ex[Y_{j, T} \mid j \in \Ic^d_T] -    b_{j, T}, \label{eq:LTV_base_blip_7}
\end{align}
where condition~(ii) of \eqref{eq:LTV_donor}, together with Assumption~\ref{assumption:LTV_factor_model}, gives
$\Ex[\varepsilon^{(\bar{0}^T)}_{j,T}\mid j\in\Ic_T^d]=0$.
Assumption~\ref{assumption:LTV_seq_exog}, applied at $t=T$ with
$\bar d^T=(\bar{0}^{T-1},d)$, gives
$\Ex[\varepsilon^{(\bar{0}^{T-1},d)}_{j,T}
-\varepsilon^{(\bar{0}^T)}_{j,T}\mid j\in\Ic_T^d]=0$.
Consequently,
$\Ex[\varepsilon^{(\bar{0}^{T-1},d)}_{j,T}\mid j\in\Ic_T^d]=0$,
which, together with the fact that the factor terms are deterministic
conditional on $\LFc$, gives \eqref{eq:LTV_base_blip_2};
\eqref{eq:LTV_base_blip_7} then follows from
Assumptions~\ref{assumption:LTV_factor_model} and
\ref{assumption:SUTVA}.

Next, \eqref{eq:base_blip_synthetic} follows by the same argument as \eqref{eq:base_synthetic}: for $i \notin \Ic^d_T$, the quantity $\ldot{\psi^{T, T}_{i}}{w_{d} - w_{0_T}}$ is deterministic conditional on $\LFc$, and expanding it via Assumption~\ref{assumption:LTV_well_supported_factors} gives $\gamma_{i, T, T}(d) = \sum_{j \in \Ic^d_T} \beta^{i, \Ic^d_T}_{j} \gamma_{j, T, T}(d) \mid \Ic^d_T$.

\noindent{\bf 4. Verifying \eqref{eq:recursive_blip_observed} and \eqref{eq:recursive_blip_synthetic}.}
We first show \eqref{eq:recursive_blip_observed}. For all $d \in [A]$, $t < T$, $j \in \Ic^d_t$, note that $\bar{D}^{t}_j = (\bar{0}^{t - 1}, d)$, so $Y^{(\bar{D}^{t-1}_j, \underline{0}^{t})}_{j, T} = Y^{(\bar{0}^T)}_{j, T}$. Using Assumption~\ref{assumption:SUTVA} and telescoping,
\begin{align}
\Ex\left[ Y_{j, T} - Y^{(\bar{0}^T)}_{j, T} \mid j \in \Ic^d_t \right]
&= \sum^T_{\ell = t} \Ex\left[ Y^{(\bar{D}^\ell_j, \underline{0}^{\ell + 1})}_{j, T} - Y^{(\bar{D}^{\ell - 1}_j, \underline{0}^{\ell})}_{j, T} \mid j \in \Ic^d_t \right] \nonumber
\\ &= \sum^T_{\ell = t} \Ex\left[ \ldot{\psi^{T, \ell}_j}{w_{D_{j, \ell}} -w_{0_\ell} } + \varepsilon^{(\bar{D}^\ell_j, \underline{0}^{\ell + 1})}_{j, T} - \varepsilon^{(\bar{D}^{\ell - 1}_j, \underline{0}^{\ell})}_{j, T} \mid j \in \Ic^d_t \right] \label{eq:LTV_recursive_blip_2}
\\ &=  \gamma_{j, T, t}(d)  + \Ex\Big[\sum^T_{\ell = t + 1}  \gamma_{j, T, \ell}(D_{j, \ell}) \,\Big|\, j \in \Ic^d_t\Big],  \label{eq:LTV_recursive_blip_6}
\end{align}
where \eqref{eq:LTV_recursive_blip_2} follows from Assumption~\ref{assumption:LTV_factor_model}.
For \eqref{eq:LTV_recursive_blip_6}: in the $\ell = t$ term, $D_{j, t} = d$ for $j \in \Ic^d_t$ and $\ldot{\psi^{T, t}_j}{w_{d} -w_{0_t}} = \gamma_{j, T, t}(d)$ is deterministic given $\LFc$; for $\ell > t$ the factor term equals $\gamma_{j, T, \ell}(D_{j, \ell})$ by definition and is kept inside the conditional expectation, since $D_{j, \ell}$ may be adaptive. The noise terms vanish: condition (i) of \eqref{eq:LTV_donor} restricts only $\bar{D}^t_j$ (and condition (ii) the law of unit $j$), so the event $\{j \in \Ic^d_t\}$ is determined by $\bar{D}^t_j$ given $\LFc$, and for $\ell \ge t$ the tower rule over $\bar{D}^{\ell}_j$ reduces each noise difference to an average over $\bar{\delta}^\ell$ of $\Ex[\varepsilon^{(\bar{\delta}^\ell, \underline{0}^{\ell + 1})}_{j, T} - \varepsilon^{(\bar{\delta}^{\ell - 1}, \underline{0}^{\ell})}_{j, T} \mid \bar{D}^\ell_j = \bar{\delta}^\ell]$, which is zero by Assumption~\ref{assumption:LTV_seq_exog}.

Re-arranging \eqref{eq:LTV_recursive_blip_6},
\begin{align*}
\gamma_{j, T, t}(d)
&= \Ex\Big[ Y_{j, T} - \sum^T_{\ell = t + 1}  \gamma_{j, T, \ell}(D_{j, \ell}) \,\Big|\, j \in \Ic^d_t \Big] -  b_{j, T},
\end{align*}
where condition~(ii) of \eqref{eq:LTV_donor} and
Assumption~\ref{assumption:LTV_factor_model} give
$\Ex[Y^{(\bar{0}^T)}_{j,T}\mid j\in\Ic_t^d]
=\Ex[Y^{(\bar{0}^T)}_{j,T}]=b_{j,T}$.

Finally, \eqref{eq:recursive_blip_synthetic} follows by the same argument as \eqref{eq:base_synthetic}: for all $d \in [A]$, $t < T$, $i \notin \Ic^d_t$, the quantity $\ldot{\psi^{T, t}_{i}}{w_{d} - w_{0_t}}$ is deterministic conditional on $\LFc$, so
\begin{align*}
\gamma_{i, T, t}(d)
=&  \ldot{\psi^{T, t}_{i}}{w_{d} - w_{0_t}} \mid \Ic^d_t
=  \sum_{j \in \Ic^d_t} \beta^{i, \Ic^d_t}_{j}\ldot{\psi_{j}^{T,t}}{w_{d} - w_{0_t}} \mid \Ic^d_t
= \sum_{j \in \Ic^d_t} \beta^{i, \Ic^d_t}_{j} \gamma_{j, T, t}(d) \mid \Ic^d_t,
\end{align*}
using Assumption~\ref{assumption:LTV_well_supported_factors} in the second equality.
\subsection{General Remarks on the LTV Setting}\label{subsection:notes-on-ltv-setting}

\subsubsection{Covariate Design}

Assumption~\ref{assumption:additional-covariates} in its base form posits $p$ covariates per unit, each with respect to the terminal-time factor $v_{n,T}$, and Theorem~\ref{thm:consistency-time-varying} requires $p \to \infty$ with $T$ fixed, which seems unlikely in practice; time-varying covariates justify $p \to \infty$ in the regime of Theorem~\ref{thm:consistency-time-varying-fixed-lags}, where $T \to \infty$ as well. A general construction is the following.

\begin{assumption}[Time-varying covariates]\label{assumption:time-varying-covariates}
    For each unit $n \in [N]$, we have covariates $X_n = (X_{n,1}^{\top}, X_{n,2}^{\top}, \dots, X_{n,T}^{\top})^{\top} \in \Rb^{pT}$ with $X_{n,t} \in \Rb^p$ for any $t \in [T]$, where $X_{n,t,k} = \langle v_{n,t}, \rho^t_k \rangle + \varepsilon^{x}_{n,t,k}$ for any $k \in [p]$, with $v_{n,t}$ the unit latent factor defined in Assumption~\ref{assumption:general_latent_factor_model} and $\varepsilon^{x}_{n,t,k}$ mean-zero noise. That is, we collect $p$ features at each time step $t \in [T]$. Denote $X = [X_1, \dots, X_N] \in \Rb^{pT \times N}$.
\end{assumption}

With this added flexibility, observed pre-treatment outcomes can serve as covariates as well: in applications with a pre-treatment period of length $T_0$ during which all units are in control, the pre-treatment outcomes supply $T_0$ additional covariates, one per pre-treatment period.

\subsubsection{Row-Space Inclusion}

Here is a sufficient condition under which row-space inclusion holds.

\begin{lemma}\label{lemma:row-space-LTV-general}
    Let the setup of Assumption~\ref{assumption:row-space} hold together with Assumptions~\ref{assumption:LTV_factor_model} and~\ref{assumption:LTV_seq_exog} and condition (ii) of \eqref{eq:LTV_donor}, denote $V_{\Ic^d_t} = ([v_{j,T}]_{j \in \Ic^d_t})^{\top} \in \Rb^{|\Ic^d_t| \times mT}$, and for $s \in \{t-1, t\}$ let $b^{(s)} = ([\langle v_{j,T}, w_{(\bar{D}_j^{\,s},\, \underline{0}^{\,s+1})}\rangle]_{j \in \Ic^d_t})^{\top} \in \Rb^{|\Ic^d_t|}$. If $\text{span}(\{\rho_i\}_{i \in [p]}) \cap \{y \in \Rb^{mT}: V_{\Ic^d_t}y = b^{(s)}\}$ is non-empty for $s \in \{t-1, t\}$, then part (ii) of Assumption~\ref{assumption:balanced-spectra} holds for the pair $(d,t)$, with $\rho_i$ as defined in Assumption~\ref{assumption:additional-covariates}.
\end{lemma}

\begin{proof}
    Fix $s \in \{t-1, t\}$. We require weights $\xi^{(d,t,s)} \in \Rb^p$ with $\Ex[Y_{j,T}^{(\bar{D}_j^{\,s},\, \underline{0}^{\,s+1})}|\LFc, j \in \Ic^d_t] = \sum_{i = 1}^p \xi_i^{(d,t,s)} \cdot \Ex[(X_{\Ic^d_t})_{ij}|\LFc, j \in \Ic^d_t]$ for any $j \in \Ic^d_t$. By Assumption~\ref{assumption:LTV_factor_model}, $Y_{j,T}^{(\bar{D}_j^{\,s},\, \underline{0}^{\,s+1})} = \langle v_{j,T}, w_{(\bar{D}_j^{\,s},\, \underline{0}^{\,s+1})}\rangle + \varepsilon^{(\bar{D}_j^{\,s},\, \underline{0}^{\,s+1})}_{j,T}$, and $\Ex[\varepsilon^{(\bar{D}_j^{\,s},\, \underline{0}^{\,s+1})}_{j,T} \mid \LFc, j \in \Ic^d_t] = 0$ by condition (ii) of \eqref{eq:LTV_donor} together with Assumption~\ref{assumption:LTV_factor_model} and, for $s = t$, Assumption~\ref{assumption:LTV_seq_exog}, as in the proof of Theorem~\ref{thm:newLTV}; hence $\Ex[Y_{j,T}^{(\bar{D}_j^{\,s},\, \underline{0}^{\,s+1})}|\LFc, j \in \Ic^d_t] = \langle v_{j,T}, w_{(\bar{D}_j^{\,s},\, \underline{0}^{\,s+1})}\rangle$. Moreover, $\Ex[(X_{\Ic^d_t})_{ij}|\LFc, j \in \Ic^d_t] = \langle v_{j,T}, \rho_i \rangle$ under Assumption~\ref{assumption:additional-covariates}, whose covariate noise is independent. Thus the requirement is exactly the linear system $V_{\Ic^d_t}[\rho_1, \dots, \rho_p]\xi^{(d,t,s)} = b^{(s)}$, which is solvable under the stated span condition.
\end{proof}

In short, row-space inclusion holds if the covariates defined by $\{\rho_i\}_{i \in [p]}$ are sufficiently expressive, as when $p \ge mT$ and $\{\rho_i\}_{i \in [p]}$ span $\Rb^{mT}$.

\subsubsection{Fixed Lag Dependence Assumption}

Assumption~\ref{assumption:outcome-fixed-lag-dep} is not restrictive. In the dynamic triangular system setting of Proposition~\ref{lemma:LTV_representation}, consider the following conditions:
(i) hard memory cutoff: $\exists q \in \Nb$ such that $\prod_{k=s-q}^{s} \bB_{n,k} = 0$ for all $s \in [q+1, T]$;
(ii) exponential forgetting (spectral decay): $\exists C > 0, \lambda \in (0,1)$ such that $\| \prod_{k=t}^{s} \bB_{n,k} \|_2 \leq C \lambda^{s-t}$ for all $t \le s \le T$;
(iii) input-channel memory cutoff: $\exists q \in \Nb$ such that $\big(\prod_{k=\ell+1}^{t}\bB_{n,k}\big)\bC_{n,\ell} = 0$ for all $t \in [q+1, T]$ and $\ell \in [t-q-1]$. Since $\psi^{t,\ell}_n = \big(\big(\prod_{k=\ell+1}^{t}\bB_{n,k}\big)\bC_{n,\ell}\big)^{\top}\theta_{n,t}$ for $\ell \in [t-1]$ (SA~\ref{app:LTV_example_details}), condition (iii) is sufficient for Assumption~\ref{assumption:outcome-fixed-lag-dep} (and necessary for it to hold irrespective of $\theta_{n,t}$), and (i) implies (iii); under (ii) the assumption holds only approximately, up to a truncation error of order $\lambda^{q}$ in the omitted lags. Condition (i) is the \textit{strongest}; in general, this shows our assumption of fixed memory is a reasonable one.

\subsubsection{Sufficient Conditions for Assumption~\ref{assumption:LTV_seq_exog}}\label{app:seq-exog-sufficient}

We give two sufficient conditions under which Assumption~\ref{assumption:LTV_seq_exog} holds.

{\em 1. Non-action-dependent noise.} Assumption~\ref{assumption:LTV_seq_exog} holds if $\varepsilon_{n, T}^{(\bar{d}^t, \underline{0}^{t + 1})} = \varepsilon_{n, T}^{(\bar{d}^{t - 1}, \underline{0}^{t})}$, which occurs if neither is a function of its respective action sequence. Example \ref{ex:dyn_triangular_model} of a classic linear time-varying dynamic triangular system satisfies this property.

{\em 2. Additive action-dependent noise.} Relaxing the above, suppose for all $\bar{d}^T \in [A]^T$ that $\varepsilon_{n, T}^{(\bar{d}^T)} = \sum^{T}_{t = 1} \eta^{(d_t)}_{n, t}$, where conditional on $\LFc$ the $\eta^{(d_t)}_{n, t}$ are mutually independent for all $t \in [T]$ and $d_t \in [A]$. Then $\varepsilon_{n, T}^{(\bar{d}^t, \underline{0}^{t + 1})} - \varepsilon_{n, T}^{(\bar{d}^{t - 1}, \underline{0}^{t})} = \eta^{(d_t)}_{n, t} - \eta^{(0_t)}_{n, t}$, and a sufficient condition for Assumption~\ref{assumption:LTV_seq_exog} is that the time-$t$ innovations $(\eta^{(a)}_{n, t})_{a \in [A]}$ are independent of the action history $\bar{D}^t_n$ given $\LFc$ (no anticipation: neither the current nor any earlier action depends on the innovation generated at time step $t$), since then $\Ex[\eta^{(d_t)}_{n, t} - \eta^{(0_t)}_{n, t} \mid \bar{D}^t_n = \bar{d}^t, \LFc] = \Ex[\eta^{(d_t)}_{n, t} - \eta^{(0_t)}_{n, t} \mid \LFc] = 0$, the last equality because $\Ex[\varepsilon^{(\bar{d}^T)}_{n, T} \mid \LFc] = 0$ for all paths (Assumption~\ref{assumption:LTV_factor_model}). Note, however, that $\varepsilon_{n, t}^{(\bar{d}^t)} \ \notindep \ D_{n, t} \mid \LFc$: the noise terms remain auto-correlated, $\varepsilon_{n, t-1}^{(\bar{d}^{t -1})} \ \notindep \varepsilon_{n, t}^{(\bar{d}^{t})} \mid \LFc$, and $\varepsilon_{n, t-1}^{(\bar{d}^{t -1})} \ \notindep \ D_{n, t} \mid \LFc$ since the action $D_{n, t}$ can be a function of the observed outcome $Y_{n, t-1}$.

\subsection{Proof of Theorem~\ref{thm:consistency-time-varying}}\label{subsection:proof-of-varying-consistency}

We first state the fixed-$T$ consistency result proved in this subsection. All rates are expressed through the function
\begin{equation}\label{eq:rate-function}
\Rc(a, b; \Cc) := \sqrt{\log(p\pi_{\Ic})}\left(\frac{k^{a}}{p^{1/4}} + k^{b}\max\left\{\frac{\sqrt{\pi_{\Ic}}}{p^{3/2}}, \frac{1}{\sqrt{\alpha_{\Ic}-1}}, \frac{1}{\sqrt{p}}\right\}\right),
\end{equation}
where $a, b > 0$ are exponents, $\Cc$ is a collection of donor-set cardinalities, $\pi_{\Ic} := \max\Cc$, and $\alpha_{\Ic} := \min\Cc$ (so that $\alpha_{\Ic} \leq \pi_{\Ic}$). Note that $\Rc$ is nondecreasing in $a$ and $b$ and, since it is nondecreasing in $\pi_{\Ic}$ and nonincreasing in $\alpha_{\Ic}$, it cannot decrease when $\Cc$ is enlarged. Each $O_p(\cdot)$ below is defined with respect to the sequence $\min\{p, \alpha_{\Ic}\}$.

\begin{theorem}\label{thm:consistency-time-varying} Let Assumptions~\ref{assumption:SUTVA}, \ref{assumption:general_latent_factor_model}, \ref{assumption:LTV_factor_model}, \ref{assumption:LTV_well_supported_factors}, \ref{assumption:LTV_seq_exog}, \ref{assumption:additional-covariates}, \ref{control-factors-support}, \ref{assumption:sub-gaussian-noise}, and \ref{assumption:balanced-spectra} hold. Consider the \texttt{SBE-PCR} estimator in Section~\ref{subsection:SBE-Varying-Estimator} and suppose $k = \max_{d \in [0,A], t\in [T]}\text{rank}(\EE[X_{\Ic_t^d}|\LFc])$. Then, conditional on $\LFc$, $\{\rho_i\}_{i \in [p]}$, and the donor-defining treatment histories $\{\bar{D}^{t}_{j} : j \in \Ic^d_t\}$ of all donor sets (which fix every donor set and its cardinality, but not the donors' subsequent, possibly adaptive, actions), we have, for any $n \in [N]$, $d \in [A]$, $t \in [T-1]$, and $\bar{d}^T \in [A]^T$, the following bounds for (i) the baseline, (ii) the terminal blip, (iii) the non-terminal blips, and (iv) the target parameter:
\begin{align*}
\text{(i)}\quad & \hat{b}_{n,T} - b_{n,T} \mid \LFc = O_p\big(\Rc(5/4, 5/2; \{|\Ic_T^0|\})\big);\\
\text{(ii)}\quad & \hat{\gamma}_{n,T,T}(d) - \gamma_{n,T,T}(d) \mid \LFc = O_p\big(\Rc(7/4, 3; \{|\Ic_T^0|, |\Ic_T^d|\})\big);\\
\text{(iii)}\quad & \hat{\gamma}_{n,T,t}(d) - \gamma_{n,T,t}(d) \mid \LFc = O_p\big((T-t)\,\Rc(T-t+7/4,\, T-t+3;\, \Cc^d_t)\big);\\
\text{(iv)}\quad & \widehat{\Ex}[Y_{n, T}^{(\bar{d}^T)}] - \EE[Y_{n, T}^{(\bar{d}^T)}\mid \LFc] = O_p\big(T\,\Rc(T+3/4,\, T+2;\, \Cc)\big),
\end{align*}
where $\Cc^d_t := \{|\Ic_T^0|, |\Ic_t^d|, (|\Ic_{q}^{D_{n,q}}|)_{n \in [N], q \in [t+1, T]}\}$ and $\Cc := \{|\Ic_T^0|, (|\Ic_{t}^{d_{t}}|)_{t \in [T]}, (|\Ic_{t}^{D_{n,t}}|)_{n \in [N], t \in [2, T]}\}$.
\end{theorem}

The remainder of this subsection proves the theorem. Two arguments recur throughout: a row-space projection identity, and a transfer argument that converts an entrywise consistency rate for a donor-set target vector into a rate for its estimated PCR combination. We state each once, in the generality in which it is invoked, and then verify parts (i)--(iv).

\noindent{\bf 0. Recurring arguments:} Throughout, $J$ denotes a donor set of the form $\Ic_t^d$ or $\Ic_t^d \setminus n$ for some $t \in [T]$ and $d \in [0,A]$. We write $X_{J} = X_{:, J} \in \Rb^{p \times |J|}$, let $k_{J} = \text{rank}(\EE[X_{J}])$, and $V \in \Rb^{|J| \times k_{J}}$ collect the right singular vectors of $\EE[X_{J}] = U \Sigma V^{\top}$. Note that $k_J \leq k$ and $|\Ic_t^d \setminus n| = |\Ic_t^d| - 1$. For a unit $n \notin J$, $\hat{\beta}^{n,J}$ denotes the regression coefficients from regressing the additional covariates $X_n \in \Rb^p$ on the rank-$k_{J}$ approximation of $X_{J}$, i.e., PCR with parameter $k_{J}$; $\beta^{n,J}$ denotes its population counterpart, $\tilde{\beta}^{n,J} = VV^{\top}\beta^{n,J}$, and $\Delta_{n, J} = \hat{\beta}^{n,J} - \tilde{\beta}^{n,J}$. When $J = \Ic_t^d \setminus n$ these coefficients are denoted $\hat{\phi}^{n, \Ic_t^d}, \phi^{n, \Ic_t^d}, \tilde{\phi}^{n, \Ic_t^d}$, as in the main text. All statements are conditional on $\LFc$, $\{\rho_i\}_{i \in [p]}$, and the donor-defining histories $\{\bar{D}^{t}_{j} : j \in \Ic^d_t\}$ of all donor sets; in some displays we omit part of this conditioning for brevity, and a statement of the form ``$Z \mid \LFc = O_p(r)$'' means that $Z = O_p(r)$ under this conditional law. Since this conditioning fixes every donor-set cardinality, an entry $|\Ic^{D_{i, \ell}}_{\ell}|$ of a collection $\Cc$ below (indexed by a realized action) is read as the maximum, in $\pi_{\Ic}$, and the minimum, in $\alpha_{\Ic}$, over the actions $a$ with $\Ic^a_\ell \neq \emptyset$, which bounds every realization. For a donor $j \in \Ic^d_t$ we write $\varepsilon^{\mathrm{obs}}_{j} := \varepsilon^{(\bar{D}^T_j)}_{j, T} = Y_{j, T} - b_{j, T} - \sum_{\ell = t}^{T} \gamma_{j, T, \ell}(D_{j, \ell})$ for the noise along its realized path (Assumption~\ref{assumption:LTV_factor_model}, using $D_{j, \ell} = 0_\ell$ for $\ell < t$). As in the proof of Theorem~\ref{thm:newLTV} (the tower rule over $\bar{D}^\ell_j$ for $\ell \ge t$, Assumption~\ref{assumption:LTV_seq_exog}, and condition (ii) of \eqref{eq:LTV_donor}), $\Ex[\varepsilon^{\mathrm{obs}}_j \mid \LFc, j \in \Ic^d_t] = 0$. We apply Assumption~\ref{assumption:sub-gaussian-noise} to these realized-path noise terms as follows. \emph{Independence.} Assumption~\ref{assumption:sub-gaussian-noise} states cross-unit independence for each fixed path, whereas realized paths differ across donors. Throughout the consistency proofs we therefore use the following cross-unit condition, which strengthens the independence clause of Assumption~\ref{assumption:sub-gaussian-noise} to unit-specific paths and to the action processes: conditional on $\LFc$, the unit-level processes $\{(\varepsilon^{(\bar{d}^s)}_{j,s})_{s \in [T],\, \bar{d}^s \in [A]^s},\ (\varepsilon_{j,k})_{k \in [p]},\ \bar{D}^{T}_j\}$ are independent across units $j \in [N]$. It holds in Example~\ref{ex:dyn_triangular_model} when the innovations are independent across units, since each unit's actions depend only on its own history, and it implies the cross-unit independence asserted in Assumption~\ref{assumption:sub-gaussian-noise} for each fixed path, since conditioning a product law on a product event (here, all units' realized histories through $s$) preserves cross-unit independence. Likewise, given $\LFc$ each donor-defining event $\{j \in \Ic^d_t\}$ is determined by $\bar{D}^{t}_j$, and $\varepsilon^{\mathrm{obs}}_j = \sum_{\bar{d}^T \in [A]^T} \mathbf{1}\{\bar{D}^T_j = \bar{d}^T\}\,\varepsilon^{(\bar{d}^T)}_{j,T}$ is a function of unit $j$'s own processes, so the realized-path noise terms are independent across units conditional on $\LFc$ and the donor-defining histories, and $\Ex[\varepsilon^{\mathrm{obs}}_j \mid \LFc, j \in \Ic^d_t] = 0$ is unaffected by conditioning on the other units' histories. \emph{Tails.} By Assumption~\ref{assumption:sub-gaussian-noise}(i) and the definition of the $\psi_2$ norm, $\Ex[\exp\{(\varepsilon^{(\bar{d}^T)}_{j,T})^2/(C\sigma)^2\} \mid \LFc, \{\bar{D}^T_n\}_{n \in [N]}] \le 2$ for every fixed path $\bar{d}^T$. Since $\bar{D}^T_j$ is measurable under this conditioning, the same bound holds for $\varepsilon^{\mathrm{obs}}_j$, and since the donor-defining histories are functions of $\{\bar{D}^T_n\}_{n \in [N]}$, the tower property yields the same bound conditional on $\LFc$ and these histories. Hence the realized-path noise is sub-Gaussian with the same constant $C\sigma$ as the fixed-path noise under the conditioning of the theorem.

\begin{lemma}[Row-space projection]\label{lemma:ltv-row-space-projection}
    Fix a donor set $J$ as above and let $\theta \in \Rb^{|J|}$ be any one of:
    (i) $\theta = \EE[Y_{J} \mid \LFc]$;
    (ii) $\theta = b_{J,T} = (b_{j,T})_{j \in J}$;
    (iii) $\theta = [(\gamma_{j, T, t}(d))_{j \in J}]^{\top}$, the blips of the donors' own action $d$ at their switching time $t$.
    Then $VV^{\top}\theta = \theta$, and consequently $\langle w, \theta \rangle = \langle VV^{\top} w, \theta \rangle$ for every $w \in \Rb^{|J|}$.
\end{lemma}
\begin{proof}
    In each case it suffices to show that $\theta^{\top}$ lies in the row space of $\EE[X_{J}]$, i.e., that there exists $\xi \in \Rb^p$ with $\theta_j = \sum_{i = 1}^p \xi_i \cdot \Ex[(X_{J})_{ij}\mid \LFc, j \in J]$ for all $j \in J$: this is equivalent to $\theta \in \text{span}(\{v_1, \dots, v_{k_{J}}\})$, whence $VV^{\top}\theta = \theta$ and $\langle VV^{\top}w, \theta \rangle = \theta^{\top}VV^{\top}w = \langle w, \theta\rangle$.

    All three cases follow from part (ii) of Assumption~\ref{assumption:balanced-spectra} (applied to the donor set $\Ic^d_t$, or to $\Ic^d_t \setminus n$ via its footnote), which supplies, for $s \in \{t-1, t\}$, a common vector $\xi^{(d,t,s)}$ with $\Ex[Y_{j,T}^{(\bar{D}_j^{\,s},\, \underline{0}^{\,s+1})} \mid \LFc, j \in J] = \sum_{i=1}^p \xi^{(d,t,s)}_i \cdot \Ex[(X_{J})_{ij} \mid \LFc, j \in J]$ for all $j \in J$.
    \textit{Case (i)} is the member $s = t = T$, since case (i) is invoked only for $J \subseteq \Ic^0_T$ or $J \subseteq \Ic^d_T$ and $Y_{j,T} = Y_{j,T}^{(\bar{D}_j^{\,T})}$.
    \textit{Case (ii)} is the member $s = t-1$, since $b_{j,T} = \EE[Y_{j,T}^{(\bar{0}^T)} \mid \LFc]$ and $(\bar{D}_j^{\,t-1}, \underline{0}^{\,t}) = \bar{0}^T$ for $j \in \Ic^d_t$.
    \textit{Case (iii):} under Assumption~\ref{assumption:LTV_factor_model}, for $j \in J \subseteq \Ic^d_t$ we have $\bar{D}^{t}_j = (\bar{0}^{t-1}, d)$ and $\gamma_{j,T,t}(d) = \langle \psi_{j}^{T,t}, w_{d} - w_{0_{t}}\rangle = \EE[Y_{j,T}^{(\bar{D}_j^{\,t},\, \underline{0}^{\,t+1})} \mid \LFc] - \EE[Y_{j,T}^{(\bar{D}_j^{\,t-1},\, \underline{0}^{\,t})} \mid \LFc]$, the difference of the members $s = t$ and $s = t - 1$; hence $\xi = \xi^{(d,t,t)} - \xi^{(d,t,t-1)}$ works, which concludes the proof.
\end{proof}

Note that Assumption~\ref{assumption:bounded-expected-potential-outcomes}(ii) guarantees $\|\theta\|_{\infty} \leq 1$ in cases (i)--(ii) (e.g., $|b_{j,T}| = |\EE[Y_{j,T}^{(\bar{0}^T)}]| \leq 1$) and $\|\theta\|_{\infty} \leq 2$ in case (iii), since each entry is an expected potential outcome or a difference of two.

We next state, without proof, three results imported from \cite{SI}.
\begin{lemma}[Appendix B.$4$, Lemma $8$ of \cite{SI}]\label{lemma:1a-interventions-results}
    Given any $n \in [N]$, $d\in [A]$, and $t \in [T]$ let all relevant assumptions hold. Then, conditioned on the latent factors, we have that
\[
\|\tilde{\beta}^{n,\Ic_t^d}\|_2 \leq  C_1 \cdot \sqrt{\frac{k_{\Ic_t^d}}{|\Ic_t^d|}}
\]
for some constant $C_1 > 0$. This immediately implies that $\|\tilde{\beta}^{n,\Ic_t^d}\|_1 \leq C_1 \sqrt{k_{\Ic_t^d}}$.
\end{lemma}
\begin{lemma}[Appendix B.$4$, Lemma $7$ of \cite{SI}]\label{lemma:1b-interventions-result}
    Let the setup of Lemma~\ref{lemma:1a-interventions-results} hold, then w.p. at least $1 - O(1/(p |\Ic_t^d|)^{10})$,
\[
\|\tilde{\beta}^{n,\Ic_t^d} - \hat{\beta}^{n,\Ic_t^d}\|_2^2 \leq C(\sigma) \cdot \log(p |\Ic_t^d|) \left( \frac{k^{3/2}}{p^{1/2} |\Ic_t^d|} + \frac{k^3}{\min\{p, |\Ic_t^d|\}^2} \right),
\]
where $C(\sigma)$ is a constant that depends only on $\sigma$ and $C$ of Assumption~\ref{assumption:sub-gaussian-noise}.
\end{lemma}
\begin{lemma}[Appendix C, Lemma $9$ of \cite{SI}]\label{lemma:1c-interventions-result}
Let the setup from Lemma~\ref{lemma:1a-interventions-results} hold, then
    $$\|VV^{\top}\Delta_{n, \Ic_t^d}\|_2 = O_p \left( \frac{\sqrt{k}}{\sqrt{|\Ic_t^d|} p^{1/4}} + \frac{k^{3/2} \sqrt{\log(p |\Ic_t^d|)}}{\sqrt{|\Ic_t^d|} \cdot \min\{\sqrt{p}, \sqrt{|\Ic_t^d|}\}} + \frac{k^2 \sqrt{\log(p |\Ic_t^d|)}}{\min\{p^{3/2}, |\Ic_t^d|^{3/2}\}} \right).$$
\end{lemma}
When these three lemmas are invoked for $J = \Ic_t^d \setminus n$ (i.e., for the coefficients $\hat{\phi}^{n,\Ic_t^d}$), the appropriate versions with $\Ic_t^d$ replaced by $\Ic_t^d \setminus n$ apply verbatim.

\begin{lemma}[Rate transfer]\label{lemma:pcr-rate-transfer}
    Fix a donor set $J$ and a unit $n \notin J$ as above. Let $\theta \in \Rb^{|J|}$ satisfy $VV^{\top}\theta = \theta$ and $\|\theta\|_{\infty} \leq C_0$ for an absolute constant $C_0$, and let $\hat{\theta}$ be an estimator of $\theta$ satisfying $\|\hat{\theta} - \theta\|_{\infty} = O_p(\Rc(a, b; \Cc))$ for some $a \geq 3/4$, $b \geq 2$, and a collection $\Cc$ of donor-set cardinalities containing $|\Ic_t^d|$. Then
    $$\left\langle \hat{\beta}^{n,J}, \hat{\theta} \right\rangle - \left\langle \beta^{n,J}, \theta \right\rangle = O_p\big(\Rc(a + 1/2, b + 1/2; \Cc)\big).$$
\end{lemma}
\begin{proof}
    Since $VV^{\top}\theta = \theta$ we have $\langle \beta^{n,J}, \theta \rangle = \langle \tilde{\beta}^{n,J}, \theta \rangle$, hence with $\eta = \hat{\theta} - \theta$ we obtain the decomposition\footnote{Notice that the analysis from \cite{SI} that resolves the donor unit case does not apply to Terms a and b since $\eta$ is not composed of independent $\sigma^2$-subgaussian random variables.}
    $$\left\langle \hat{\beta}^{n,J}, \hat{\theta} \right\rangle - \left\langle \beta^{n,J}, \theta \right\rangle =  \underbrace{\langle \tilde{\beta}^{n,J}, \eta\rangle}_{\text{Term a}} + \underbrace{\langle \Delta_{n, J} , \eta\rangle}_{\text{Term b}} +  \underbrace{\langle \Delta_{n, J}, \theta\rangle}_{\text{Term c}}.$$

    \textit{Term a:} By the H\"older and Cauchy-Schwarz Inequalities, $\langle \tilde{\beta}^{n,J}, \eta\rangle \leq \|\tilde{\beta}^{n,J}\|_1 \cdot \|\eta\|_{\infty} \leq \sqrt{|J|}\cdot\|\tilde{\beta}^{n,J}\|_2 \cdot \|\eta\|_{\infty}$, and Lemma~\ref{lemma:1a-interventions-results} gives $\|\tilde{\beta}^{n,J}\|_2 \leq C_1\sqrt{k/|J|}$, so $\text{Term a} = O_p(\sqrt{k}\,\|\eta\|_{\infty}) = O_p(\Rc(a + 1/2, b + 1/2; \Cc))$, which is exactly the claimed rate.

    \textit{Term b:} Similarly $\langle \Delta_{n,J}, \eta \rangle \leq \sqrt{|J|}\cdot\|\Delta_{n,J}\|_2\cdot\|\eta\|_{\infty}$, where Lemma~\ref{lemma:1b-interventions-result} (with $\sqrt{x + y} \leq \sqrt{x} + \sqrt{y}$, and using that a bound holding w.h.p. yields an $O_p$ bound) gives $\sqrt{|J|}\cdot\|\Delta_{n,J}\|_2 = O_p\big(\sqrt{\log(p|J|)}\big(\tfrac{k^{3/4}}{p^{1/4}} + \tfrac{k^{3/2}\sqrt{|J|}}{\min\{p, |J|\}}\big)\big)$, so Term b is $O_p$ of the product of this rate with $\|\eta\|_{\infty}$.

    \textit{Term c:} Since $VV^{\top}\theta = \theta$ and $VV^{\top}$ is symmetric, $\langle \Delta_{n,J}, \theta \rangle = \langle \theta, VV^{\top}\Delta_{n,J}\rangle \leq \|\theta\|_2 \cdot \|VV^{\top}\Delta_{n,J}\|_2$ by Cauchy-Schwarz, with $\|\theta\|_2 \leq C_0\sqrt{|J|}$. Lemma~\ref{lemma:1c-interventions-result} then yields
    $$\text{Term c} = O_p \left( \frac{\sqrt{k}}{ p^{1/4}} + \frac{k^{3/2} \sqrt{\log(p |J|)}}{\min\{\sqrt{p}, \sqrt{|J|}\}} + \frac{k^2 \sqrt{|J|\log(p |J|)}}{\min\{p^{3/2}, |J|^{3/2}\}} \right).$$

    Since $a \geq 3/4$ and $b \geq 2$, Terms b and c are dominated by the rate of Term a,\footnote{In order to get this final rate we made some assumptions on how the donor-set cardinalities and $p$ grow relative to each other.} which proves the claim.
\end{proof}

\noindent{\bf 1. Verifying Baseline Consistency:}

\textit{Donor Set Baseline Consistency:} Consider unit $n \in \Ic_T^0$. We know the baseline outcome admits the representation $\hat{b}_{n,T} - b_{n,T} \mid \LFc  = \left\langle \hat{\phi}^{n, \Ic_T^0}, Y_{\Ic_T^0 \setminus n} \right\rangle - \left\langle \phi^{n, \Ic_T^0}, \mathbb{E}[Y_{\Ic_T^0 \setminus n} \mid \LFc] \right\rangle$, and by Lemma~\ref{lemma:ltv-row-space-projection}(i) with $J = \Ic_T^0 \setminus n$ we may replace $\phi^{n, \Ic_T^0}$ by $\tilde{\phi}^{n, \Ic_T^0} = VV^{\top}\phi^{n, \Ic_T^0}$. We can then lift the proof technique in \citet[Appendix C, Theorem 2]{SI} to show consistency for $n \in \Ic_T^0$:
\begin{align}\label{eq:donor-baseline-consistency}
    \hat{b}_{n,T} - b_{n,T} \mid \LFc = O_p\big(\Rc(3/4, 2; \{|\Ic_T^0|\})\big),
\end{align}
where we set $T_1 = 1$, $\tilde{w}^{(i,d)} = \tilde{\phi}^{n, \Ic_T^0}$, $\hat{w}^{(i,d)} = \hat{\phi}^{n, \Ic_T^0}$, $Y_{t,\mathcal{I}^{(d)}} = Y_{\Ic_T^0 \setminus n}$, $\EE[Y_{t,\mathcal{I}^{(d)}}] = \mathbb{E}[Y_{\Ic_T^0 \setminus n} \mid \LFc] $, and $\mathcal{P}_{V_{\text{pre}}} = VV^{\top}$. Furthermore, in the final rate we set $T_0 = p$, $N_d = |\Ic_T^0 \setminus n|$, and $r_{\text{pre}} = k_{\Ic_T^0 \setminus n}$. To conclude, we used that $|\Ic_T^0 \setminus n| = |\Ic_T^0| - 1$ and $k_{\Ic_T^0 \setminus n} \leq k$.

\textit{Non-Donor Set Baseline Consistency:} Consider unit $n \notin \Ic_T^0$, for which $\hat{b}_{n,T} - b_{n,T} \mid \LFc  = \left\langle \hat{\beta}^{n, \Ic_T^0}, \hat{b}_{\Ic_T^0 } \right\rangle - \left\langle \beta^{n, \Ic_T^0}, b_{\Ic_T^0 } \right\rangle$. Lemma~\ref{lemma:ltv-row-space-projection}(ii) with $J = \Ic_T^0$ gives $VV^{\top}b_{\Ic_T^0} = b_{\Ic_T^0}$, and Equation~\eqref{eq:donor-baseline-consistency}, which holds for every unit in $\Ic_T^0$, provides the entrywise rate for $\|\hat{b}_{\Ic_T^0} - b_{\Ic_T^0}\|_{\infty}$ at the same order.\footnote{We use the probability bounds underlying the PCR arguments in
\citet[Appendices B.4 and C]{SI}, with failure probability
$O((p\pi_{\Ic})^{-10})$ for each PCR regression and constants
independent of the unit.
For fixed $A,T$ (or fixed $A,q$ in the fixed-lag case),
estimating the target parameter for one unit requires
$O(\pi_{\Ic})$ such regressions.
A union bound therefore ensures that these bounds hold
simultaneously with probability $1-o(1)$, with the required
logarithmic factor already included in
$\sqrt{\log(p\pi_{\Ic})}$.
The inequalities in the proof of
Lemma~\ref{lemma:pcr-rate-transfer} and the induction are applied
on this same event, justifying the bounds on the maximum
baseline or blip estimation error over donors used in this proof.} The assumptions of Lemma~\ref{lemma:pcr-rate-transfer} therefore hold with $\theta = b_{\Ic_T^0}$, $a = 3/4$, $b = 2$, and $\Cc = \{|\Ic_T^0|\}$, yielding for $n \notin \Ic_T^0$ the rate $\Rc(5/4, 5/2; \{|\Ic_T^0|\})$. Since this dominates the donor rate \eqref{eq:donor-baseline-consistency}, we conclude for any $n \in [N]$
\begin{align}\label{eq:baseline-consistency-rate}
    \hat{b}_{n,T} - b_{n,T} \mid \LFc = O_p\big(\Rc(5/4, 5/2; \{|\Ic_T^0|\})\big).
\end{align}

\noindent{\bf 2. Verifying Terminal Blip Consistency:}

For any $d \in [A]$:

\textit{Donor Set Consistency:} Consider unit $n \in \Ic_T^d$. We know the estimator admits the representation
$$\hat{\gamma}_{n,T,T}(d) - \gamma_{n,T,T}(d) \mid \LFc  = \underbrace{\left\langle \hat{\phi}^{n, \Ic_T^d}, Y_{\Ic_T^d \setminus n} \right\rangle - \left\langle \phi^{n, \Ic_T^d}, \mathbb{E}[Y_{\Ic_T^d \setminus n} \mid \LFc] \right\rangle}_{\text{Term 1}} + \underbrace{b_{n,T}\mid \LFc - \hat{b}_{n,T}}_{\text{Term 2}}.$$
By Lemma~\ref{lemma:ltv-row-space-projection}(i) with $J = \Ic_T^d \setminus n$ and the argument that yielded Equation~\eqref{eq:donor-baseline-consistency}, Term $1$ is $O_p(\Rc(3/4, 2; \{|\Ic_T^d|\}))$. Term $2$ is (minus) the baseline error of unit $n$, bounded in Equation~\eqref{eq:baseline-consistency-rate}. Combining both rates, we find for any $n \in \Ic^d_T$
\begin{equation}\label{eq:donor-terminal-rate}
\hat{\gamma}_{n,T,T}(d) - \gamma_{n,T,T}(d) \mid \LFc = O_p\big(\Rc(5/4, 5/2; \{|\Ic_T^0|, |\Ic_T^d|\})\big).
\end{equation}

\textit{Non-Donor Set Consistency:} Consider unit $n \notin \Ic_T^d$, for which $\hat{\gamma}_{n,T,T}(d) - \gamma_{n,T, T}(d) \mid \LFc  = \left\langle \hat{\beta}^{n, \Ic_T^d}, \hat{\gamma}_{\Ic_T^d, T, T}(d) \right\rangle - \left\langle \beta^{n, \Ic_T^d}, \gamma_{\Ic_T^d, T, T}(d) \right\rangle$. Lemma~\ref{lemma:ltv-row-space-projection}(iii) with $J = \Ic_T^d$ gives $VV^{\top}\gamma_{\Ic_T^d, T, T}(d) = \gamma_{\Ic_T^d, T, T}(d)$, and Equation~\eqref{eq:donor-terminal-rate} provides the entrywise rate for $\hat{\gamma}_{\Ic_T^d, T, T}(d)$. Lemma~\ref{lemma:pcr-rate-transfer} (with $a = 5/4$, $b = 5/2$, $\Cc = \{|\Ic_T^0|, |\Ic_T^d|\}$) then yields for $n \notin \Ic_T^d$ the rate $\Rc(7/4, 3; \Cc)$, which dominates the donor rate \eqref{eq:donor-terminal-rate}; hence, for any $n \in [N]$,
\begin{equation}\label{eq:terminal-rate}
\hat{\gamma}_{n,T,T}(d) - \gamma_{n,T,T}(d) \mid \LFc = O_p\big(\Rc(7/4, 3; \{|\Ic_T^0|, |\Ic_T^d|\})\big).
\end{equation}

\noindent{\bf 3. Verifying Non-Terminal Blip Consistency:}

For any unit $n \in [N]$, treatment $d \in [A]$, and $t \in [1, T-1]$, consider the statement $P_{d,n}(t)$: $\hat{\gamma}_{n, T, t}(d) - \gamma_{n, T, t}(d) \mid \LFc = O_p\big((T-t)\,\Rc(T-t+7/4, T-t+3; \Cc^d_t)\big)$, with $\Cc^d_t = \{|\Ic_T^0|, |\Ic_t^d|, (|\Ic_{q}^{D_{n,q}}|)_{n\in [N],q \in [t+1, T]}\}$ as in the theorem.

We proceed by strong induction, downward from $t = T-1$. The donor-set decomposition and the bounds on its first three terms are common to the base case and the inductive step, so we derive them once for general $t \in [T-1]$ and any $d \in [A]$.

\textit{Donor Set Decomposition:} Consider unit $n \in \Ic_t^d$ and write $\theta := [(\gamma_{j, T, t}(d))_{j \in \Ic^d_t \setminus n}]^{\top}$ and $\varepsilon^{\mathrm{obs}} := (\varepsilon^{\mathrm{obs}}_j)_{j \in \Ic^d_t \setminus n}$ for the realized-path noise introduced in item 0, so that the blipped-down donor outcomes used in Step~3 of \texttt{SBE-PCR} satisfy $Y_{\Ic^d_t \setminus n} - \hat{b}_{\Ic^d_t \setminus n} - \sum_{\ell = t+1}^{T} \hat{\gamma}_{\Ic^d_t \setminus n, T, \ell}(D_{\Ic^d_t \setminus n, \ell}) = \theta + \varepsilon^{\mathrm{obs}} - \eta_b - \sum_{\ell = t+1}^{T} \eta_\ell$, where $\eta_b := \hat{b}_{\Ic^d_t \setminus n} - b_{\Ic^d_t \setminus n}$, $\eta_\ell := \hat{\gamma}_{\Ic^d_t \setminus n, T, \ell}(D_{\Ic^d_t \setminus n, \ell}) - \gamma_{\Ic^d_t \setminus n, T, \ell}(D_{\Ic^d_t \setminus n, \ell})$, and $\gamma_{\Ic^d_t \setminus n,T,\ell}(D_{\Ic^d_t \setminus n,\ell}) = [(\gamma_{j, T, \ell}(D_{j, \ell}))_{j \in \Ic^d_t \setminus n}]^{\top}$ (and analogously for $\hat{\gamma}$). Since $\gamma_{n,T,t}(d) = \langle \phi^{n, \Ic_t^d}, \theta \rangle$ by Assumption~\ref{control-factors-support}, the estimator admits the representation
\begin{align*}
\hat{\gamma}_{n,T,t}(d) - \gamma_{n,T,t}(d) \mid \LFc  &= \underbrace{\left\langle \hat{\phi}^{n, \Ic_t^d}, \theta + \varepsilon^{\mathrm{obs}} \right\rangle - \left\langle \phi^{n, \Ic_t^d}, \theta \right\rangle}_{\text{Term 1}} \\
&\quad - \underbrace{\left\langle \hat{\phi}^{n, \Ic^d_t}, \eta_b \right\rangle}_{\text{Term 2}}
- \underbrace{\left\langle \hat{\phi}^{n, \Ic^d_t}, \eta_T \right\rangle}_{\text{Term 3}}
- \sum_{\ell = t+1}^{T-1} \underbrace{\left\langle \hat{\phi}^{n, \Ic^d_t}, \eta_\ell \right\rangle}_{\text{Term } \ell},
\end{align*}
where the summation is empty when $t = T-1$.

\textit{Bounding Term 1:} Lemma~\ref{lemma:ltv-row-space-projection}(iii) with $J = \Ic_t^d \setminus n$ gives $VV^{\top}\theta = \theta$, so $\langle \phi^{n, \Ic_t^d}, \theta \rangle = \langle \tilde{\phi}^{n, \Ic_t^d}, \theta \rangle$; since $\theta + \varepsilon^{\mathrm{obs}}$ has conditional mean $\theta$ in the row space and $\varepsilon^{\mathrm{obs}}$ consists of independent mean-zero sub-Gaussian noise (item 0), the technique of \citet[Appendix C, Theorem 2]{SI} applies with $Y_{\Ic^0_T \setminus n}$ replaced by $\theta + \varepsilon^{\mathrm{obs}}$, and Term $1$ is $O_p(\Rc(3/4, 2; \{|\Ic_t^d|\}))$, the bound of Equation~\eqref{eq:donor-baseline-consistency} with $|\Ic_T^0|$ replaced by $|\Ic_t^d|$.

\textit{Bounding Term 2:} Writing $\hat{\phi}^{n, \Ic_t^d} = \tilde{\phi}^{n, \Ic_t^d} + \Delta_{n, \Ic^d_t \setminus n}$, Term $2$ equals Terms a and b of Lemma~\ref{lemma:pcr-rate-transfer} with $\eta = \eta_b$, whose entrywise rate $\Rc(5/4, 5/2; \{|\Ic_T^0|\})$ is given by Equation~\eqref{eq:baseline-consistency-rate} (and hence, enlarging the collection, by $\Rc(5/4, 5/2; \{|\Ic_T^0|, |\Ic_t^d|\})$); no row-space condition is needed since Term c does not arise. Hence Term $2$ is $O_p(\Rc(7/4, 3; \{|\Ic_T^0|, |\Ic_t^d|\}))$.

\textit{Bounding Term 3:} Likewise, Term $3$ equals Terms a and b of Lemma~\ref{lemma:pcr-rate-transfer} with $\eta = \eta_T$, whose entries are bounded by $\max_{a \in \{D_{i, T}\}_{i \in [N]}} |\hat{\gamma}_{j,T,T}(a) - \gamma_{j,T,T}(a)|$, a maximum over at most $A$ terms each obeying Equation~\eqref{eq:terminal-rate}. Hence
\[
\text{Term }3 = O_p\big(\Rc(9/4, 7/2; \{|\Ic_T^0|,|\Ic_t^d|, (|\Ic_T^{D_{n,T}}|)_{n \in [N]}\})\big),
\]
which dominates the rates for Terms $1$ and $2$.

\textbf{Base case $t = T-1$, i.e., proving $P_{d,n}(T-1)$:}

\textit{Donor Set Consistency:} Here the summation is empty, so combining Terms $1$--$3$ yields
\begin{align}\label{eq:donor-base-blip-rate}
\hat{\gamma}_{n,T,T-1}(d) - \gamma_{n,T,T-1}(d) \mid \LFc = O_p\big(\Rc(9/4, 7/2; \Cc^d_{T-1})\big)
\end{align}
for any $n \in \Ic_{T-1}^d$, since $\Cc^d_{T-1} = \{|\Ic_T^0|,|\Ic_{T-1}^d|, (|\Ic_T^{D_{n,T}}|)_{n \in [N]}\}$.

\textit{Non-Donor Set Consistency:} Consider any $t \in [T-1]$ and unit $n \notin \Ic_t^d$ (we give the argument for general $t$ as it is reused in the inductive step). We know the estimator admits the representation $\hat{\gamma}_{n,T,t}(d) - \gamma_{n,T,t}(d) \mid \LFc  = \left\langle \hat{\beta}^{n, \Ic_t^d}, \hat{\gamma}_{\Ic_t^d, T,t}(d) \right\rangle - \left\langle \beta^{n, \Ic_t^d}, \gamma_{\Ic_t^d, T,t}(d) \right\rangle$. Lemma~\ref{lemma:ltv-row-space-projection}(iii) with $J = \Ic_t^d$ gives $VV^{\top}\gamma_{\Ic_t^d, T,t}(d) = \gamma_{\Ic_t^d, T,t}(d)$, so Lemma~\ref{lemma:pcr-rate-transfer} applies with $\theta = \gamma_{\Ic_t^d, T,t}(d)$ once an entrywise rate for $\hat{\gamma}_{\Ic_t^d, T,t}(d)$ is available, raising its exponents $(a, b)$ to $(a + 1/2, b + 1/2)$. For $t = T-1$, Equation~\eqref{eq:donor-base-blip-rate} supplies that rate ($a = 9/4$, $b = 7/2$), yielding for $n \notin \Ic_{T-1}^d$ the rate $\Rc(11/4, 4; \Cc^d_{T-1})$. Combining the two cases yields the base case.

\textbf{Inductive Step:} We assume $P_{d,n}(\ell)$ for $\ell \in [t+1,T-1]$ and prove $P_{d,n}(t)$.

\textit{Donor Set Consistency:} Consider unit $n \in \Ic_t^d$. In the decomposition above, Terms $1$--$3$ are already bounded, so only the summation remains. For any $\ell \in [t+1,T-1]$, Term $\ell = \langle \hat{\phi}^{n, \Ic_t^d}, \eta_\ell \rangle$ equals Terms a and b of Lemma~\ref{lemma:pcr-rate-transfer} with $\eta = \eta_\ell$, whose entrywise rate is supplied by the inductive hypothesis $P_{a, j}(\ell)$ for $a \in \{D_{i, \ell}\}_{i \in [N]}$ (a maximum over at most $A$ actions), so that $\text{Term }\ell = O_p\big((T-\ell)\,\Rc(T-\ell+9/4, T-\ell+7/2; \Cc_\ell)\big)$ with $\Cc_\ell := \{|\Ic_T^0|, |\Ic_{t}^d|, (|\Ic_{q}^{D_{n,q}}|)_{n\in [N],q \in [\ell, T]}\}$. Since $\Cc_\ell \subseteq \Cc^d_t$ for every such $\ell$, we may replace $\Cc_\ell$ by $\Cc^d_t$ (see the remark after \eqref{eq:rate-function}). Writing $C := \sqrt{\log(p\pi_{\mathcal{I}})}$ and $C'$ for the maximum in \eqref{eq:rate-function}, both evaluated at $\Cc^d_t$, the substitution $m = T - \ell$ gives
\begin{align*}
\sum_{\ell = t+1}^{T-1}\text{Term }\ell
&= O_p\left(C\left(\frac{k^{9/4}}{p^{1/4}}+C'k^{7/2}\right)\sum_{m=1}^{T-t-1}mk^m\right) \\
&= O_p\big((T-t)\,\Rc(T-t+5/4, T-t+5/2; \Cc^d_t)\big),
\end{align*}
where the last equality uses, with $M = T-t-1$, the bound $\sum_{m=1}^{M}mk^m \le M \sum_{m=1}^{M} k^m \le \frac{k}{k-1}Mk^{M} \le 2Mk^{M}$ for $k \geq 2$, taken without loss of generality.\footnote{$k$ enters every rate only as an upper bound on the ranks $k_J$ and all bounds are nondecreasing in $k$, so for $k = 1$ the statements hold with $k$ replaced by $2$ at the cost of an absolute constant.} Since Terms $1$--$3$ are dominated by the summation, for any $n \in \Ic^d_t$ the donor-set rate is $O_p\big((T-t)\,\Rc(T-t+5/4, T-t+5/2; \Cc^d_t)\big)$.

\textit{Non-Donor Set Consistency:} Applying the \textit{Non-Donor Set Consistency} argument written above for general $t$---with the entrywise rate supplied by the donor-set bound just derived---raises both exponents by $1/2$ and therefore proves $P_{d,n}(t)$.

\noindent{\bf 4. Verifying Target Causal Parameter Consistency:}
For any unit $n \in [N]$ and $\bar{d}^T \in [A]^T$ we recall the \texttt{SBE-PCR} estimator and the corresponding causal estimand: $\hEx[Y_{n,T}^{(\bar{d}^T)}] = \sum_{t = 1}^T \hat{\gamma}_{n,T,t}(d_t) + \hat{b}_{n,T}$ and $\mathbb{E}[Y_{n,T}^{(\bar{d}^T)} \mid \LFc ] = \sum_{t=1}^T \gamma_{n,T,t}(d_t) + b_{n,T} \mid \LFc$, whose difference is exactly $( \hat{b}_{n,T} - b_{n,T} \mid \LFc ) + \sum_{t=1}^T (\hat{\gamma}_{n,T,t}(d_t) - \gamma_{n,T,t}(d_t) \mid \LFc)$. We apply the known bound for each term, specifically Equation~\eqref{eq:baseline-consistency-rate}, Equation~\eqref{eq:terminal-rate} with $d = d_T$, and $P_{d_{t},n}(t)$ for every $t \in [T-1]$, enlarging every collection to the collection $\Cc$ of part (iv). The same geometric sum as above then gives $O_p(T\,\Rc(T+3/4, T+2; \Cc))$, the baseline rate being dominated by that of the sum.

\subsection{Proof of Theorem~\ref{thm:consistency-time-varying-fixed-lags}}\label{subsection:proof-of-varying-consistency-fixed-lags}

We recall that for any unit $n \in [N]$ and $\bar{d}^T \in [A]^T$
$$\mathbb{E}\left[Y_{n,T}^{(\bar{d}^T)} \mid \LFc \right] = \sum_{t=1}^T \gamma_{n,T,t}(d_t) + b_{n,T} \mid \LFc = \sum_{t=1}^T \langle \psi_n^{T,t}, w_{d_t} - w_{0_t}\rangle + b_{n,T} \mid \LFc.$$

Given Assumption~\ref{assumption:outcome-fixed-lag-dep} we know that $\psi_{n}^{T,T-q-i} = 0$ for all $i \geq 1$. As such, 
$$\mathbb{E}\left[Y_{n,T}^{(\bar{d}^T)} \mid \LFc \right]  = \sum_{t=T-q}^T \langle \psi_n^{T,t}, w_{d_t} - w_{0_t}\rangle + b_{n,T} \mid \LFc = \sum_{t=T-q}^T \gamma_{n,T,t}(d_t) + b_{n,T} \mid \LFc.$$

We modify the \texttt{SBE-PCR} estimator accordingly: $\hEx\left[Y_{n,T}^{(\bar{d}^T)}\right] := \sum_{t = T-q}^T \hat{\gamma}_{n,T,t}(d_t) + \hat{b}_{n,T}$.

Applying the analysis from the proof of Theorem~\ref{thm:consistency-time-varying} yields the desired result.

\section{Summary Statistics for the Full Sample (Section~\ref{sec:application})}\label{app:summary_full}

Table~\ref{tab:summary_stats_full} reports pre-treatment averages for the
2,052 firms that received no export credit support during 2006--2010, before
the matching and trimming described in Section~\ref{subsec:data}. Table
\ref{tab:summary_stats_pretreatment} in the main text reports the same
statistics for the estimation sample of 661 firms.

\begin{table}[!htbp]
\centering
\caption{Pre-treatment Summary Statistics: Full Sample}
\label{tab:summary_stats_full}
\footnotesize
\begin{tabular}{lcccc}
\toprule
Variable & Supported & Never-supported & $t$-statistic & $p$-value \\
\midrule
Export share & 0.204 & 0.156 & -5.349 & 0.000 \\
$\ln$(Sales) & 10.605 & 10.496 & -2.694 & 0.007 \\
Workers & 209.138 & 212.792 & 0.198 & 0.843 \\
$\ln$(Capital) & 9.328 & 9.195 & -2.879 & 0.004 \\
$\ln$(Value added) & 9.024 & 9.021 & -0.081 & 0.935 \\
TFP & 235.707 & 343.420 & 0.692 & 0.489 \\
$\ln$(Wage bill) & 8.507 & 8.543 & 1.105 & 0.269 \\
$\ln$(R\&D) & 5.307 & 4.347 & -9.040 & 0.000 \\
Debt-asset ratio & 0.555 & 0.515 & -4.481 & 0.000 \\
Liquidity ratio & 0.085 & 0.130 & 4.682 & 0.000 \\
FDI indicator & 0.378 & 0.267 & -6.945 & 0.000 \\
Parent affiliation & 0.102 & 0.212 & 7.619 & 0.000 \\
Establishment year & 1987 & 1985 & -3.752 & 0.000 \\
Industry capital intensity & 0.274 & 0.294 & 1.209 & 0.227 \\
Industry wage indicator & 0.451 & 0.400 & -2.889 & 0.004 \\
\bottomrule
\end{tabular}

\vspace{0.5em}
\begin{minipage}{\linewidth}
\footnotesize \textit{Note:} Pre-treatment averages over 2006--10 for the full sample of 2,052 firms. \emph{Supported} firms receive at least one form of export credit support during 2011--15 and none during 2006--10 (167 firms); \emph{never-supported} firms receive none throughout (1,885 firms). The last two columns report two-sample $t$-tests.
\end{minipage}
\end{table}


\clearpage
\phantomsection
\addcontentsline{toc}{part}{Additional Online Appendix}
\begin{center}
{\LARGE Additional Online Appendix}
\end{center}
\vspace{1em}

\section{Alternative Targeting Rules for the Application (Section \ref{sec:application})}\label{sec:appendix_five_period}

This section of the SA reports the prospective policy learning exercise of Section~\ref{subsubsec:optimal_new_firms} over five periods, where the feasible set also varies the timing and spacing of support. Welfare here is truncated because the carry-over of a fifth-period action is unobserved.

\begin{figure}[htbp]
\centering
\subfloat[Early and late treatment options]{\includegraphics[width=0.495\textwidth]{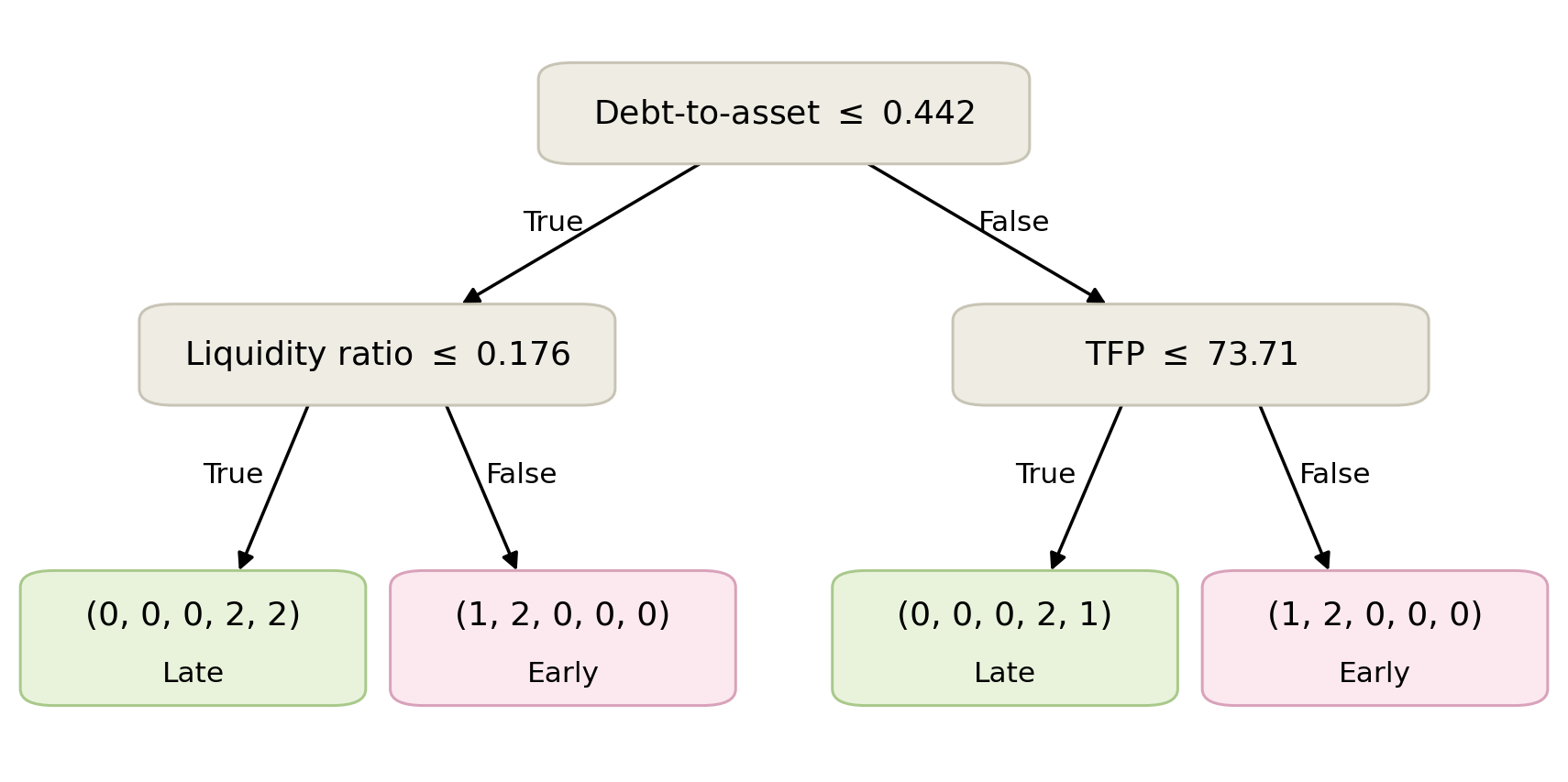}\label{fig:early_late_tree}}\hfill
\subfloat[Front-, even- and back-loading treatment options]{\includegraphics[width=0.495\textwidth]{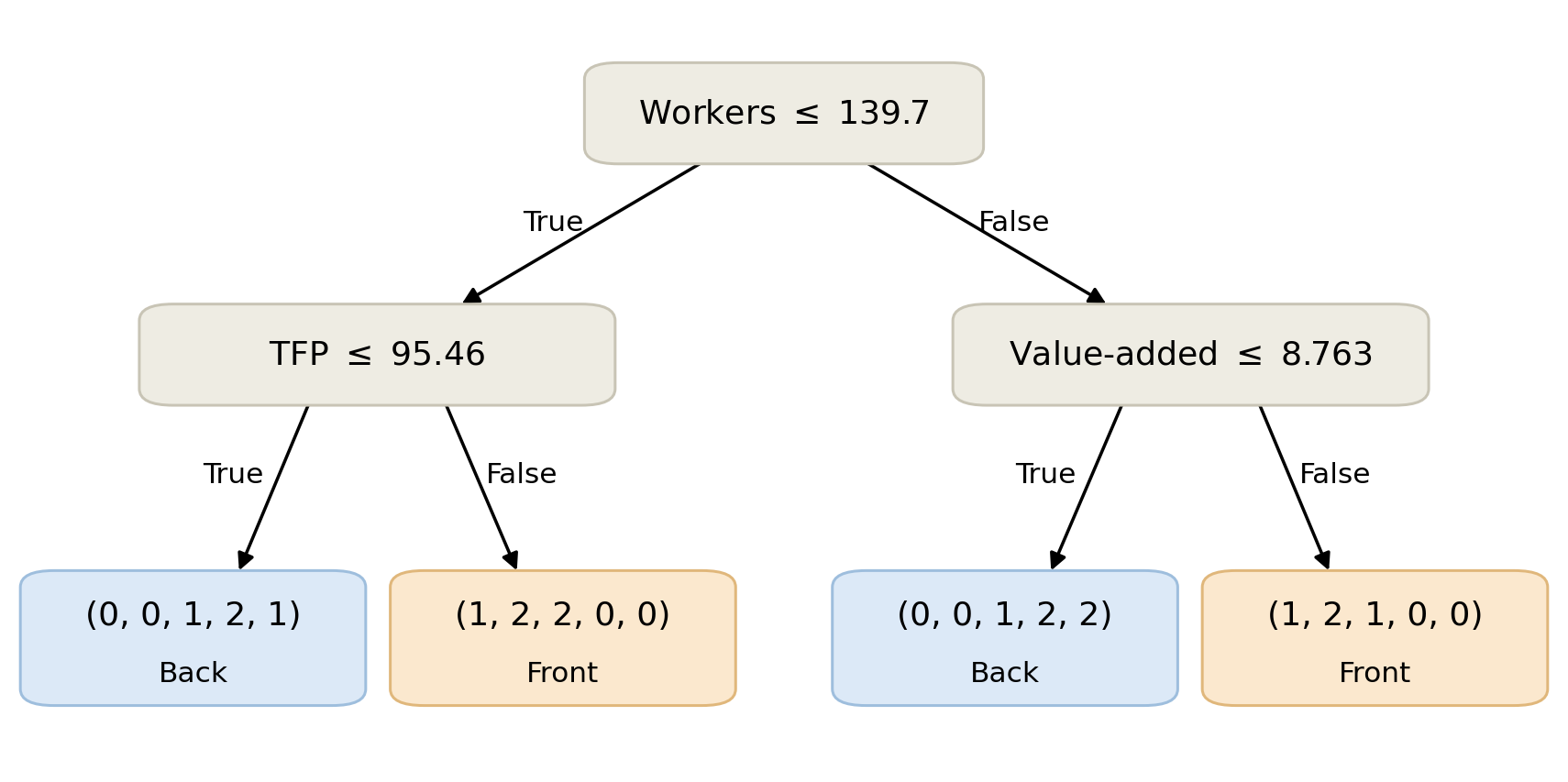}\label{fig:front_even_back_tree}}
\caption{Decision Trees for Five-Period Targeting Rules}
\vspace{0.5em}
\noindent\begin{minipage}[t]{1\columnwidth}%
\small Notes: In the optimal treatment sequences, $d_t = 1$ indicates insurance and $2$ indicates loans support. The threshold for value added is in natural logarithms.
\end{minipage}
\label{fig:decision_trees_five}
\end{figure}

First, we consider $\tilde{\mathcal{D}}$ to be the set of early treatment $(d_1,d_2,0,0,0)$ and late treatment $(0,0,0,d_4,d_5)$ for $d_t\in\{1,2\}$, together with the no-support schedule $(0,0,0,0,0)$, yielding nine possible schedules in total. We restrict attention to insurance or loans only. Figure~\ref{fig:decision_trees_five}(a) shows the optimal policy tree. The tree splits first on the debt-to-asset ratio, and then on the liquidity ratio among less leveraged firms and on productivity among more leveraged ones. Leverage does not itself determine timing. Within each leverage group, it is the second split that separates early from late support, and the instrument mix moves with it. Firms that are more liquid or more productive are assigned insurance first and early, while the rest receive loans first and later. This is consistent with a difference in which constraint binds: stronger firms need cover against non-payment in order to expand, whereas weaker ones need working capital.

Next, we consider targeting rules that not only concern timing but also spacing over time. In particular, we restrict $\tilde{\mathcal{D}}$ to be the set of front-loaded $(d_1,d_2,d_3,0,0)$, evenly-spaced $(d_1,0,d_3,0,d_5)$, and back-loaded $(0,0,d_3,d_4,d_5)$ treatments for $d_t\in\{1,2\}$, again with no support, yielding twenty-five sequences in total. Figure~\ref{fig:decision_trees_five}(b) shows the resulting decision tree, which splits first on employment and then on productivity among smaller firms and on value added among larger ones. Firms with higher productivity or higher value added receive support concentrated in early periods, while the rest receive it concentrated in later periods (back-loaded), and insurance comes first in every selected schedule, front- or back-loaded.\footnote{The no-support schedule is available in both exercises but is selected in no leaf: for every covariate group some positive schedule attains a higher estimated outcome.}

\section{Proofs and Remarks for LTI Factor Models}\label{app:LTI_proofs}

\subsection{Motivating Example}\label{sec:LTI_example}
We show that the linear time-invariant dynamic triangular system satisfies the factor-model representation in Assumption~\ref{assumption:LTI_factor_model}.

\begin{example}[LTI dynamic triangular models with adaptive treatments]\label{ex:LTI_dyn_triangular_model}
Suppose for all $t \in [T]$, all units $n \in [N]$ obey the following dynamic triangular model for a sequence of actions $\bar{D}^t_{n}$ and counterfactual outcomes $Y^{(\bar{D}_n^t)}_{n, t}$:
\begin{align}
    Y^{(\bar{D}_n^t)}_{n, t} &= \ldot{\theta_{n}}{z^{(\bar{D}_n^t)}_{n, t}} + \ldot{\tilde{\theta}_{n}}{w_{D_{n, t}}} + \tilde{\eta}_{n, t},\label{eq:dynamic_triangular_4}
    \\
    D_{n, t} &= h_n(w_{D_{n, t-1}}, \ z^{(\bar{D}_n^{t-1})}_{n, t - 1}),\label{eq:dynamic_triangular_3}
\end{align}
where $z^{(\bar{D}_n^t)}_{n, t} = \bB_{n} \ z^{(\bar{D}_n^{t-1})}_{n, t - 1} + \bC_{n} \ w_{D_{n, t}} + \eta_{n, t}$ and $z_{n, 0} = w_{D_{n, 0}} = 0$. Here, $z^{(\bar{D}_n^t)}_{n, t} \in \Rb^{m}$ is the latent state associated with unit $n$ at time $t$ and $w_{D_{n, t-1}} \in \Rb^{m}$ is the chosen action at time $t - 1$.
$\eta_{n, t} \in \Rb^{m}$ and $\tilde{\eta}_{n, t} \in \Rb$ represent independent mean-zero random innovations at each time step $t$.
$\bB_{n}, \bC_{n} \in \Rb^{m \times m}$ are matrices governing the linear dynamics of $z^{(\bar{D}_n^t)}_{n, t}$.
In contrast to the linear time-varying dynamic triangular system described in Example~\ref{ex:dyn_triangular_model} of the main text, these transition matrices are invariant across all $t \in [T]$.
$\theta_{n}, \tilde{\theta}_{n} \in \Rb^{m}$ are  parameters governing how the outcome of interest $Y^{(\bar{D}_n^t)}_{n, t}$ is a linear function of $z^{(\bar{D}_n^t)}_{n, t} $ and $w_{D_{n, t}}$, respectively.
$h_n(\cdot)$ is a function which decides how the next action $w_{D_{n, t}}$ is chosen as a function of the previous action $w_{D_{n, t - 1}}$ and the previous state $z^{(\bar{D}_n^{t-1})}_{n, t - 1}$.
We see that due to the input of $z^{(\bar{D}_n^{t-1})}_{n, t - 1}$ in $h_n(\cdot)$, the action sequence is {\em adaptive}.
As a result, $\eta_{n, \ell}$ is correlated with $D_{n, t}$ for $\ell < t$.

\begin{proposition}\label{lemma:LTI_representation}
Suppose the dynamic triangular model \eqref{eq:dynamic_triangular_4}--\eqref{eq:dynamic_triangular_3} holds.
Then we have the following representation,
\begin{align}
    Y^{(\bar{d}^t)}_{n, t} = \sum^{t}_{\ell = 1} \Big(\ldot{\psi^{t - \ell}_{n}}{w_{d_\ell}} + \varepsilon_{n, t, \ell} \Big), 
    \label{eq:LTI_factor_model_representation}
\end{align}
where $\psi^{t - \ell}_{n},  w_{d_\ell} \in \Rb^m$ for $\ell \in [t]$; here,
\begin{align*}
    \psi^{t - \ell}_{n} &=  \left( \bB^{t - \ell}_n \bC_n \right)^{\top} \theta_n \quad \text{for} \quad \ell \in [t-1], 
    \\ \psi^{0}_{n} &= (\bC_n)^{\top} \theta_n + \tilde{\theta}_n,
    \\ \varepsilon_{n, t, \ell} &=  \left( \bB^{t - \ell}_n \eta_{n, \ell} \right)^{\top} \theta_n \quad \text{for} \quad \ell \in [t-1],
    \\ \varepsilon_{n, t, t} &= \left( \eta_{n, t} \right)^{\top} \theta_n + \tilde{\eta}_{n, t}.
\end{align*}
Therefore, the factor-model representation in Assumption~\ref{assumption:LTI_factor_model} holds (its time-invariant control sequence clause is a separate condition on the environment), with the additional structure that $\varepsilon^{(\bar{d}^t)}_{n, t}$ has an additive decomposition as $\sum^{t}_{\ell = 1} \varepsilon_{n, t, \ell}$, and it is not a function of $d_\ell$.
\end{proposition}
In this example, our target parameter $\Ex[Y^{(\bar{d}^T)}_{n, T} ~|~ \LFc]$ defined in \eqref{eq:target_causal_param} translates to the expected potential outcome once we condition on the latent parameters $\psi^{t - \ell}_{n},  w_{d_\ell}$, which are functions of $\bB_{n}, \bC_{n}, \theta_{n}, \tilde{\theta}_{n}$, and we take the average over the per-step independent mean-zero random innovations, $\varepsilon_{n, t, \ell}$, which are functions of $\eta_{n, \ell}$ (and of $\tilde{\eta}_{n, t}$ when $\ell = t$), $\bB_{n}$, and $\theta_{n}$.
\end{example}

\subsection{\texttt{SBE-PCR} Algorithm in the LTI Setting}\label{app:lti-algorithm}

Assumption~\ref{control-factors-support-LTI} allows the weights to be estimated exactly as in Section~\ref{subsection:SBE-Varying-Estimator}: for each donor set $\Ic \in \{\Ic^d : d \in [A]\} \cup \{\Ic^0_t : t \in [T]\}$, we apply PCR by regressing $X_n \in \Rb^p$ on the rank-$k_{\Ic \setminus n}$ approximation of $X_{\Ic \setminus n}$ with $k_{\Ic \setminus n} = \text{rank}(\EE[X_{\Ic \setminus n}])$, which yields weights $\hat{\phi}^{n, \Ic} \in \Rb^{|\Ic| - 1}$ if $n \in \Ic$ and $\hat{\beta}^{n, \Ic} \in \Rb^{|\Ic|}$ if $n \notin \Ic$. Given these weights, the \texttt{SBE-PCR} algorithm in the LTI setting is as follows.

\noindent{\bf Step 1: Estimate baseline outcomes.} For $t \in [T]$:
\begin{align*}
	\forall \ j \in \Ic^0_t~:~&\hat{b}_{j,t} = \sum_{k \in \Ic^0_t \setminus j}\hat{\phi}_k^{j, \Ic^0_t} Y_{k,t},\\
	\forall \ i \notin \Ic^0_t~:~&\hat{b}_{i,t} = \sum_{j \in \Ic^0_t} \hat{\beta}_{j}^{i,\Ic^0_t} \hat{b}_{j,t}.
\end{align*}

\noindent{\bf Step 2: Estimate blip effects for lag $0$.} For $d \in [A]$:
\begin{align*}
	\forall \ j \in \Ic^d~:~&\hat{\gamma}_{j, 0}(d) = \sum_{k \in \Ic^d \setminus j}\hat{\phi}_k^{j, \Ic^d}\left(Y_{k, t_k^*} - \hat{b}_{k, t_k^*}\right),\\
	\forall \ i \notin \Ic^d~:~&\hat{\gamma}_{i, 0}(d) = \sum_{j \in \Ic^d} \hat{\beta}_{j}^{i,\Ic^d} \hat{\gamma}_{j, 0}(d).
\end{align*}

\noindent{\bf Step 3: Recursively estimate blip effects for lag $t > 0$.} For $d \in [A]$ and $t \in [1, T-1]$, recursively estimate as follows:
\begin{align*}
	\forall \ j \in \Ic^d~:~&\hat{\gamma}_{j, t}(d) = \sum_{k \in \Ic^d \setminus j}\hat{\phi}_k^{j, \Ic^d}\left(Y_{k, t_k^* + t} - \hat{b}_{k, t_k^* + t} - \sum_{\ell = 0}^{t - 1} \hat{\gamma}_{k, \ell}(D_{k, t_k^* + t - \ell})\right),\\
	\forall \ i \notin \Ic^d~:~&\hat{\gamma}_{i, t}(d) = \sum_{j \in \Ic^d} \hat{\beta}_{j}^{i,\Ic^d} \hat{\gamma}_{j, t}(d).
\end{align*}

\noindent{\bf Step 4: Estimate target causal parameter.} For $n \in [N]$ and $\bar{d}^T \in [A]^T$, estimate the causal parameter as follows:
\begin{align}\label{eq:causal_estimator2}
\widehat{\Ex}[Y_{n, T}^{(\bar{d}^T)}\mid \LFc] =~& \sum_{t =1}^T \hat{\gamma}_{n, T - t}(d_t) + \hat{b}_{n,T} .
\end{align}

\subsection{Proof of Proposition~\ref{lemma:LTI_representation}}
Recall $z^{(\bar{d}^t)}_{n, t}$ is the latent state of unit $n$ if it undergoes action sequence $\bar{d}^t$.
By a simple recursion we have
\begin{align*}
    z^{(\bar{d}^t)}_{n, t}
    &= \sum^{t}_{\ell = 1} \bB^{t - \ell}_{n}  \bC_{n} \ w_{d_\ell} 
    + \sum^{t}_{\ell = 1} \bB^{t - \ell}_{n} \eta_{n, \ell}.
\end{align*}
Hence,
\begin{align*}
    Y^{(\bar{d}^t)}_{n, t}
    &= 
    \left\langle
        \theta_{n}, \
        \sum^{t}_{\ell = 1} \bB^{t - \ell}_{n}  \bC_{n} \ w_{d_\ell} 
        + \sum^{t}_{\ell = 1} \bB^{t - \ell}_{n} \eta_{n, \ell} 
    \right\rangle 
    + \langle \tilde{\theta}_{n},  w_{d_t} \rangle + \tilde{\eta}_{n, t} 
    \\ &= \sum^{t}_{\ell = 1} \Big(\ldot{\psi^{t - \ell}_{n}}{w_{d_\ell}} + \varepsilon_{n, t, \ell} \Big),
\end{align*}
where in the last line we use the definitions of $\psi^{t - \ell}_{n}$ and $\varepsilon_{n, t, \ell}$ in the proposition statement.
This completes the proof.

\subsection{Proof of Theorem~\ref{thm:newLTI}}
Throughout the proof, all expectations are conditioned on $\LFc$; for simplicity, we omit this conditioning in all derivations.

\noindent{\bf 1. Verifying \eqref{eq:LTI_identification}:}
First, we verify \eqref{eq:LTI_identification} holds, which allows us to express the counterfactual outcomes in terms of the blips and the baseline. 
For all $n \in [N]$, using Assumption~\ref{assumption:LTI_factor_model} we have:
\begin{align}
\Ex[Y^{(\bar{d}^T)}_{n, T}] \nonumber
&= \Ex[Y^{(\bar{d}^T)}_{n, T} - Y^{(\bar{0}^T)}_{n, T}] + \Ex[Y^{(\bar{0}^T)}_{n, T}] \nonumber
\\ &= \Ex\left[
\sum^{T}_{t = 1} \ldot{\psi^{T - t}_{n}}{w_{d_t} - w_{\tZero}} + \varepsilon^{(\bar{d}^T)}_{n, T} - \varepsilon^{(\bar{0}^T)}_{n, T}\right] + \Ex\left[\sum^{T}_{t = 1} \ldot{\psi^{T - t}_{n}}{w_{\tZero}} + \varepsilon^{(\bar{0}^T)}_{n, T}\right] \nonumber
\\ &= \sum^{T}_{t = 1} \gamma_{n, T- t}(d_t)  + b_{n, T}. \nonumber
\end{align}

\noindent{\bf 2. Verifying \eqref{eq:LTI_base_observed} and \eqref{eq:LTI_base_synthetic}:} 

We first show \eqref{eq:LTI_base_observed} holds.
For $j \in \Ic^{0}_t$:
\begin{align}
b_{j, t} 
&= \sum^t_{\ell = 1} \ldot{\psi^{t - \ell}_{j}}{w_{\tZero}} 
= \Ex\left[ \sum^t_{\ell = 1} \ldot{\psi^{t - \ell}_{j}}{w_{\tZero}} + \varepsilon^{(\bar{0}^t)}_{j, t} \right] \label{eq:LTI_base_1}
\\ &= \Ex\left[ \sum^t_{\ell = 1} \ldot{\psi^{t - \ell}_{j}}{w_{\tZero}} + \varepsilon^{(\bar{0}^t)}_{j, t}  \mid \Ic^{0}_t \right] \label{eq:LTI_base_2}
\\ &= \Ex\left[Y_{j,t}^{(\bar{0}^t)} \mid j \in \Ic^{0}_t \right] \label{eq:LTI_base_4}
\\&= \Ex\left[Y_{j,t} \mid j \in \Ic^{0}_t \right], \label{eq:LTI_base_5}
\end{align}
where \eqref{eq:LTI_base_1} and \eqref{eq:LTI_base_4} follow from Assumption~\ref{assumption:LTI_factor_model}; 
\eqref{eq:LTI_base_2} follows from the fact that $\ldot{\psi^{t - \ell}_{j}}{w_{\tZero}}$ is deterministic conditional on $\LFc$, and that $\Ex[\varepsilon^{(\bar{0}^t)}_{j, t} \mid \Ic^{0}_t] = \Ex[\varepsilon^{(\bar{0}^t)}_{j, t}]$ as seen in the definition of $\Ic^{0}_t$;
\eqref{eq:LTI_base_5} follows from Assumption~\ref{assumption:SUTVA}.

Next we show \eqref{eq:LTI_base_synthetic} holds.
For $i \notin \Ic^{0}_t$:
\begin{align}
b_{i, t} 
&=  \sum^{t}_{\ell = 1} \ldot{\psi^{t - \ell}_{i}}{w_{\tZero}}  \nonumber
\\&= \sum^{t}_{\ell = 1} \ldot{\psi^{t - \ell}_{i}}{w_{\tZero}} \mid \Ic^{0}_t  \label{eq:LTI_base_synth_1}
\\&= \sum^{t}_{\ell = 1} \sum_{j \in \Ic^{0}_t} \beta_j^{i,\Ic^{0}_t} \ldot{\psi^{t - \ell}_{j}}{w_{\tZero}} \mid \Ic^{0}_t \label{eq:LTI_base_synth_2}
\\&=  \sum_{j \in \Ic^{0}_t} \beta_j^{i,\Ic^{0}_t}   b_{j, t} \mid \Ic^{0}_t,  \nonumber
\end{align}
where \eqref{eq:LTI_base_synth_1} follows from the fact that $\ldot{\psi^{t - \ell}_{i}}{w_{\tZero}}$ is deterministic conditional on $\LFc$;
\eqref{eq:LTI_base_synth_2} follows from Assumption~\ref{assumption:LTI_well_supported_factors}.

\noindent{\bf 3. Verifying \eqref{eq:LTI_base_blip_observed} and \eqref{eq:LTI_base_blip_synthetic}:}

We first show \eqref{eq:LTI_base_blip_observed} holds.
For all $d\in[A]$ and $j\in\Ic^d$,
\begin{align}
\gamma_{j,0}(d)
&=
\Ex\!\left[
Y_{j,t_j^*}^{(\bar{0}^{t_j^*-1},d)}
-
Y_{j,t_j^*}^{(\bar{0}^{t_j^*})}
\,\middle|\,
j\in\Ic^d
\right]
\label{eq:LTI_base_blip_2}
\\
&=
\Ex\!\left[
Y_{j,t_j^*}
-
Y_{j,t_j^*}^{(\bar{0}^{t_j^*})}
\,\middle|\,
j\in\Ic^d
\right]
\label{eq:LTI_base_blip_5}
\\
&=
\Ex\!\left[
Y_{j,t_j^*}
\mid j\in\Ic^d
\right]
-
b_{j,t_j^*}.
\label{eq:LTI_base_blip_7}
\end{align}
Here \eqref{eq:LTI_base_blip_2} follows from
Assumptions~\ref{assumption:LTI_factor_model} and
\ref{assumption:LTI_seq_exog}, applied with both the outcome date and
the switch period equal to $t_j^*$.
Equation~\eqref{eq:LTI_base_blip_5} follows from condition~(i) of
\eqref{eq:LTI_donor} and Assumption~\ref{assumption:SUTVA}.
Finally, \eqref{eq:LTI_base_blip_7} follows from condition~(ii) of
\eqref{eq:LTI_donor}, evaluated at outcome date $\ell=t_j^*$, and
Assumption~\ref{assumption:LTI_factor_model}.

Next we show \eqref{eq:LTI_base_blip_synthetic} holds.
For $i \notin \Ic^d$:
\begin{align}
\gamma_{i, 0}(d)
&=  \ldot{\psi^{0}_{i}}{w_{d} - w_{\tZero}} 
= \ldot{\psi^{0}_{i}}{w_{d} - w_{\tZero}} \mid \Ic^d \label{eq:LTI_base_blip_synth_3}
\\ &= \sum_{j \in \Ic^d} \beta^{i, \Ic^d}_{j}\ldot{\psi_{j}^{0}}{w_{d} - w_{\tZero}} \mid \Ic^d  \label{eq:LTI_base_blip_synth_4}
\\ &= \sum_{j \in \Ic^d} \beta^{i, \Ic^d}_{j} \gamma_{j, 0}(d) \mid \Ic^d, \nonumber
\end{align} 
where \eqref{eq:LTI_base_blip_synth_3} follows from the fact that $\ldot{\psi^{0}_{i}}{w_{d} - w_{\tZero}}$ is deterministic conditional on $\LFc$;
\eqref{eq:LTI_base_blip_synth_4} follows from Assumption~\ref{assumption:LTI_well_supported_factors}.

\noindent{\bf 4. Verifying \eqref{eq:LTI_recursive_blip_observed} and \eqref{eq:LTI_recursive_blip_synthetic}:}

We first show \eqref{eq:LTI_recursive_blip_observed} holds.
For all $d\in[A]$, $t\in[T-1]$, and $j\in\Ic^d$ such that
$t_j^*+t\leq T$:
\begin{align}
&\Ex\left[ Y_{j, t^*_j + t} - Y^{(\bar{0}^{t^*_j + t})}_{j, t^*_j + t} \mid j \in \Ic^d \right] 
= \Ex\left[ Y^{(\bar{D}^{t^*_j + t}_j)}_{j, t^*_j + t} - Y^{(\bar{0}^{t^*_j + t})}_{j, t^*_j + t} \mid j \in \Ic^d \right] \label{eq:LTI_recursive_blip_0}
\\ &= \Ex\left[ Y^{(\bar{D}^{t^*_j + t}_j)}_{j, t^*_j + t} - Y^{(\bar{D}^{t^*_j - 1}_j, \underline{0}^{t^*_j})}_{j, t^*_j + t} \mid j \in \Ic^d \right] \label{eq:LTI_recursive_blip_1}
\\ &= \sum^{t}_{\ell = 0} \Ex\left[ Y^{(\bar{D}^{t^*_j + t - \ell}_j, \underline{0}^{t^*_j + t - \ell + 1})}_{j, t^*_j + t} - Y^{(\bar{D}^{t^*_j + t - \ell - 1}_j, \underline{0}^{t^*_j + t - \ell})}_{j, t^*_j + t} \mid j \in \Ic^d \right] \label{eq:LTI_recursive_blip_1.1}
\end{align}
where \eqref{eq:LTI_recursive_blip_0} follows from Assumption~\ref{assumption:SUTVA};
\eqref{eq:LTI_recursive_blip_1} uses that for $j \in \Ic^d$,  $\bar{D}^{t^*_j - 1}_j = (\tZero, \dots, \tZero)$, and Assumption~\ref{assumption:SUTVA}.
Then,
\begin{align}
&\sum^{t}_{\ell = 0} \Ex\left[ Y^{(\bar{D}^{t^*_j + t - \ell}_j, \underline{0}^{t^*_j + t - \ell + 1})}_{j, t^*_j + t} - Y^{(\bar{D}^{t^*_j + t - \ell - 1}_j, \underline{0}^{t^*_j + t - \ell})}_{j, t^*_j + t} \mid j \in \Ic^d \right]  \nonumber
\\ &= \sum^{t}_{\ell = 0} \Ex\left[ \ldot{\psi^{\ell}_j}{w_{D_{j, t^*_j + t - \ell}} -w_{\tZero} } + \varepsilon^{(\bar{D}^{t^*_j + t - \ell}_j, \underline{0}^{t^*_j + t - \ell + 1})}_{j, t^*_j + t} - \varepsilon^{(\bar{D}^{t^*_j + t - \ell - 1}_j, \underline{0}^{t^*_j + t - \ell})}_{j, t^*_j + t} \mid j \in \Ic^d \right] \label{eq:LTI_recursive_blip_2}
%
%
\\ &= \Ex\left[\ldot{\psi^{t}_j}{w_{D_{j, t^*_j}} -w_{\tZero} } + \varepsilon^{(\bar{D}^{t^*_j}_j, \underline{0}^{t^*_j + 1})}_{j, t^*_j + t} - \varepsilon^{(\bar{D}^{t^*_j - 1}_j, \underline{0}^{t^*_j})}_{j, t^*_j + t} \mid j \in \Ic^d \right] \nonumber
\\ & \quad \quad + \sum^{t - 1}_{\ell = 0} \Ex\left[ \ldot{\psi^{\ell}_j}{w_{D_{j, t^*_j + t - \ell}} -w_{\tZero} } + \varepsilon^{(\bar{D}^{t^*_j + t - \ell}_j, \underline{0}^{t^*_j + t - \ell + 1})}_{j, t^*_j + t} - \varepsilon^{(\bar{D}^{t^*_j + t - \ell - 1}_j, \underline{0}^{t^*_j + t - \ell})}_{j, t^*_j + t} \mid j \in \Ic^d \right] \nonumber
\\ &= \ldot{\psi^{t}_j}{w_{d} -w_{\tZero} } + \Ex\left[ \varepsilon^{(\bar{D}^{t^*_j}_j, \underline{0}^{t^*_j + 1})}_{j, t^*_j + t} - \varepsilon^{(\bar{D}^{t^*_j - 1}_j, \underline{0}^{t^*_j})}_{j, t^*_j + t} \mid j \in \Ic^d \right] \nonumber
\\ & \quad \quad + \sum^{t - 1}_{\ell = 0} \Ex\left[ \ldot{\psi^{\ell}_j}{w_{D_{j, t^*_j + t - \ell}} -w_{\tZero} } + \varepsilon^{(\bar{D}^{t^*_j + t - \ell}_j, \underline{0}^{t^*_j + t - \ell + 1})}_{j, t^*_j + t} - \varepsilon^{(\bar{D}^{t^*_j + t - \ell - 1}_j, \underline{0}^{t^*_j + t - \ell})}_{j, t^*_j + t} \mid j \in \Ic^d \right] \label{eq:LTI_recursive_blip_4}
\\ &=  \ldot{\psi^{t}_j}{w_{d} -w_{\tZero} }  + \sum^{t - 1}_{\ell = 0} \Ex\left[ \ldot{\psi^{\ell}_j}{w_{D_{j, t^*_j + t - \ell}} -w_{\tZero} }  \mid j \in \Ic^d \right] \nonumber
\\ &  \quad \quad + \sum^{t}_{\ell = 0} \Ex\left[  \Ex\left[ \varepsilon^{(\bar{\delta}^{t^*_j + t - \ell}, \underline{0}^{t^*_j + t - \ell + 1})}_{j, t^*_j + t} - \varepsilon^{(\bar{\delta}^{t^*_j + t - \ell - 1}, \underline{0}^{t^*_j + t - \ell})}_{j, t^*_j + t} \mid \bar{D}_j^{t^*_j + t - \ell} = \bar{\delta}^{t^*_j + t - \ell} \right]  \mid j \in \Ic^d \right] \nonumber
\\ &= \ldot{\psi^{t}_j}{w_{d} -w_{\tZero} }  
+  \Ex\Big[\sum^{t - 1}_{\ell = 0}  \ldot{\psi^{\ell}_j}{w_{D_{j, t^*_j + t - \ell}} -w_{\tZero} } \,\Big|\, j \in \Ic^d\Big]  \label{eq:LTI_recursive_blip_5}
\\ &=  \gamma_{j, t}(d)  
+ \Ex\Big[\sum^{t - 1}_{\ell = 0}  \gamma_{j, \ell}(D_{j, t^*_j + t - \ell}) \,\Big|\, j \in \Ic^d\Big], \label{eq:LTI_recursive_blip_6}
\end{align}
where \eqref{eq:LTI_recursive_blip_2} follows from Assumption~\ref{assumption:LTI_factor_model};
\eqref{eq:LTI_recursive_blip_4} follows from the definition of $\Ic^d$, i.e., for $j \in \Ic^d$, $\bar{D}^{t^*_j}_j = (\bar{0}^{t^*_j - 1}, d)$;
\eqref{eq:LTI_recursive_blip_5} follows from Assumption~\ref{assumption:LTI_seq_exog}, applied for each $\ell \in [0, t]$ with outcome date $t^*_j + t$ and switch period $t^*_j + t - \ell$; in particular, the $\ell = t$ noise difference, carried through the two preceding displays, vanishes by the same argument. Note that $t^*_j + t \le T$ under the restriction stated in part (iii) of Theorem~\ref{thm:newLTI}, so the quantification $t \in [T]$ in the assumption suffices. Throughout this part, conditioning on $j \in \Ic^d$ fixes the donor's first treatment time, i.e., all expectations are given $\{j \in \Ic^d, t^*_j = s\}$ for a fixed $s$ with $s + t \le T$, as stipulated after part (iii) of Theorem~\ref{thm:newLTI}; this is what makes $b_{j, s + t}$ deterministic given $\LFc$. The tower rule applies because condition (i) of \eqref{eq:LTI_donor} restricts only $\bar{D}^{s}_j$ (and condition (ii) the law of unit $j$), so the event $\{j \in \Ic^d, t^*_j = s\}$ is determined by $\bar{D}^{s}_j$ and hence by $\bar{D}^{s + t - \ell}_j$ for every $\ell \in [0, t]$; the lower-lag factor terms in \eqref{eq:LTI_recursive_blip_5} are kept inside the conditional expectation since the actions $D_{j, s + t - \ell}$, $\ell < t$, may be adaptive.
Re-arranging \eqref{eq:LTI_recursive_blip_6}, we have that
\begin{align}
\gamma_{j, t}(d)  
&= \Ex\Big[ Y_{j, t^*_j + t} - Y^{(\bar{0}^{t^*_j + t})}_{j, t^*_j + t} - \sum^{t - 1}_{\ell = 0}  \gamma_{j, \ell}(D_{j, t^*_j + t - \ell}) \,\Big|\, j \in \Ic^d \Big] \nonumber
\\ &= \Ex\Big[ Y_{j, t^*_j + t} - \sum^{t - 1}_{\ell = 0}  \gamma_{j, \ell}(D_{j, t^*_j + t - \ell}) \,\Big|\, j \in \Ic^d \Big] - \Ex\left[ Y^{(\bar{0}^{t^*_j + t})}_{j, t^*_j + t}  \right]  \label{eq:LTI_recursive_blip_7}
\\ &= \Ex\Big[ Y_{j, t^*_j + t} - \sum^{t - 1}_{\ell = 0}  \gamma_{j, \ell}(D_{j, t^*_j + t - \ell}) \,\Big|\, j \in \Ic^d \Big] -  b_{j, t^*_j + t}, \label{eq:LTI_recursive_blip_8}
\end{align}
where \eqref{eq:LTI_recursive_blip_7} follows from condition~(ii) of
\eqref{eq:LTI_donor}, evaluated at outcome date
$\ell=t_j^*+t$;
\eqref{eq:LTI_recursive_blip_8} follows from Assumption~\ref{assumption:LTI_factor_model}.

Next we show \eqref{eq:LTI_recursive_blip_synthetic} holds.
For all $d \in [A]$, $t \in [T-1]$, $i \notin \Ic^d$:
\begin{align}
\gamma_{i, t}(d) 
&=  \ldot{\psi^{t}_{i}}{w_{d} - w_{\tZero}} \label{eq:LTI_recursive_blip_synthetic_2}
%
%
\\ &= \ldot{\psi^{t}_{i}}{w_{d} - w_{\tZero}} \mid \Ic^d  \label{eq:LTI_recursive_blip_synthetic_4}
\\ &=  \sum_{j \in \Ic^d} \beta^{i, \Ic^d}_{j}\ldot{\psi_{j}^{t}}{w_{d} - w_{\tZero}} \mid \Ic^d   \label{eq:LTI_recursive_blip_synthetic_5}
\\ &= \sum_{j \in \Ic^d} \beta^{i, \Ic^d}_{j} \gamma_{j, t}(d) \mid \Ic^d, \nonumber
\end{align}
where \eqref{eq:LTI_recursive_blip_synthetic_4} follows from the fact that $\ldot{\psi^{t}_{i}}{w_{d} - w_{\tZero}}$ is deterministic conditional on $\LFc$;
\eqref{eq:LTI_recursive_blip_synthetic_5} follows from Assumption~\ref{assumption:LTI_well_supported_factors}.
%

\subsection{General Remarks on the LTI Setting}\label{subsection:notes-on-lti-setting}

\subsubsection{Fixed Lag Dependence Assumption}

Assumption~\ref{assumption:outcome-fixed-lag-dep-invariant} is not restrictive. Recalling Example~\ref{ex:LTI_dyn_triangular_model} (LTI dynamic triangular models) and Proposition~\ref{lemma:LTI_representation}, we consider the following conditions: (i) hard memory cutoff: $\exists q \in \mathbb{N},  \bB_{n}^{q+1} = 0$; (ii) exponential forgetting (spectral decay condition): $\exists C > 0, \lambda \in (0,1)$, such that for all $t$, $\left\| \bB_{n}^{t} \right\|_2 \leq C \lambda^{t}$; (iii) input-channel memory cutoff: $\exists q \in \mathbb{N}, \ \bB_{n}^{q+1} \bC_n = 0$. Since $\psi^{t-\ell}_n = (\bB_n^{t-\ell}\bC_n)^{\top}\theta_n$ for $\ell \in [t-1]$, condition (iii) is sufficient for Assumption~\ref{assumption:outcome-fixed-lag-dep-invariant} (and necessary for it to hold irrespective of $\theta_n$), and (i) implies (iii); under (ii) the assumption holds only approximately, up to a truncation error of order $\lambda^{q}$ in the omitted lags. Clearly, the first condition is the \textit{strongest}. In general, this shows that our assumption of fixed memory is a reasonable one, supporting the effectiveness of our methodology within the dynamic treatment regime from a statistical perspective.

\subsection{Preliminary Results for the Proof of Theorem~\ref{thm:consistency-time-invariant-fixed-lags}}\label{subsection:proof-of-invariant-consistency}

We present consistency results (and their proofs) that serve as preliminaries for proving Theorem~\ref{thm:consistency-time-invariant-fixed-lags}. This is analogous to Theorem~\ref{thm:consistency-time-varying}, which serves as a preliminary result for Theorem~\ref{thm:consistency-time-varying-fixed-lags}.

\begin{theorem}\label{thm:consistency-time-invariant} Let Assumptions~\ref{assumption:SUTVA}, \ref{assumption:general_latent_factor_model}, \ref{assumption:LTI_factor_model}, \ref{assumption:LTI_well_supported_factors}, \ref{assumption:LTI_seq_exog}, \ref{assumption:additional-covariates}, \ref{control-factors-support-LTI}, \ref{assumption:sub-gaussian-noise}, and \ref{assumption:lti-balanced-spectra} hold, including part (ii) of each of the last two (bounded conditional means and row-space inclusion, respectively). Consider the LTI variant of the \texttt{SBE-PCR} estimator (SA~\ref{app:lti-algorithm}) and suppose $k = \max_{\Ic \in \{\Ic^d\} \cup \{\Ic^0_t\}}\text{rank}(\EE[X_{\Ic}])$. Then, conditional on $\LFc$, $\{\rho_i\}_{i \in [p]}$, and the donor-defining treatment histories $\{\bar{D}^{t^*_j}_{j} : j \in \Ic^d\}$ and $\{\bar{D}^{t}_{j} : j \in \Ic^0_t\}$ of all donor sets, we have:

\noindent\textbf{(i) Baseline Consistency:} For any $n \in [N]$ and $t \in [T]$
$$\hat{b}_{n,t} - b_{n,t} \mid \LFc
    = O_p\left(\sqrt{\log(p|\Ic_t^0|)}\left(\frac{k^{5/4}}{p^{1/4}} +k^{5/2}\max\left\{\frac{\sqrt{|\Ic_t^0|}}{p^{3/2}}, \frac{1}{\sqrt{|\Ic_t^0|-1}}, \frac{1}{\sqrt{p}}\right\}\right)\right).$$

\noindent\textbf{(ii) Lag-$0$ Blip Consistency:} For any $d \in [A]$ and unit $n \in [N]$
$$
\hat{\gamma}_{n,0}(d) - \gamma_{n,0}(d) \mid \LFc = O_p\left(\sqrt{\log(p\pi_{\Ic})}\left(\frac{k^{9/4}}{p^{1/4}} +k^{7/2}\max\left\{\frac{\sqrt{\pi_{\Ic}}}{p^{3/2}}, \frac{1}{\sqrt{\alpha_{\Ic}-1}}, \frac{1}{\sqrt{p}}\right\}\right)\right),
$$
where $\pi_{\Ic} = \max\{|\Ic_1^0|,|\Ic^d|\}$ and $\alpha_{\Ic} = \min\{|\Ic_T^0|,|\Ic^d|\}$.

\noindent\textbf{(iii) Higher-Lag Blip Consistency:} For any $d \in [A]$, unit $n \in [N]$, and $t \in [1, T-1]$:
\begin{align*}
    &\hat{\gamma}_{n, t}(d) - \gamma_{n, t}(d) \mid \LFc\\
    &= O_p\left(t\sqrt{\log(p\pi_{\Ic})}\left(\frac{k^{t+9/4}}{p^{1/4}} + k^{t+7/2}\max\left\{\frac{\sqrt{\pi_{\Ic}}}{p^{3/2}}, \frac{1}{\sqrt{\alpha_{\Ic}-1}}, \frac{1}{\sqrt{p}}\right\}\right)\right),
\end{align*}
where $\Cc = \{|\Ic^d|, |\Ic_T^0|, |\Ic_1^0|, (|\Ic^{D_{n, t_n^* + q}}|)_{n \in [N], q\in[1, t]} \}$ with $\pi_{\Ic} = \max\Cc, \alpha_{\Ic} = \min\Cc$.

\noindent\textbf{(iv) Target Causal Parameter Consistency:} For any $n \in [N]$ and $\bar{d}^T \in [A]^T$:
$$
\widehat{\Ex}[Y_{n, T}^{(\bar{d}^T)}] - \EE[Y_{n, T}^{(\bar{d}^T)}\mid \LFc]=O_p\left(T\sqrt{\log(p\pi_{\Ic})}\left(\frac{k^{T+5/4}}{p^{1/4}} + k^{T+5/2}\max\left\{\frac{\sqrt{\pi_{\Ic}}}{p^{3/2}}, \frac{1}{\sqrt{\alpha_{\Ic}-1}}, \frac{1}{\sqrt{p}}\right\}\right)\right),
$$
where $\Cc = \{|\Ic_T^0|, |\Ic^0_1|, (|\Ic^{d_{t}}|)_{t \in [T]} ,(|\Ic^{D_{n, t_n^* + t}}|)_{n \in [N], t \in [1, T-1]}\}$ with $\pi_{\Ic} = \max\Cc$ and $\alpha_{\Ic} = \min\Cc$. Here, each $O_p(\cdot)$ is defined with respect to the sequence $\min\{p, \alpha_{\Ic}\}$.\footnote{Notice that $\alpha_{\Ic} \leq \pi_{\Ic}$ by definition.}
\end{theorem}
Below we provide a full proof of Theorem~\ref{thm:consistency-time-invariant}, which is quite similar to that of Theorem~\ref{thm:consistency-time-varying}. 

\noindent{\bf 0. Recurring arguments:} As in SA~\ref{subsection:proof-of-varying-consistency}, all statements are conditional on $\LFc$, $\{\rho_i\}_{i \in [p]}$, and the donor-defining histories $\{\bar{D}^{t^*_j}_{j} : j \in \Ic^d\}$ and $\{\bar{D}^{t}_{j} : j \in \Ic^0_t\}$ of all donor sets; in some displays we omit part of this conditioning for brevity, and a statement of the form ``$Z \mid \LFc = O_p(r)$'' means that $Z = O_p(r)$ under this conditional law. Since this conditioning fixes every donor-set cardinality, an entry $|\Ic^{D_{i, s}}|$ of a collection $\Cc$ below (indexed by a realized action) is to be read as the maximum, in $\pi_{\Ic}$, and the minimum, in $\alpha_{\Ic}$, over the actions $a$ with $\Ic^a \neq \emptyset$, which bounds every realization. For a donor $j \in \Ic^d$ and a lag $t \in [0, T-1]$ we write $\varepsilon^{(t)}_{j} := \varepsilon^{(\bar{D}^{t^*_j + t}_j)}_{j, t^*_j + t} = Y_{j, t^*_j + t} - b_{j, t^*_j + t} - \sum_{\ell = 0}^{t} \gamma_{j, \ell}(D_{j, t^*_j + t - \ell})$ for the noise along its realized path (Assumption~\ref{assumption:LTI_factor_model}, using $D_{j, t^*_j} = d$ and $D_{j, s} = \tZero$ for $s < t^*_j$). By the argument in the proof of Theorem~\ref{thm:newLTI} (the tower rule over $\bar{D}^{t^*_j + t - \ell}_j$ for $\ell \in [0, t]$, Assumption~\ref{assumption:LTI_seq_exog}, and condition (ii) of \eqref{eq:LTI_donor} at outcome date $t^*_j + t$, with $t^*_j + t \le T$ as stipulated in part (iii) of that theorem), $\Ex[\varepsilon^{(t)}_j \mid \LFc, j \in \Ic^d] = 0$. As in item~0 of SA~\ref{subsection:proof-of-varying-consistency}, we apply Assumption~\ref{assumption:sub-gaussian-noise} to these realized-path noise terms together with the cross-unit product condition stated there (conditional on $\LFc$, the unit-level noise, covariate-error, and action processes are independent across units): given $\LFc$, each donor-defining event, $\{j \in \Ic^d\}$ or $\{j \in \Ic^0_t\}$, is determined by unit $j$'s own history, and each $\varepsilon^{(t)}_j$ is a function of unit $j$'s own processes, so the $\varepsilon^{(t)}_j$ are independent across units conditional on $\LFc$ and the donor-defining histories even though the donors' switching dates $t^*_j$ and realized paths differ; and they are sub-Gaussian with the same constant $C\sigma$ as the fixed-path noise: on $\{t^*_j = \tau\}$, Assumption~\ref{assumption:sub-gaussian-noise}(i) at outcome date $\tau + t$ gives $\Ex[\exp\{(\varepsilon^{(t)}_j)^2/(C\sigma)^2\} \mid \LFc, \{\bar{D}^{\tau + t}_n\}_{n \in [N]}] \le 2$ as in item~0 of SA~\ref{subsection:proof-of-varying-consistency}, and under the cross-unit product condition the conditional law of unit $j$'s processes given the donor-defining histories of all units depends only on $\LFc$ and its own donor-defining history $\bar{D}^{\tau}_j$, a function of $\bar{D}^{\tau + t}_j$, so the tower property transfers the bound to the conditioning of the theorem.

\noindent{\bf 1. Verifying Baseline Consistency:} 

For any $t \in [T]$:

\textit{Donor Set Baseline Consistency:} Consider unit $n \in \Ic_t^0$. Denote $X_{\Ic_t^0 \setminus n} = X_{:, \Ic^0_t\setminus n} \in \Rb^{ p \times |\Ic_t^0 \setminus n|}$. We know the baseline outcome admits the representation
$$\hat{b}_{n,t} - b_{n,t} \mid \LFc  = \left\langle \hat{\phi}^{n, \Ic_t^0}, Y_{\Ic_t^0 \setminus n,t} \right\rangle - \left\langle \phi^{n, \Ic_t^0}, \mathbb{E}[Y_{\Ic_t^0 \setminus n, t} \mid \LFc] \right\rangle,$$
where $\hat{\phi}^{n, \Ic_t^0}$ are the regression coefficients from regressing additional covariates  $X_n \in \Rb^p$ on the rank-$k_{\Ic_t^0 \setminus n}$ approximation of $X_{\Ic_t^0 \setminus n}$ with $k_{\Ic_t^0 \setminus n} = \text{rank}(\EE[X_{\Ic_t^0 \setminus n}])$, i.e., doing PCR with parameter $k_{\Ic_t^0 \setminus n}$.

\begin{lemma}\label{lemma:tilde-baseline-invariant}
    We claim the following
     $$\left\langle \phi^{n, \Ic_t^0}, \mathbb{E}[Y_{\Ic_t^0 \setminus n, t}] \right\rangle = \left\langle \tilde{\phi}^{n, \Ic_t^0}, \mathbb{E}[Y_{\Ic_t^0 \setminus n,t} ] \right\rangle,$$
     where $\tilde{\phi}^{n, \Ic_t^0} = VV^{\top}\phi^{n, \Ic_t^0}$, with $V \in \Rb^{|\Ic_t^0 \setminus n| \times k_{\Ic_t^0 \setminus n}}$ denoting the right singular vectors of $\EE[X_{\Ic_t^0 \setminus n}]$.
\end{lemma}

\begin{proof}
By Assumption~\ref{assumption:lti-row-space} there exists $\xi^{(0,t)}$ such that for any $j \in \Ic^0_t \setminus n$
$$\Ex[Y_{j,t}|\LFc, j \in \Ic^0_t \setminus n] = \sum_{i = 1}^p \xi^{(0,t)}_i \cdot \Ex[(X_{\Ic^0_t\setminus n})_{ij}|\LFc, j \in \Ic^0_t \setminus n].$$

As such,
$$\EE[Y_{\Ic_t^0 \setminus n, t}] = VV^{\top} \EE[Y_{\Ic_t^0 \setminus n, t}],$$
which gives us
$$\left\langle \tilde{\phi}^{n, \Ic_t^0}, \mathbb{E}[Y_{\Ic_t^0 \setminus n, t} ] \right\rangle = \left\langle VV^{\top}\phi^{n, \Ic_t^0}, \mathbb{E}[Y_{\Ic_t^0 \setminus n, t}] \right\rangle =  \mathbb{E}[Y_{\Ic_t^0 \setminus n, t} ]^{\top}VV^{\top} \cdot \phi^{n, \Ic_t^0} =  \left\langle \phi^{n, \Ic_t^0}, \mathbb{E}[Y_{\Ic_t^0 \setminus n, t} ] \right\rangle,$$
proving the desired result.
\end{proof}

Using Lemma~\ref{lemma:tilde-baseline-invariant}, we can now lift the proof technique in \citet[Appendix C, Theorem 2]{SI} to show consistency for $n \in \Ic_t^0$
\begin{align}\label{eq:donor-baseline-consistency-invariant}
    \hat{b}_{n,t} - b_{n,t} \mid \LFc &= \left\langle \hat{\phi}^{n, \Ic_t^0}, Y_{\Ic_t^0 \setminus n, t} \right\rangle - \left\langle \tilde{\phi}^{n, \Ic_t^0}, \mathbb{E}[Y_{\Ic_t^0 \setminus n, t}] \right\rangle \nonumber\\
    &= O_p \left( \sqrt{\log(p |\Ic_t^0|)} \left[ \frac{{k}^{3/4}}{p^{1/4}} + {k}^2 \max \left\{ \frac{\sqrt{|\Ic_t^0|}}{p^{3/2}}, \frac{1}{\sqrt{p}}, \frac{1}{\sqrt{|\Ic_t^0|-1}} \right\} \right] \right),
\end{align}
where we set $T_1 = 1$, $\tilde{w}^{(i,d)} = \tilde{\phi}^{n, \Ic_t^0}$, $\hat{w}^{(i,d)} = \hat{\phi}^{n, \Ic_t^0}$, $Y_{t,\mathcal{I}^{(d)}} = Y_{\Ic_t^0 \setminus n, t}$, $\EE[Y_{t,\mathcal{I}^{(d)}}] = \mathbb{E}[Y_{\Ic_t^0 \setminus n, t} \mid \LFc] $, and $\mathcal{P}_{V_{\text{pre}}} = VV^{\top}$. Furthermore, in the final rate we set $T_0 = p$, $N_d = |\Ic_t^0 \setminus n|$, and $r_{\text{pre}} = k_{\Ic_t^0 \setminus n}$. To conclude, we used that $|\Ic_t^0 \setminus n| = |\Ic_t^0| - 1$ and $k_{\Ic_t^0 \setminus n} \leq k$ where $k$ is the uniform upper bound on the rank of all possible expected covariate matrices, i.e.,  $k = \max_{\Ic \in \{\Ic^d\} \cup \{\Ic^0_t\}}\text{rank}(\EE[X_{\Ic}]).$

\textit{Non-Donor Set Baseline Consistency:} Consider unit $n \notin \Ic_t^0$. Denote $X_{\Ic_t^0} = X_{:, \Ic^0_t}\in \Rb^{p \times |\Ic_t^0|}$. We know the baseline outcome admits the representation
$$\hat{b}_{n,t} - b_{n,t} \mid \LFc  = \left\langle \hat{\beta}^{n, \Ic_t^0}, \hat{b}_{\Ic_t^0,t } \right\rangle - \left\langle \beta^{n, \Ic_t^0}, b_{\Ic_t^0,t } \right\rangle,$$
where $\hat{\beta}^{n, \Ic_t^0}$ are the regression coefficients from regressing additional covariates  $X_n \in \Rb^p$ on the rank-$k_{\Ic_t^0 }$ approximation of  $X_{\Ic_t^0 }$ with $k_{\Ic_t^0 } = \text{rank}(\EE[X_{\Ic_t^0}])$, i.e., doing PCR with parameter $k_{\Ic_t^0 }$.

\begin{lemma}\label{lemma:project-beta-b-invariant}
    We have that
     $$\left\langle \beta^{n, \Ic_t^0}, b_{\Ic_t^0,t} \right\rangle = \left\langle \tilde{\beta}^{n, \Ic_t^0}, b_{\Ic_t^0,t} \right\rangle$$
     with $\tilde{\beta}^{n, \Ic_t^0} = VV^{\top}{\beta}^{n, \Ic_t^0}$, where $V$ denotes the right singular vectors of $\EE[X_{ \Ic_t^0}]$. 
\end{lemma}
\begin{proof}
    It suffices to prove that
    $$VV^{\top}b_{\Ic_t^0,t} = b_{\Ic_t^0,t},$$
    which is equivalent to $(b_{\Ic_t^0, t})^{\top}$ being in the row space of $\EE[X_{\Ic_t^0}]$. By definition, for any $j \in \Ic^0_t$ we know $ b_{j,t} = \EE[Y_{j,t}|\LFc, j \in \Ic_t^0]$. Lastly, by Assumption~\ref{assumption:lti-row-space} there exists $\xi^{(0,t)} \in \Rb^p$ such that 
    $$\EE[Y_{j,t}|\LFc, j \in \Ic_t^0] = \sum_{i = 1}^p\xi_i^{(0,t)} \cdot  \EE[(X_{\Ic_t^0})_{ij}|\LFc, j \in \Ic_t^0].$$
    This concludes the proof.
\end{proof}

Lemma~\ref{lemma:project-beta-b-invariant} allows us to write
\begin{align*}
    \hat{b}_{n,t} - b_{n,t} \mid \LFc  &= \left\langle \hat{\beta}^{n, \Ic_t^0}, \hat{b}_{\Ic_t^0,t } \right\rangle - \left\langle \tilde{\beta}^{n, \Ic_t^0}, b_{\Ic_t^0 ,t} \right\rangle\\
    &=  \underbrace{\langle \tilde{\beta}^{n,\Ic_t^0}, \eta_{\Ic_t^0}\rangle}_{\text{Term 1a}} + \underbrace{\langle \Delta_{n, \Ic_t^0} , \eta_{\Ic_t^0}\rangle}_{\text{Term 1b}} +  \underbrace{\langle \Delta_{n, \Ic_t^0}, b_{\Ic_t^0,t}\rangle}_{\text{Term 1c}},
\end{align*}
where $\eta_{\Ic_t^0} = \hat{b}_{\Ic_t^0,t} - b_{\Ic_t^0,t}$ and $\Delta_{n, \Ic_t^0} = \hat{\beta}^{n,\Ic_t^0} - \tilde{\beta}^{n,\Ic_t^0}$. Using the argument of Lemma~\ref{lemma:pcr-rate-transfer} in SA~\ref{subsection:proof-of-varying-consistency} (with $J = \Ic_t^0$, $\theta = b_{\Ic_t^0,t}$, $a = 3/4$, $b = 2$, $\Cc = \{|\Ic_t^0|\}$; its Terms a, b, c are Terms 1a, 1b, 1c here) and applying the appropriate versions of Lemmas~\ref{lemma:1a-interventions-results}, \ref{lemma:1b-interventions-result}, and \ref{lemma:1c-interventions-result} allows us to claim for $n \notin \Ic_t^0$
\begin{align*}
    \hat{b}_{n,t} - b_{n,t} \mid \LFc &= O_p\left(\sqrt{k\log(p|\Ic_t^0|)}\left(\frac{k^{3/4}}{p^{1/4}} + k^2 \max\left\{\frac{\sqrt{|\Ic_t^0|}}{p^{3/2}}, \frac{1}{\sqrt{|\Ic_t^0|-1}}, \frac{1}{\sqrt{p}}\right\}\right)\right).
\end{align*}

\textit{Baseline Consistency:} The donor and non-donor cases together imply that for any $n \in [N]$
\begin{align}
    \label{eq:baseline-consistency-rate-invariant}
    \hat{b}_{n,t} - b_{n,t} \mid \LFc &= O_p\left(\sqrt{\log(p|\Ic_t^0|)}\left(\frac{k^{5/4}}{p^{1/4}} +k^{5/2}\max\left\{\frac{\sqrt{|\Ic_t^0|}}{p^{3/2}}, \frac{1}{\sqrt{|\Ic_t^0|-1}}, \frac{1}{\sqrt{p}}\right\}\right)\right).
\end{align}

\noindent{\bf 2. Verifying Lag-$0$ Blip Consistency:} 

For any $d \in [A]$:

\textit{Donor Set Consistency:} Consider unit $n \in \Ic^d$. Denote $X_{\Ic^d \setminus n} = X_{:, \Ic^d \setminus n}\in \Rb^{p \times |\Ic^d \setminus n|}$. We know the blip effect admits the representation
\begin{align*}
    \hat{\gamma}_{n,0}(d) - \gamma_{n,0}(d) \mid \LFc  &= \underbrace{\left\langle \hat{\phi}^{n, \Ic^d}, Y_{\Ic^d \setminus n, t^*_{\Ic^d}} \right\rangle - \left\langle \phi^{n, \Ic^d}, \mathbb{E}[Y_{\Ic^d \setminus n, t^*_{\Ic^d}} \mid \LFc] \right\rangle}_{\text{Term 1}}\\
    &+ \underbrace{\left\langle \phi^{n, \Ic^d}, b_{\Ic^d \setminus n, t^*_{\Ic^d}} \right\rangle - \left\langle\hat{\phi}^{n, \Ic^d}, \hat{b}_{\Ic^d \setminus n, t^*_{\Ic^d}}\right\rangle}_{\text{Term 2}},
\end{align*}
where $Y_{\Ic^d \setminus n, t^*_{\Ic^d}} = [(Y_{j, t_j^*})_{j \in \Ic^d \setminus n}]^{\top}$ and $\hat{\phi}^{n, \Ic^d}$ are the regression coefficients from regressing additional covariates  $X_n \in \Rb^p$ on the rank-$k_{\Ic^d \setminus n}$ approximation of $X_{\Ic^d \setminus n}$ with $k_{\Ic^d \setminus n} = \text{rank}(\EE[X_{\Ic^d \setminus n}])$, i.e., doing PCR with parameter $k_{\Ic^d \setminus n}$.\footnote{The vectorized baseline term is defined similarly to the outcome, as shown above.}

\textit{Bounding Term $1$:} This argument is nearly identical to that for \textit{Donor Set Baseline Consistency}.

\begin{lemma}\label{lemma:project-beta-b-donot-terminal-invariant}
    We have that
     $$\left\langle \phi^{n, \Ic^d}, \mathbb{E}[Y_{\Ic^d \setminus n, t^*_{\Ic^d}}] \right\rangle = \left\langle \tilde{\phi}^{n, \Ic^d}, \mathbb{E}[Y_{\Ic^d \setminus n, t^*_{\Ic^d}}] \right\rangle$$
     with $\tilde{\phi}^{n, \Ic^d} = VV^{\top}{\phi}^{n, \Ic^d}$, where $V$ denotes the right singular vectors of $\EE[X_{ \Ic^d\setminus n}]$.
\end{lemma}
\begin{proof}
    It would suffice to prove that
    $$VV^{\top}\EE[Y_{\Ic^d \setminus n, t^*_{\Ic^d}} ] = \EE[Y_{\Ic^d \setminus n, t^*_{\Ic^d}} ],$$
    which is equivalent to $\EE[Y_{\Ic^d \setminus n, t^*_{\Ic^d}}]^{\top}$ being in the row space of $\EE[X_{\Ic^d \setminus n}]$. By Assumption~\ref{assumption:lti-row-space} there exists $\xi^{(d,0,0)}$ (the member $t = 0$, $\ell = 0$ of its second display) such that for any $j \in \Ic^d \setminus n$
    $$\Ex[Y_{j,t_j^*}|\LFc, j \in \Ic^d \setminus n] = \sum_{i = 1}^p \xi^{(d,0,0)}_i \cdot \Ex[(X_{\Ic^d\setminus n})_{ij}|\LFc, j \in \Ic^d \setminus n].$$
    This concludes the proof.
\end{proof}
Using Lemma~\ref{lemma:project-beta-b-donot-terminal-invariant}, we can once again use the proof technique in \citet[Appendix C, Theorem 2]{SI} to show consistency of
\begin{align}\label{eq:donor-terminal-blip-consistency-term-1-invariant}
    \text{Term 1} &= \left\langle \hat{\phi}^{n, \Ic^d}, Y_{\Ic^d \setminus n, t^*_{\Ic^d}} \right\rangle - \left\langle \tilde{\phi}^{n, \Ic^d}, \mathbb{E}[Y_{\Ic^d \setminus n, t^*_{\Ic^d}}] \right\rangle\\
    &= O_p \left( \sqrt{\log(p |\Ic^d|)} \left[ \frac{{k}^{3/4}}{p^{1/4}} + {k}^2 \max \left\{ \frac{\sqrt{|\Ic^d|}}{p^{3/2}}, \frac{1}{\sqrt{p}}, \frac{1}{\sqrt{|\Ic^d|-1}} \right\} \right] \right).\nonumber
\end{align}

\textit{Bounding Term 2:} 
\begin{lemma}\label{lemma:baseline_terminal_row_space_donor-invariant}
We have 
$$\left\langle \phi^{n, \Ic^d},b_{\Ic^d\setminus n,t^*_{\Ic^d}} \right\rangle = \left\langle \tilde{\phi}^{n, \Ic^d},b_{\Ic^d\setminus n,t^*_{\Ic^d}}\right\rangle$$
with $\tilde{\phi}^{n, \Ic^d} = VV^{\top}\phi^{n, \Ic^d}$, where $V$ denotes the right singular vectors of $\EE[X_{ \Ic^d\setminus n}]$. 
\end{lemma}
\begin{proof}
     It would suffice to prove that
    $$VV^{\top}b_{\Ic^d\setminus n,t^*_{\Ic^d}}= b_{\Ic^d\setminus n,t^*_{\Ic^d}},$$
    which is equivalent to $(b_{\Ic^d\setminus n,t^*_{\Ic^d}})^{\top}$ being in the row space of $\EE[X_{\Ic^d \setminus n}]$. Applying the second display of Assumption~\ref{assumption:lti-row-space} with $t = 0$ and $\ell = t + 1 = 1$---which is the all-control path, since $\bar{D}^{t_j^*-1}_j = \overbar{\tilde{0}}^{t_j^*-1}$ for $j \in \Ic^d$---we know for any $j \in \Ic^d \setminus n$
    $$b_{j,t_j^* } =  \EE\left[Y_{j, t_j^*}^{(\overbar{\tilde{0}}^{t_j^*})}\big|\LFc, j \in \Ic^d \setminus n\right] = \sum_{i = 1}^p \xi_i^{(d,0,1)}\cdot \EE[(X_{\Ic^d \setminus n})_{ij}\mid \LFc, j \in \Ic^d \setminus n].$$
    This concludes the proof.

\end{proof}

Using Lemma~\ref{lemma:baseline_terminal_row_space_donor-invariant} we can write
$$\left\langle \phi^{n, \Ic^d},b_{\Ic^d \setminus n, t^*_{\Ic^d}}\right\rangle - \left\langle \hat{\phi}^{n, \Ic^d},\hat{b}_{\Ic^d \setminus n, t^*_{\Ic^d}} \right \rangle = \left\langle \tilde{\phi}^{n, \Ic^d},b_{\Ic^d \setminus n, t^*_{\Ic^d}} \right\rangle - \left\langle \hat{\phi}^{n, \Ic^d},\hat{b}_{\Ic^d \setminus n, t^*_{\Ic^d}} \right \rangle.$$
Next we negate the RHS and decompose as follows:\footnote{The negation is used primarily for convenience as it makes no difference in the final rate.}
\begin{align*}
    \left\langle \hat{\phi}^{n, \Ic^d},\hat{b}_{\Ic^d \setminus n, t^*_{\Ic^d}} \right \rangle& - \left\langle \tilde{\phi}^{n, \Ic^d},b_{\Ic^d \setminus n, t^*_{\Ic^d}} \right\rangle  \\
    &= \underbrace{\left\langle  \tilde{\phi}^{n, \Ic^d}, \eta_{\Ic^d \setminus n}\right\rangle}_{\text{Term 1a}} + \underbrace{\left\langle \Delta_{n, \Ic^d}, \eta_{\Ic^d \setminus n}\right\rangle}_{\text{Term 1b}} + \underbrace{\left\langle \Delta_{n, \Ic^d} , b_{\Ic^d \setminus n, t^*_{\Ic^d}}\right\rangle}_{\text{Term 1c}},
\end{align*}
where $\eta_{\Ic^d \setminus n}  = \hat{b}_{\Ic^d \setminus n, t^*_{\Ic^d}} - b_{\Ic^d \setminus n, t^*_{\Ic^d}}$ and $\Delta_{n, \Ic^d} = \hat{\phi}^{n, \Ic^d} - \tilde{\phi}^{n, \Ic^d}$. Using the argument of Lemma~\ref{lemma:pcr-rate-transfer} by applying the appropriate versions of Lemmas~\ref{lemma:1a-interventions-results}, \ref{lemma:1b-interventions-result}, and \ref{lemma:1c-interventions-result} alongside Equation~\ref{eq:baseline-consistency-rate-invariant} for Terms 1a, 1b, and 1c respectively allows us to claim
\begin{align}\label{eq:donor-terminal-consistency-term-2}
    -\text{Term 2} &=  \left\langle \hat{\phi}^{n, \Ic^d},\hat{b}_{\Ic^d \setminus n, t^*_{\Ic^d}} \right \rangle - \left\langle \tilde{\phi}^{n, \Ic^d},b_{\Ic^d \setminus n, t^*_{\Ic^d}} \right\rangle\\
    &= O_p \left( \sqrt{\log(p \pi_{\Ic})} \left[ \frac{{k}^{7/4}}{p^{1/4}} + {k}^3 \max \left\{ \frac{\sqrt{\pi_{\Ic}}}{p^{3/2}}, \frac{1}{\sqrt{p}}, \frac{1}{\sqrt{\alpha_{\Ic}-1}} \right\} \right] \right),\nonumber  
\end{align}
where $\pi_{\Ic} = \max\{|\Ic_1^0|,|\Ic^d|\}$ and $\alpha_{\Ic} = \min\{|\Ic_T^0|,|\Ic^d|\}$. To be precise, both collections of donor sets above should include $(\Ic_t^0)_{t\in [T]}$, but note that $\Ic^0_T \subseteq \dots \subseteq \Ic^0_1$.

Combining Term $1$ and $2$ rates, we find for any $n \in \Ic^d$
\begin{equation}\label{eq:donor-terminal-rate-invariant}
\hat{\gamma}_{n,0}(d) - \gamma_{n,0}(d) \mid \LFc = O_p\left(\sqrt{\log(p\pi_{\Ic})}\left(\frac{k^{7/4}}{p^{1/4}} +k^{3}\max\left\{\frac{\sqrt{\pi_{\Ic}}}{p^{3/2}}, \frac{1}{\sqrt{\alpha_{\Ic}-1}}, \frac{1}{\sqrt{p}}\right\}\right)\right),
\end{equation}
where $\pi_{\Ic} = \max\{|\Ic_1^0|,|\Ic^d|\}$ and $\alpha_{\Ic} = \min\{|\Ic_T^0|,|\Ic^d|\}$. 

\textit{Non-Donor Set Consistency:} Consider unit $n \notin \Ic^d$. Denote $X_{\Ic^d} = X_{:, \Ic^d} \in \Rb^{p \times |\Ic^d|}$. We know the blip effect admits the representation
$$\hat{\gamma}_{n,0}(d) - \gamma_{n,0}(d) \mid \LFc  = \left\langle \hat{\beta}^{n, \Ic^d}, \hat{\gamma}_{\Ic^d, 0}(d) \right\rangle - \left\langle \beta^{n, \Ic^d}, \gamma_{\Ic^d, 0}(d) \right\rangle,$$
where $\hat{\beta}^{n, \Ic^d}$ are the regression coefficients from regressing additional covariates  $X_n \in \Rb^p$ on the rank-$k_{\Ic^d }$ approximation of $X_{\Ic^d }$ with $k_{\Ic^d } = \text{rank}(\EE[X_{\Ic^d}])$, i.e., doing PCR with parameter $k_{\Ic^d }$.

We use an essentially identical argument to that established in \textit{Non-Donor Set Baseline Consistency}.

\begin{lemma}\label{lemma:project-beta-terminal-blip-non-donor-invariant}
    We have that
     $$\left\langle \beta^{n, \Ic^d}, \gamma_{\Ic^d, 0}(d) \right\rangle = \left\langle \tilde{\beta}^{n, \Ic^d}, \gamma_{\Ic^d, 0}(d) \right\rangle$$
     with $\tilde{\beta}^{n, \Ic^d} = VV^{\top}{\beta}^{n, \Ic^d}$, where $V$ denotes the right singular vectors of $\EE[X_{ \Ic^d}]$. 
\end{lemma}
\begin{proof}
    It would suffice to prove that
    $$VV^{\top}\gamma_{\Ic^d, 0}(d)  = \gamma_{\Ic^d, 0}(d) ,$$
    which is equivalent to $\gamma_{\Ic^d, 0}(d) ^{\top}$ being in the row space of $\EE[X_{\Ic^d}]$. 
    To that end, recall for any $j \in \Ic^d$
    \begin{align*}
        \gamma_{j, 0}(d)  &= \langle \psi_j^0, w_d - w_{\tilde{0}}\rangle\\
        &= \EE\left[Y_{j, t_j^* }^{(\bar{D}^{t_j^*}_j)}\right] - \EE\left[Y_{j, t_j^*}^{(\bar{D}^{t_j^*-1}_j, \overbar{\tilde{0}}^{\,1})}\right]\\
        &= \EE\left[Y_{j, t_j^*} - Y_{j, t_j^*}^{(\overbar{\tilde{0}}^{t_j^*})} \big| j \in \Ic^d\right]\\
        &= \sum_{i = 1}^p (\xi_i^{(d,0,0)} - \xi_i^{(d,0,1)})\cdot\EE[(X_{\Ic^d})_{ij}|\LFc,j \in \Ic^d].
    \end{align*}
    The first two equalities follow by the definition of blips; the third follows from $\bar{D}^{t_j^*}_j = (\tZero, \dots, \tZero, d)$ where $d$ occurs in the $t_j^*$ index. The last equality is due to the second display of Assumption~\ref{assumption:lti-row-space} with $t = 0$, applied with $\ell = 0$ (the observed outcome) and $\ell = 1$ (the all-control path) to each term respectively.
\end{proof}
Lemma~\ref{lemma:project-beta-terminal-blip-non-donor-invariant} allows us to write
\begin{align*}
    \hat{\gamma}_{n,0}(d) - \gamma_{n,0}(d) \mid \LFc  &= \left\langle \hat{\beta}^{n, \Ic^d}, \hat{\gamma}_{\Ic^d, 0}(d) \right\rangle - \left\langle \beta^{n, \Ic^d}, \gamma_{\Ic^d, 0}(d) \right\rangle\\
    &=  \underbrace{\langle \tilde{\beta}^{n,\Ic^d}, \eta_{\Ic^d}(d)\rangle}_{\text{Term 1a}} + \underbrace{\langle \Delta_{n, \Ic^d} , \eta_{\Ic^d}(d)\rangle}_{\text{Term 1b}} +  \underbrace{\langle \Delta_{n, \Ic^d}, \gamma_{\Ic^d, 0}(d)\rangle}_{\text{Term 1c}},
\end{align*}
where $\eta_{\Ic^d}(d) = \hat{\gamma}_{\Ic^d, 0}(d)  - \gamma_{\Ic^d, 0}(d) $ and $\Delta_{n, \Ic^d} = \hat{\beta}^{n,\Ic^d} - \tilde{\beta}^{n,\Ic^d}$. Using the argument of Lemma~\ref{lemma:pcr-rate-transfer} and applying the appropriate versions of Lemmas~\ref{lemma:1a-interventions-results}, \ref{lemma:1b-interventions-result}, and \ref{lemma:1c-interventions-result} allows us to claim for $n \notin \Ic^d$
\begin{equation}\label{eq:non-donor-terminal-rate-invariant}
\hat{\gamma}_{n,0}(d) - \gamma_{n,0}(d) \mid \LFc = O_p\left(\sqrt{\log(p\pi_{\Ic})}\left(\frac{k^{9/4}}{p^{1/4}} +k^{7/2}\max\left\{\frac{\sqrt{\pi_{\Ic}}}{p^{3/2}}, \frac{1}{\sqrt{\alpha_{\Ic}-1}}, \frac{1}{\sqrt{p}}\right\}\right)\right),
\end{equation}
where $\pi_{\Ic} = \max\{|\Ic_1^0|,|\Ic^d|\}$ and $\alpha_{\Ic} = \min\{|\Ic_T^0|,|\Ic^d|\}$.

\textit{Lag-$0$ Blip Consistency:} The above two cases allow us to conclude that for any $n \in [N]$
\begin{equation}\label{eq:terminal-rate-invariant}
\hat{\gamma}_{n,0}(d) - \gamma_{n,0}(d) \mid \LFc = O_p\left(\sqrt{\log(p\pi_{\Ic})}\left(\frac{k^{9/4}}{p^{1/4}} +k^{7/2}\max\left\{\frac{\sqrt{\pi_{\Ic}}}{p^{3/2}}, \frac{1}{\sqrt{\alpha_{\Ic}-1}}, \frac{1}{\sqrt{p}}\right\}\right)\right),
\end{equation}
where $\pi_{\Ic} = \max\{|\Ic_1^0|,|\Ic^d|\}$ and $\alpha_{\Ic} = \min\{|\Ic_T^0|,|\Ic^d|\}$.

\noindent{\bf 3. Verifying Higher-Lag Blip Consistency:} 

For any unit $n \in [N]$, treatment $d \in [A]$, and $t \in [1, T-1]$, consider the statement $P_{d,n}(t)$:
\begin{align*}
    &\hat{\gamma}_{n, t}(d) - \gamma_{n, t}(d) \mid \LFc\\
    &= O_p\left(t\sqrt{\log(p\pi_{\Ic})}\left(\frac{k^{t+9/4}}{p^{1/4}} + k^{t+7/2}\max\left\{\frac{\sqrt{\pi_{\Ic}}}{p^{3/2}}, \frac{1}{\sqrt{\alpha_{\Ic}-1}}, \frac{1}{\sqrt{p}}\right\}\right)\right),
\end{align*}
where $\Cc = \{|\Ic^d|, |\Ic_T^0|, |\Ic_1^0|, (|\Ic^{D_{n, t_n^* + q}}|)_{n \in [N], q\in[1, t]} \}$ with $\pi_{\Ic} = \max\Cc, \alpha_{\Ic} = \min\Cc$.

We proceed by strong induction. The donor-set decomposition and the bounds on its first three terms are common to the base case and the inductive step, so we derive them once for general $t \in [T-1]$ and any $d \in [A]$.

\textit{Donor Set Decomposition:} Consider unit $n \in \Ic^d$ and write $\theta := [(\gamma_{j, t}(d))_{j \in \Ic^d \setminus n}]^{\top}$ and $\varepsilon^{(t)} := (\varepsilon^{(t)}_j)_{j \in \Ic^d \setminus n}$ for the realized-path noise introduced after Theorem~\ref{thm:consistency-time-invariant}, so that the blipped-down donor outcomes used in Step~3 of \texttt{SBE-PCR} (SA~\ref{app:lti-algorithm}) satisfy $Y_{\Ic^d \setminus n, t^*_{\Ic^d} + t} - \hat{b}_{\Ic^d \setminus n, t^*_{\Ic^d} + t} - \sum_{\ell = 0}^{t-1} \hat{\gamma}_{\Ic^d \setminus n, \ell}(D_{\Ic^d \setminus n, t^*_{\Ic^d} + t - \ell}) = \theta + \varepsilon^{(t)} - \eta_b - \sum_{\ell = 0}^{t-1} \eta_\ell$, where $\eta_b := \hat{b}_{\Ic^d \setminus n, t^*_{\Ic^d} + t} - b_{\Ic^d \setminus n, t^*_{\Ic^d} + t}$, $\eta_\ell := \hat{\gamma}_{\Ic^d \setminus n, \ell}(D_{\Ic^d \setminus n, t^*_{\Ic^d} + t - \ell}) - \gamma_{\Ic^d \setminus n, \ell}(D_{\Ic^d \setminus n, t^*_{\Ic^d} + t - \ell})$, and $\gamma_{\Ic^d \setminus n, \ell}(D_{\Ic^d \setminus n, t^*_{\Ic^d} + t - \ell}) = [(\gamma_{j, \ell}(D_{j, t^*_j + t - \ell}))_{j \in \Ic^d \setminus n}]^{\top}$ (and analogously for $\hat{\gamma}$). Since $\gamma_{n,t}(d) = \langle \phi^{n, \Ic^d}, \theta \rangle$ by Assumption~\ref{control-factors-support-LTI}, the estimator admits the representation
\begin{align*}
\hat{\gamma}_{n,t}(d) - \gamma_{n,t}(d) \mid \LFc  &= \underbrace{\left\langle \hat{\phi}^{n, \Ic^d}, \theta + \varepsilon^{(t)} \right\rangle - \left\langle \phi^{n, \Ic^d}, \theta \right\rangle}_{\text{Term 1}} \\
&\quad - \underbrace{\left\langle \hat{\phi}^{n, \Ic^d}, \eta_b \right\rangle}_{\text{Term 2}}
- \underbrace{\left\langle \hat{\phi}^{n, \Ic^d}, \eta_0 \right\rangle}_{\text{Term 3}}
- \sum_{\ell = 1}^{t-1} \underbrace{\left\langle \hat{\phi}^{n, \Ic^d}, \eta_\ell \right\rangle}_{\text{Term } \ell},
\end{align*}
where $\hat{\phi}^{n, \Ic^d}$ are the regression coefficients from regressing additional covariates $X_n \in \Rb^p$ on the rank-$k_{\Ic^d \setminus n}$ approximation of $X_{\Ic^d \setminus n} = X_{:, \Ic^d \setminus n}\in \Rb^{p \times |\Ic^d \setminus n|}$, with $k_{\Ic^d \setminus n} = \text{rank}(\EE[X_{\Ic^d \setminus n}])$, and the summation is empty when $t = 1$.

\textit{Bounding Term 1:} We first prove a row space result.
\begin{lemma}\label{eq:project-phi-t-1-blip-donor-invariant}
    We have for any $t \in [T-1]$
     $$\left\langle \phi^{n, \Ic^d}, \gamma_{\Ic^d \setminus n, t}(d) \right\rangle = \left\langle \tilde{\phi}^{n, \Ic^d}, \gamma_{\Ic^d \setminus n, t}(d) \right\rangle$$
     with $\tilde{\phi}^{n, \Ic^d} = VV^{\top}\phi^{n, \Ic^d}$, where $V$ denotes the right singular vectors of $\EE[X_{ \Ic^d\setminus n}]$.
\end{lemma}
\begin{proof}
    It suffices to prove that $VV^{\top}\gamma_{\Ic^d \setminus n, t}(d) = \gamma_{\Ic^d \setminus n, t}(d)$, i.e., that $\gamma_{\Ic^d \setminus n, t}(d)^{\top}$ lies in the row space of $\EE[X_{\Ic^d \setminus n}]$. This is shown exactly as in the proof of Lemma~\ref{lemma:project-beta-blip-non-donor-invariant} below, with $\Ic^d$ replaced by $\Ic^d \setminus n$ (for which Assumption~\ref{assumption:lti-row-space} also holds, by its footnote): for $j \in \Ic^d \setminus n$, $\gamma_{j, t}(d)$ is the difference of the members $\ell = t$ and $\ell = t + 1$ of the second display of that assumption.
\end{proof}
By Lemma~\ref{eq:project-phi-t-1-blip-donor-invariant}, $\langle \phi^{n, \Ic^d}, \theta \rangle = \langle \tilde{\phi}^{n, \Ic^d}, \theta \rangle$; since $\theta + \varepsilon^{(t)}$ has conditional mean $\theta$ in the row space, with $\|\theta\|_{\infty} \leq 2$ by Assumption~\ref{assumption:bounded-expected-potential-outcomes}(ii) (each entry is a difference of two expected potential outcomes), and $\varepsilon^{(t)}$ consists of independent mean-zero sub-Gaussian noise (see the paragraph following Theorem~\ref{thm:consistency-time-invariant}), the argument that yielded Equation~\eqref{eq:donor-baseline-consistency-invariant} applies with $Y_{\Ic_t^0 \setminus n, t}$ replaced by $\theta + \varepsilon^{(t)}$, and
\begin{align}\label{eq:donor-non-terminal-consistency-term-1-invariant}
    \text{Term 1} &= \left\langle \hat{\phi}^{n, \Ic^d}, \theta + \varepsilon^{(t)} \right\rangle - \left\langle \tilde{\phi}^{n, \Ic^d}, \theta \right\rangle\\
    &= O_p \left( \sqrt{\log(p |\Ic^d|)} \left[ \frac{{k}^{3/4}}{p^{1/4}} + {k}^2 \max \left\{ \frac{\sqrt{|\Ic^d|}}{p^{3/2}}, \frac{1}{\sqrt{p}}, \frac{1}{\sqrt{|\Ic^d|-1}} \right\} \right] \right).\nonumber
\end{align}

\textit{Bounding Term 2:} Writing $\hat{\phi}^{n, \Ic^d} = \tilde{\phi}^{n, \Ic^d} + \Delta_{n, \Ic^d}$ with $\Delta_{n, \Ic^d} = \hat{\phi}^{n, \Ic^d} - \tilde{\phi}^{n, \Ic^d}$, Term $2$ equals Terms a and b of Lemma~\ref{lemma:pcr-rate-transfer} (in its version for the donor set $\Ic^d \setminus n$) with $\eta = \eta_b$, whose entrywise rate is given by Equation~\eqref{eq:baseline-consistency-rate-invariant} ($a = 5/4$, $b = 5/2$; the cardinalities $|\Ic^0_{t^*_j + t}|$ at the different dates are bounded using $\Ic^0_T \subseteq \dots \subseteq \Ic^0_1$); no row-space condition is needed since Term c does not arise. Hence
\begin{align}\label{eq:donor-non-terminal-consistency-term-2-invariant}
    \text{Term 2} &= O_p \left( \sqrt{\log(p \pi_{\Ic})} \left[ \frac{{k}^{7/4}}{p^{1/4}} + {k}^3 \max \left\{ \frac{\sqrt{\pi_{\Ic}}}{p^{3/2}}, \frac{1}{\sqrt{p}}, \frac{1}{\sqrt{\alpha_{\Ic}-1}} \right\} \right] \right),
\end{align}
where $\pi_{\Ic} = \max\{|\Ic_1^0|,|\Ic^d|\}$ and $\alpha_{\Ic} = \min\{|\Ic_T^0|,|\Ic^d|\}$.

\textit{Bounding Term 3:} Likewise, Term $3$ equals Terms a and b of Lemma~\ref{lemma:pcr-rate-transfer} with $\eta = \eta_0$, whose entries are bounded by $\max_{a \in \{D_{i, t^*_i + t}\}_{i \in [N]}} |\hat{\gamma}_{j,0}(a) - \gamma_{j,0}(a)|$, a maximum over at most $A$ terms each obeying Equation~\eqref{eq:terminal-rate-invariant} ($a = 9/4$, $b = 7/2$). Hence
\begin{align*}
\text{Term }3
&= O_p\left(\sqrt{\log(p\pi_{\Ic})}\left(\frac{k^{11/4}}{p^{1/4}} +k^{4}\max\left\{\frac{\sqrt{\pi_{\Ic}}}{p^{3/2}}, \frac{1}{\sqrt{\alpha_{\Ic}-1}}, \frac{1}{\sqrt{p}}\right\}\right)\right),
\end{align*}
where $\pi_{\Ic} = \max\{|\Ic_1^0|,|\Ic^d|, (|\Ic^{D_{n,t^*_n + t}}|)_{n \in [N]}\}$ and $\alpha_{\Ic} = \min\{|\Ic_T^0|,|\Ic^d|, (|\Ic^{D_{n,t^*_n + t}}|)_{n \in [N]}\}$; this dominates the rates for Terms $1$ and $2$.

\textbf{Base case $t = 1$, i.e., proving $P_{d,n}(1)$:}

\textit{Donor Set Consistency:} Here the summation is empty, so combining Terms $1$--$3$ yields for any $n \in \Ic^d$
\begin{align}\label{eq:donor-base-blip-rate-invariant}
\hat{\gamma}_{n,1}(d) - \gamma_{n,1}(d) \mid \LFc
&= O_p\left(\sqrt{\log(p\pi_{\Ic})}\left(\frac{k^{11/4}}{p^{1/4}} +k^{4}\max\left\{\frac{\sqrt{\pi_{\Ic}}}{p^{3/2}}, \frac{1}{\sqrt{\alpha_{\Ic}-1}}, \frac{1}{\sqrt{p}}\right\}\right)\right),
\end{align}
where $\pi_{\Ic} = \max\{|\Ic_1^0|,|\Ic^d|, (|\Ic^{D_{n,t^*_n + 1}}|)_{n \in [N]}\}$ and $\alpha_{\Ic} = \min\{|\Ic_T^0|,|\Ic^d|, (|\Ic^{D_{n,t^*_n + 1}}|)_{n \in [N]}\}$.

\textit{Non-Donor Set Consistency:} Consider any $t \in [T-1]$ and unit $n \notin \Ic^d$. Denote $X_{\Ic^d} = X_{:,\Ic^d} \in \Rb^{p \times |\Ic^d|}$. We know the blip effect admits the representation
$$\hat{\gamma}_{n,t}(d) - \gamma_{n,t}(d) \mid \LFc  = \left\langle \hat{\beta}^{n, \Ic^d}, \hat{\gamma}_{\Ic^d, t}(d) \right\rangle - \left\langle \beta^{n, \Ic^d}, \gamma_{\Ic^d, t}(d) \right\rangle,$$
where $\hat{\beta}^{n, \Ic^d}$ are the regression coefficients from regressing additional covariates  $X_n \in \Rb^p$ on the rank-$k_{\Ic^d }$ approximation of $X_{\Ic^d }$ with $k_{\Ic^d } = \text{rank}(\EE[X_{\Ic^d}])$, i.e., doing PCR with parameter $k_{\Ic^d }$.

We use an identical argument to that established in \textit{Non-Donor Set Baseline Consistency}.

\begin{lemma}\label{lemma:project-beta-blip-non-donor-invariant}
    We have for any $t \in [T-1]$ that
     $$\left\langle \beta^{n, \Ic^d}, \gamma_{\Ic^d, t}(d) \right\rangle = \left\langle \tilde{\beta}^{n, \Ic^d}, \gamma_{\Ic^d, t}(d) \right\rangle$$
     with $\tilde{\beta}^{n, \Ic^d} = VV^{\top}{\beta}^{n, \Ic^d}$, where $V$ denotes the right singular vectors of $\EE[X_{ \Ic^d}]$.
\end{lemma}
\begin{proof} 
    It would suffice to prove that
    $$VV^{\top} \gamma_{\Ic^d, t}(d)  =  \gamma_{\Ic^d, t}(d),$$
    which is equivalent to $\gamma_{\Ic^d, t}(d)^{\top}$ being in the row space of $\EE[X_{\Ic^d}]$. To that end, recall for any $j \in \Ic^d$,
    \begin{align*}
        \gamma_{j, t}(d)
        &= \langle \psi_j^t, w_d - w_{\tilde{0}} \rangle\\
        &= \EE\left[
        Y_{j,t_j^*+t}^{(\bar D_j^{t_j^*},
        \overbar{\tilde{0}}^{\,t})}
        \,\middle|\, \LFc, j \in \Ic^d
        \right]
        -
        \EE\left[
        Y_{j,t_j^*+t}^{(\bar D_j^{t_j^*-1},
        \overbar{\tilde{0}}^{\,t+1})}
        \,\middle|\, \LFc, j \in \Ic^d
        \right]\\
        &= \sum_{i = 1}^p
        \left(
        \xi_i^{(d,t,t)}
        -
        \xi_i^{(d,t,t+1)}
        \right)
        \EE[(X_{\Ic^d})_{ij}|\LFc,j \in \Ic^d].
    \end{align*}
    The first two equalities follow by the definition of blips. The last equality follows from the second display of Assumption~\ref{assumption:lti-row-space}, applied with $\ell = t$ and $\ell = t+1$. Since $\bar D_j^{t_j^*}=(\tZero,\dots,\tZero,d)$ for $j\in\Ic^d$, the first counterfactual corresponds to treatment $d$ at $t_j^*$ followed by control, while the second replaces the treatment at $t_j^*$ by control and continues with control thereafter. Because $\xi^{(d,t,t)}$ and $\xi^{(d,t,t+1)}$ are common across all $j\in\Ic^d$, it follows that $\gamma_{\Ic^d,t}(d)^{\top}$ lies in the row space of $\EE[X_{\Ic^d}]$, proving the desired result.
\end{proof}
Using the above framework and Lemma~\ref{lemma:project-beta-blip-non-donor-invariant} with $t = 1$ allows us to write
\begin{align*}
    \hat{\gamma}_{n,1}(d) - \gamma_{n,1}(d) \mid \LFc  &= \left\langle \hat{\beta}^{n, \Ic^d}, \hat{\gamma}_{\Ic^d, 1}(d) \right\rangle - \left\langle \tilde{\beta}^{n, \Ic^d}, \gamma_{\Ic^d, 1}(d) \right\rangle\\
    &=  \underbrace{\langle \tilde{\beta}^{n, \Ic^d}, \eta_{\Ic^d}(d)\rangle}_{\text{Term 1a}} + \underbrace{\langle \Delta_{n, \Ic^d} , \eta_{\Ic^d}(d)\rangle}_{\text{Term 1b}} +  \underbrace{\langle \Delta_{n, \Ic^d}, \gamma_{\Ic^d, 1}(d)\rangle}_{\text{Term 1c}},
\end{align*}
where $\eta_{\Ic^d}(d) = \hat{\gamma}_{\Ic^d, 1}(d)  - \gamma_{\Ic^d, 1}(d) $ and $\Delta_{n, \Ic^d} = \hat{\beta}^{n,\Ic^d} - \tilde{\beta}^{n,\Ic^d}$. Using the argument of Lemma~\ref{lemma:pcr-rate-transfer} by applying the appropriate versions of Lemmas~\ref{lemma:1a-interventions-results}, \ref{lemma:1b-interventions-result}, and \ref{lemma:1c-interventions-result} allows us to claim for $n \notin \Ic^d$
\begin{equation}\label{eq:non-donor-second-to-terminal-rate-invariant}
\hat{\gamma}_{n,1}(d) - \gamma_{n,1}(d) \mid \LFc = O_p\left(\sqrt{\log(p\pi_{\Ic})}\left(\frac{k^{13/4}}{p^{1/4}} +k^{9/2}\max\left\{\frac{\sqrt{\pi_{\Ic}}}{p^{3/2}}, \frac{1}{\sqrt{\alpha_{\Ic}-1}}, \frac{1}{\sqrt{p}}\right\}\right)\right),
\end{equation}
where $\pi_{\Ic} = \max\{|\Ic_1^0|,|\Ic^d|, (|\Ic^{D_{n,t^*_n + 1}}|)_{n \in [N]}\}$ and $\alpha_{\Ic} = \min\{|\Ic_T^0|,|\Ic^d|, (|\Ic^{D_{n,t^*_n + 1}}|)_{n \in [N]}\}$. Combining Equations~\ref{eq:donor-base-blip-rate-invariant} and \ref{eq:non-donor-second-to-terminal-rate-invariant} yields the base case.

\textbf{Inductive Step:} We assume $P_{d,n}(\ell)$ for $\ell \in [1, t-1]$ and prove $P_{d,n}(t)$.

For any $d \in [A]$:

\textit{Donor Set Consistency:} Consider unit $n \in \Ic^d$. In the decomposition above, Terms $1$--$3$ are already bounded, so only the summation remains. For any $\ell \in [1, t-1]$, Term $\ell = \langle \hat{\phi}^{n, \Ic^d}, \eta_\ell \rangle$ equals Terms a and b of Lemma~\ref{lemma:pcr-rate-transfer} with $\eta = \eta_\ell$, whose entrywise rate is supplied by the inductive hypothesis $P_{a, j}(\ell)$ for $a \in \{D_{i, t^*_i + t - \ell}\}_{i \in [N]}$ (a maximum over at most $A$ actions), so that
\begin{align}\label{eq:donor-terml-invariant}
\text{Term }\ell
&= O_p\left(\ell\sqrt{\log(p\pi_{\Ic})}\left(\frac{k^{\ell+11/4}}{p^{1/4}} + k^{\ell+4}\max\left\{\frac{\sqrt{\pi_{\Ic}}}{p^{3/2}}, \frac{1}{\sqrt{\alpha_{\Ic}-1}}, \frac{1}{\sqrt{p}}\right\}\right)\right),
\end{align}
where $\Cc = \{|\Ic_T^0|, |\Ic_1^0|, |\Ic^d| ,(|\Ic^{D_{n,t_n^* + q}}|)_{n\in [N],q \in [1, \ell]}, (|\Ic^{D_{n,t_n^* + t - \ell}}|)_{n\in [N]}\}$ with $\pi_{\Ic} = \max\Cc, \alpha_{\Ic} = \min\Cc$.

Note that Terms $1$ and $2$ are dominated by the summation; as such, it suffices to analyze the latter and Term $3$. To that end for the summation,
$$\sum_{\ell = 1}^{t-1}\text{Term }\ell = O_p\left(\sum_{\ell = 1}^{t-1}\ell\sqrt{\log(p\pi_{\Ic})}\left(\frac{k^{\ell+11/4}}{p^{1/4}} + k^{\ell+4}\max\left\{\frac{\sqrt{\pi_{\Ic}}}{p^{3/2}}, \frac{1}{\sqrt{\alpha_{\Ic}-1}}, \frac{1}{\sqrt{p}}\right\}\right)\right),$$

 where $\Cc = \{|\Ic_T^0|, |\Ic_1^0|, |\Ic^d| ,(|\Ic^{D_{n,t_n^* + q}}|)_{n\in [N],q \in [1, t-1]}\}$ with $\pi_{\Ic} = \max\Cc, \alpha_{\Ic} = \min\Cc$. Notice we bounded the smaller donor set cardinalities by the largest one, i.e., when $\ell = t-1$. We analyze the time-dependent terms and denote
\[
C := \sqrt{\log(p\pi_{\mathcal{I}})}, \quad C' := \max\left\{\frac{\sqrt{\pi_{\mathcal{I}}}}{p^{3/2}}, \frac{1}{\sqrt{\alpha_{\mathcal{I}}-1}}, \frac{1}{\sqrt{p}}\right\}.
\]
Upon substitution we have
\[
C \sum_{m = 1}^{t - 1} m \left( \frac{k^{m+11/4}}{p^{1/4}} + C'k^{m+4} \right)
= C \left( \frac{k^{11/4}}{p^{1/4}} + C'k^4 \right) \sum_{m = 1}^{t - 1} m k^m.
\]
As in the proof of Theorem~\ref{thm:consistency-time-varying}, without loss of generality $k \geq 2$, and then, uniformly in $M$,
\[
\sum_{m=1}^{M} m k^m \;\le\; M\sum_{m=1}^{M}k^m \;=\; M\,\frac{k(k^{M}-1)}{k-1} \;\le\; \frac{k}{k-1}\,Mk^{M} \;\le\; 2Mk^{M}.
\]
Taking $M = t - 1$, we conclude
\[
\sum_{\ell = 1}^{t-1}\text{Term }\ell = O_p\left(t\sqrt{\log(p\pi_{\mathcal{I}})}\left( \frac{k^{t+7/4}}{p^{1/4}} + k^{t+3}\max\left\{ \frac{\sqrt{\pi_{\mathcal{I}}}}{p^{3/2}}, \frac{1}{\sqrt{\alpha_{\mathcal{I}} - 1}}, \frac{1}{\sqrt{p}} \right\} \right)  \right).
\]

Combining this with Term $3$ yields for any $n \in \Ic^d$
\begin{align}\label{eq:inductive-step-donor-blip-rate-invariant}
&\hat{\gamma}_{n,t}(d) - \gamma_{n,t}(d) \mid \LFc\\
&= O_p\left(t\sqrt{\log(p\pi_{\Ic})}\left(\frac{k^{t+7/4}}{p^{1/4}} + k^{t+3}\max\left\{\frac{\sqrt{\pi_{\Ic}}}{p^{3/2}}, \frac{1}{\sqrt{\alpha_{\Ic}-1}}, \frac{1}{\sqrt{p}}\right\}\right)\right),\nonumber
\end{align}
where $\Cc = \{|\Ic_T^0|, |\Ic_1^0|, |\Ic^d| ,(|\Ic^{D_{n,t_n^* + q}}|)_{n\in [N],q \in [1, t]}\}$ with $\pi_{\Ic} = \max\Cc, \alpha_{\Ic} = \min\Cc$.

\textit{Non-Donor Set Consistency:} Applying the \textit{Non-Donor Set Consistency} argument written for the base case for general $t$, specifically Lemma~\ref{lemma:project-beta-blip-non-donor-invariant} for any $t \in [T-1]$, introduces the same additional factor $k^{1/2}$ as in the base case and therefore proves $P_{d,n}(t)$.

\noindent{\bf 4. Verifying Target Causal Parameter Consistency:} 
For any unit $n \in [N]$ and $\bar{d}^T \in [A]^T$ we recall the \texttt{SBE-PCR} estimator and the corresponding causal estimand.
$$\hEx\left[Y_{n,T}^{(\bar{d}^T)}\right] = \sum_{t = 1}^T \hat{\gamma}_{n,T-t}(d_t) + \hat{b}_{n,T} \quad \text{and} \quad \mathbb{E}\left[Y_{n,T}^{(\bar{d}^T)} \mid \LFc \right] = \sum_{t=1}^T \gamma_{n,T-t}(d_t) + b_{n,T} \mid \LFc.$$
The difference is exactly
$$\hEx\left[Y_{n,T}^{(\bar{d}^T)}\right] - \mathbb{E}\left[Y_{n,T}^{(\bar{d}^T)} \mid \LFc \right] = \left( \hat{b}_{n,T} - b_{n,T} \mid \LFc \right) +
\sum_{t=1}^T \left(\hat{\gamma}_{n,T-t}(d_t) - \gamma_{n,T-t}(d_t) \mid \LFc\right).$$

We apply the known bound for each term, specifically Equation~\ref{eq:baseline-consistency-rate-invariant}, Equation~\ref{eq:terminal-rate-invariant} with $d = d_T$, and $P_{d_{t},n}(T-t)$ for every $t \in [T-1]$. Once again we encounter the same geometric sum, which gives the desired result upon noting that the baseline rate
is dominated by that of the sum.

\subsection{Proof of Theorem~\ref{thm:consistency-time-invariant-fixed-lags}}\label{subsection:proof-of-invariant-consistency-fixed-lags}

We recall that for any unit $n \in [N]$ and $\bar{d}^T \in [A]^T$

$$\mathbb{E}\left[Y_{n,T}^{(\bar{d}^T)} \mid \LFc \right] = \sum_{t=1}^T \gamma_{n,T-t}(d_t) + b_{n,T} \mid \LFc = \sum_{t=1}^T \langle \psi_n^{T-t}, w_{d_t} - w_{\tZero}\rangle + b_{n,T} \mid \LFc.$$

Given Assumption~\ref{assumption:outcome-fixed-lag-dep-invariant} we know that $\psi_{n}^{q+i} = 0$ for all $i \in [T-q-1]$. As such, 
$$\mathbb{E}\left[Y_{n,T}^{(\bar{d}^T)} \mid \LFc \right]  = \sum_{t=T-q}^T \langle \psi_n^{T-t}, w_{d_t} - w_{\tZero}\rangle + b_{n,T} \mid \LFc = \sum_{t=T-q}^T \gamma_{n,T-t}(d_t) + b_{n,T} \mid \LFc.$$

We modify the \texttt{SBE-PCR} estimator accordingly:
$$\hEx\left[Y_{n,T}^{(\bar{d}^T)}\right] := \sum_{t = T-q}^T \hat{\gamma}_{n,T-t}(d_t) + \hat{b}_{n,T}.$$

Applying the analysis from the proof of Theorem~\ref{thm:consistency-time-invariant} yields the desired result.


\subsection{From Pointwise to Uniform and Average Consistency}\label{app:uniform-consistency}

This subsection concerns uniformity over target units, which is
a further extension beyond the simultaneous bounds over donors
needed within the proofs of pointwise consistency.

For either Theorem~\ref{thm:consistency-time-varying-fixed-lags}
or~\ref{thm:consistency-time-invariant-fixed-lags}, fix a feasible
treatment schedule $\bar d^T$ and define
\[
e_n(\bar d^T)
:=
\hEx[Y_{n,T}^{(\bar d^T)}]
-\mathbb E[Y_{n,T}^{(\bar d^T)}\mid\LFc].
\]
With probabilities conditional as in these theorems, a union
bound gives, for every $\varepsilon>0$,
\[
\mathbb P\!\left(
\max_{n\le N}|e_n(\bar d^T)|>\varepsilon
\right)
\le
N\sup_{n\le N}
\mathbb P\!\left(|e_n(\bar d^T)|>\varepsilon\right).
\]
Thus, if the supremum on the right is $o(1/N)$ for every fixed
$\varepsilon>0$, then
$\max_{n\le N}|e_n(\bar d^T)|=o_p(1)$.
This also implies average consistency, since
\[
\left|\frac1N\sum_{n=1}^N e_n(\bar d^T)\right|
\le
\frac1N\sum_{n=1}^N|e_n(\bar d^T)|
\le
\max_{n\le N}|e_n(\bar d^T)|
=o_p(1).
\]
Uniform consistency is sufficient, but not necessary, for
average consistency.

The argument above gives sufficient additional tail conditions
under which the pointwise results extend to uniform and average
consistency. These extensions require additional tail bounds beyond the
stated pointwise $O_p$ rates.
One way to establish these additional bounds is to use
the explicit probability bounds in
\citet[Appendices B.4 and C]{SI} throughout the inequalities
in the proof of Lemma~\ref{lemma:pcr-rate-transfer} and the
recursive blip argument, with constants uniform over the
relevant indices.
The union bound above, or Lemma~\ref{lem:primitive_union}
for treatment gains, then yields uniform consistency provided
the sum of the failure probabilities and the resulting bound
on the maximum estimation error both vanish.
Remark~\ref{rem:tails} gives sufficient tail and growth
conditions for the treatment-gain case.

\end{document}